\documentclass[useAMS,usenatbib,paper]{mn2e}
\usepackage{times,epsfig,psfig,natbib,amssymb,amsmath,graphicx,rotating,lscape}
\usepackage[bookmarks,bookmarksnumbered,colorlinks=true, citecolor=blue, linkcolor=black]{hyperref}
\usepackage[switch]{lineno}
\usepackage{booktabs}
\usepackage{color,natbib}
\newcommand{\ltsima} {$\; \buildrel < \over \sim \;$}  
\newcommand{\gtsima} {$\; \buildrel > \over \sim \;$}  
\newcommand{\lta} {\lower.5ex\hbox{\ltsima}}  
\newcommand{\gta} {\lower.5ex\hbox{\gtsima}}  
\newcommand{\Ha} {H$\alpha$}

\def\aj{AJ}%
\def\araa{ARA\&A}%
\def\apj{ApJ}%
\def\apjl{ApJ}%
\def\apjs{ApJS}%
\def\aap{A\&A}
\def\mnras{MNRAS}%
\def\pasp{PASP}%
\def\nat{Nature}%
\newcommand{\ergs}{\>{\rm erg}\,{\rm s}^{-1}}
\newcommand{\kms}{$\rm{\,km \,s}^{-1}$}
\newcommand{\forb}[2]{\mbox{$[{\rm #1\, #2}]$}}
\newcommand{\nii}{\forb{N}{II}\,}
\title[]{Spectropolarimetry of low redshift
  Quasars: origin of the polarization and implications for black hole mass estimates.}
\author[]{Alessandro
  Capetti$^{1}$\thanks{E-mail:alessandro.capetti@inaf.it}, Ari
  Laor$^{2}$, Ranieri D. Baldi$^{3,4}$, Andrew Robinson$^{5}$,
  Alessandro Marconi$^{6}$ \\ $^{1}$INAF-Osservatorio Astrofisico di
  Torino, Strada Osservatorio 20, I-10025, Pino Torinese, Italy
  \\ $^{2}$Physics Department, The Technion, 32000, Haifa,
  Israel\\ $^{3}$ INAF- Istituto di Radioastronomia, Via Gobetti 101,
  I-40129 Bologna, Italy \\ $^4$ Department of
Physics and Astronomy, University of Southampton, Highfield, SO17 1BJ, UK \\
  $^{5}$School of Physics and Astronomy,
  Rochester Institute of Technology, 84 Lomb Memorial Drive,
  Rochester, NY 14623-5603, USA \\$^6$ Dipartimento di Fisica e
  Astronomia, Universit\`a di Firenze, via G. Sansone 1, 50019 Sesto
  Fiorentino, Italy}

\begin{document}



\maketitle

\label{firstpage}

\begin{abstract}
We present the results of high signal-to-noise ratio VLT
spectropolarimetry of a representative sample of 25 bright type 1 AGN
at $z<0.37$, of which nine are radio-loud. The sample covers uniformly
the 5100 \AA\ optical luminosity at $L_{5100}\sim 10^{44}-10^{46}$ erg
s$^{-1}$, and H$\alpha$ width at FWHM$\sim 1000-10,000$~\kms. We
derive the continuum and the H$\alpha$ polarization amplitude,
polarization angle, and angle swing across the line, together with the
radio properties. We find the following: 1. The broad line region
(BLR) and continuum polarization are both produced by a single
scattering medium. 2. The scattering medium is equatorial, and at
right angle to the system axis. 3. The scattering medium is located at
or just outside the BLR. The continuum polarization and the H$\alpha$
polarization angle swing, can both serve as an inclination
indicator. The observed line width is found to be affected by
inclination, which can lead to an underestimate of the black hole mass
by a factor of $\sim 5$ for a close-to face-on view. The line width
measured in the polarized flux overcomes the inclination bias, and
provides a close-to equatorial view of the BLR in all AGN, which
allows to reduce the inclination bias in the BLR based black hole mass
estimates.

\end{abstract}


\section{Introduction}

\begin{figure*}
\psfig{figure=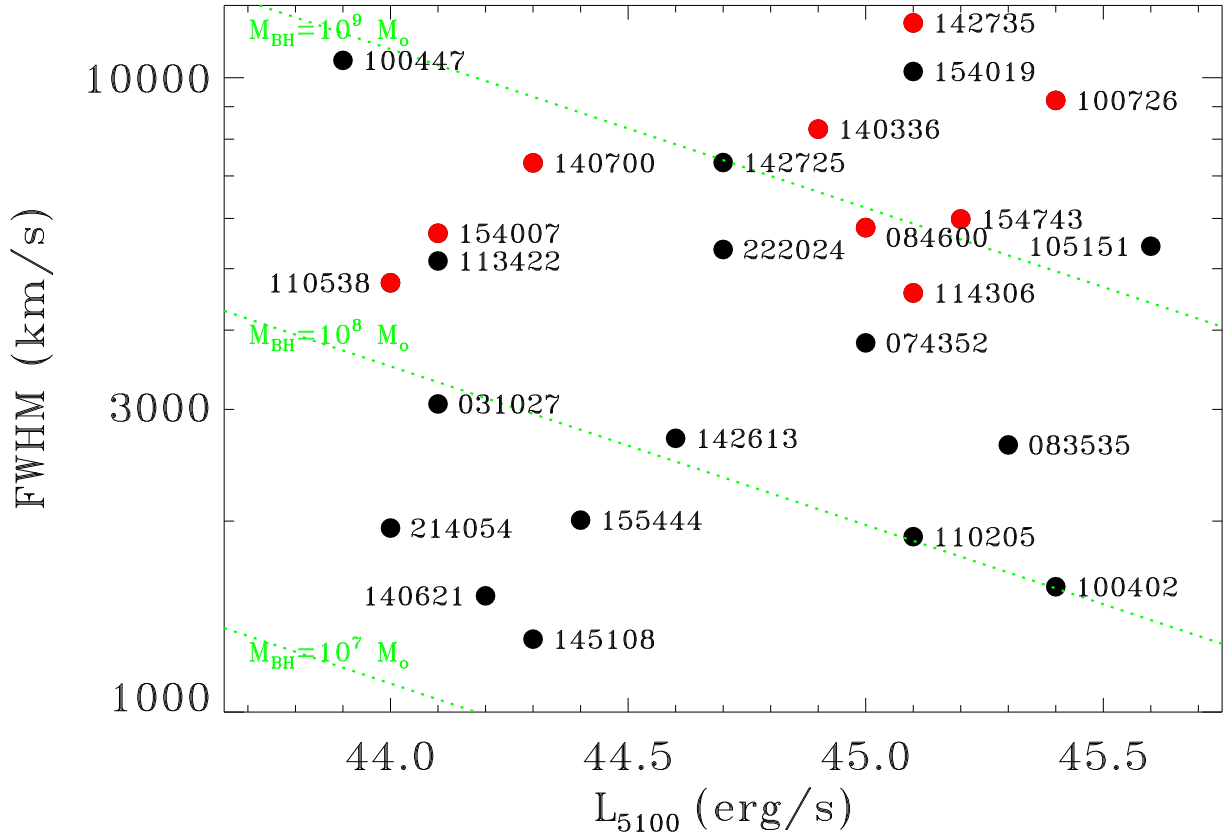,width=0.99\linewidth}
\caption{The distribution of the 5100 \AA\ luminosity $L_{5100}$
  versus the broad H$\alpha$ FWHM of the 25 quasars selected for the
  VLT spectropolarimetry. The $L_{5100}$ vs. FWHM plain was divided
  into a (logarithmically) uniform grid, and the sample consists of
  the brightest SDSS/DR7 quasar in each rectangle, which satisfies $z
  < 0.12$ or $0.21 < z < 0.38$. This process is repeated for radio
  quiet QSO (RQQ, black) and for radio loud QSO (RLQ, red). This
  selection scheme ensures the widest possible coverage in $L_{5100}$,
  FWHM, and $M_{\rm BH}$ estimated with the virial method (marked by
  diagonal dashed lines).}
\label{sel}
\end{figure*}

\begin{table*}
\caption{Sample properties}
\begin{tabular}{l l c c c @{\vline} c c c @{\vline} c c r c l}
\hline
Name    & Alt. name & z & m$_r$ & log L & \multicolumn{3}{c}{H$\alpha$} & $|$ log M$_{\rm BH} $ & log L/L$_{\rm Edd.}$ & \multicolumn{2}{c}{R} & $\alpha_{\rm opt}$ \\ 
                     &             &        &        &5100\AA\,&\,\,log L&  FWHM       &  EW      & [M$_\odot$] &    &        & & \\ 
(1)                     &   (2)          &   (3)     &    (4)    & (5) & (6) & (7)  &  (8) & (9)  & (10)   &   (11)     &  & (12) \\ 
\hline                                                                          
J031027.83-004950.8  & 	           & 0.080  & 15.81  &  44.1 & \, 42.6 &  \,\,\,2494 &  246\,\, &    7.93 & -0.97 & $<$0.8 &       & -1.09\\
J074352.02+271239.5  & 	           & 0.252  & 15.71  &  45.0 & \, 43.9 &  \,\,\,3049 &  509\,\, &    8.59 & -0.69 &    0.4 &       & -1.41\\
J083535.81+245940.2  & PG~0832+251 & 0.329  & 15.93  &  45.3 & \, 44.2 &  \,\,\,2287 &  485\,\, &    8.41 & -0.23 &    1.0 &    & -1.52\\
J084600.37+070424.7  & 	           & 0.342  & 16.99  &  45.0 & \, 44.0 &  \,\,\,5325 &  660\,\, &    8.92 & -1.09 &  430.0 &  RL   & -1.34\\
J100402.62+285535.4  & PG~1001+291 & 0.328  & 15.72  &  45.4 & \, 44.1 &  \,\,\,1660 &  330\,\, &    8.03 &  0.28 & $<$0.4 &    & -1.82\\
J100447.61+144645.6  & MRK~0715    & 0.084  & 16.35  &  43.9 & \, 42.5 &  \,\,\,8121 &  179\,\, &    8.94 & -2.13 &   10.0 &    & -0.06\\
J100726.10+124856.2  & PG~1004+130 & 0.240  & 15.47  &  45.4 & \, 43.8 &  \,\,\,5996 &  213\,\, &    9.54 & -1.27 &  220.0 &  RL& -1.79\\
J105151.44-005117.7  & PG~1049-005 & 0.359  & 15.69  &  45.6 & \, 44.4 &  \,\,\,4794 &  439\,\, &    9.17 & -0.72 & $<$0.7 &    & -1.55\\
J110205.92+084435.7  & 	           & 0.294  & 16.35  &  45.1 & \, 43.7 &  \,\,\,1692 &  285\,\, &    8.00 & -0.06 & $<$1.1 &       & -1.56\\
J110538.99+020257.3  & 	           & 0.105  & 16.53  &  44.0 & \, 42.8 &  \,\,\,3734 &  367\,\, &    8.27 & -1.39 &  520.0 &  RL   & -0.24\\
J113422.47+041127.7  & 	           & 0.108  & 16.41  &  44.1 & \, 42.9 &  \,\,\,3896 &  408\,\, &    8.40 & -1.40 & $<$1.1 &       & -0.77\\
J114306.02+184342.9  & 	           & 0.374  & 16.90  &  45.1 & \, 43.9 &  \,\,\,3873 &  506\,\, &    8.79 & -0.80 &  180.0 &  RL   & -1.37\\
J140336.43+174136.1  & 	           & 0.222  & 16.60  &  44.9 & \, 43.6 &  \,   11004 &  334\,\, &    9.21 & -1.41 &   54.0 &  RL   & -1.62\\
J140621.89+222346.5  & PG~1404+226 & 0.097  & 16.09  &  44.2 & \, 42.8 &  \,\,\,1442 &  318\,\, &    7.36 & -0.32 &    1.5 &    & -1.57\\
J140700.40+282714.6  & MRK~0668    & 0.076  & 15.61  &  44.3 & \, 43.1 &  \,\,\,5943 &  366\,\, &    8.79 & -1.63 &  580.0 &  RL& +0.04\\
J142613.32+195524.7  & 	           & 0.213  & 16.67  &  44.6 & \, 43.3 &  \,\,\,2462 &  386\,\, &    8.06 & -0.62 & $<$1.4 &       & -1.16\\
J142725.05+194952.2  & MRK~0813    & 0.110  & 15.51  &  44.7 & \, 43.4 &  \,\,\,6248 &  385\,\, &    8.98 & -1.44 &    0.2 &    & -1.42\\
J142735.61+263214.5  & PG~1425+267 & 0.364  & 16.61  &  45.1 & \, 44.1 &  \,\,\,8226 &  498\,\, &    9.66 & -1.64 &  710.0 &  RL& -1.43\\
J145108.76+270926.9  & PG~1448+273 & 0.064  & 15.41  &  44.3 & \, 42.7 &  \,\,\,1395 &  198\,\, &    7.29 & -0.12 &    0.6 &    & -1.27\\
J154007.84+141137.0  & 	           & 0.119  & 16.95  &  44.1 & \, 42.9 &  \,\,\,4496 &  375\,\, &    8.46 & -1.51 &  150.0 &  RL   & -0.54\\
J154019.57-020505.4  &    	   & 0.320  & 16.78  &  45.1 & \, 43.9 &  \,\,\,9217 &  387\,\, &    9.47 & -1.52 &    4.1 &       & -1.20\\
J154743.54+205216.7  & PG~1545+210 & 0.264  & 15.60  &  45.2 & \, 44.0 &  \,\,\,6828 &  498\,\, &    9.07 & -1.00 &  810.0 &  RL& -1.42\\
J155444.58+082221.5  & PG~1552+085 & 0.119  & 15.81  &  44.4 & \, 43.1 &  \,\,\,1553 &  313\,\, &    7.73 & -0.43 & $<$0.9 &    & -1.21\\
J214054.56+002538.2  & 	           & 0.083  & 16.45  &  44.0 & \, 42.5 &  \,\,\,1689 &  212\,\, &    7.49 & -0.63 &    0.8 &       & -0.62\\
J222024.59+010931.3  &             & 0.212  & 16.61  &  44.7 & \, 43.4 &  \,\,\,5018 &  309\,\, &    8.73 & -1.14 & $<$1.7 &    & -0.89\\
\hline                                                                          
\end{tabular}
\label{tab1}

\medskip
Column description: 1) name, 2) alternative name, 3) redshift, 4) $r$
band SDSS magnitude, 5) logarithm of the luminosity at 5100 \AA\ in
$\ergs$ (the values from column 5 to 10 are from \citealt{shen11}), 6
- 8) luminosity ($\ergs$), full width at half maximum (FWHM, \kms),
and equivalent width (EW, \AA) of the broad \Ha\ line, 9) logarithm of
the black hole mass in solar masses, estimated with the virial method,
10) logarithm of the bolometric luminosity with respect to the
Eddington luminosity, 11) radio loudness parameter (we also indicate
with RL the radio-loud sources), 12) slope of the optical continuum,
$I_\lambda$.
\end{table*}

\begin{table}
\caption{Observations log.}
\begin{tabular}{|l  c c c c }
\hline
Name    & Date & Obs. time & Cal. & S/N \\ 
\hline                                                                          
J031027.83--004950.8  & 2016-09-18 & 41 &  0.96 &     267\\
J074352.02+271239.5   & 2016-11-24 & 41 &  1.05 &     408\\
J083535.81+245940.2   & 2016-12-20 & 15 &  0.89 &     139\\
J084600.37+070424.7   & 2016-12-04 & 41 &  1.04 &     205\\
J100402.62+285535.4   & 2016-12-16 & 41 &       &     291\\
J100447.61+144645.6   & 2016-12-22 & 41 &  1.20 &     330\\
J100726.10+124856.2   & 2016-12-23 & 41 &  1.17 &     357\\
J105151.44--005117.7  & 2017-01-11 & 41 &  0.93 &     202\\
J110205.92+084435.7   & 2016-12-23 & 41 &       &     226\\
J110538.99+020257.3   & 2017-01-04 & 41 &  0.81 &     298\\
J113422.47+041127.7   & 2017-01-10 & 41 &  1.17 &     340\\
J114306.02+184342.9   & 2017-01-10 & 41 &  0.99 &     155\\
J140336.43+174136.1   & 2017-03-03 & 41 &       &     308\\
J140621.89+222346.5   & 2017-03-07 & 56 &       &     277\\
J140700.40+282714.6   & 2017-02-09 & 41 &  0.89 &     345\\
                      & 2017-03-06 & 41 &  0.76 &     413\\
J142613.32+195524.7   & 2017-03-07 & 41 &       &     233\\
J142725.05+194952.2   & 2017-03-12 & 54 &       &     361\\
J142735.61+263214.5   & 2017-03-03 & 41 &  1.20 &     185\\
J145108.76+270926.9   & 2017-03-06 & 41 &  0.88 &     593\\
J154007.84+141137.0   & 2017-03-13 & 41 &  0.77 &     219\\
J154019.57--020505.4  & 2017-02-10 & 41 &  1.60 &     161\\
J154743.54+205216.7   & 2017-03-12 & 41 &  1.01 &     327\\
J155444.58+082221.5   & 2017-03-14 & 33 &       &     298\\
J214054.56+002538.2   & 2016-09-18 & 41 &  0.76 &     265\\
J222024.59+010931.3   & 2016-09-18 & 41 &       &     238\\
\hline                                                                          
\end{tabular}
\label{tab2}

\medskip
Column description: 1) source name, 2) date of observations, 3) exposure time
in minutes, 4) ratio between the VLT and SDSS continuum flux, 5) S/N per resolution 
element at 6750 \AA.
\end{table}

Most AGN are powered by accretion through a disk, and often display a
jet. Thus, their emission properties likely have an axial symmetry and
can be inclination dependent. If the emission from such a system is
polarized, the polarization is inevitably inclination dependent, and
should approach zero when the inclination approaches zero (i.e., a
face-on view). Thus, polarization may provide a useful inclination
indicator in AGN, a point which we explore in this study.

Indeed, spectropolarimetry of type 2 AGN revealed that they harbor an
obscured type 1 AGN, which led to the AGN inclination unification
scheme \citep{miller83,antonucci83,antonucci85,antonucci93}, where a
planar torus-like structure obscures the direct light from the Broad
Line Region (BLR) and the continuum source at a closer-to edge-on
view. A scattering medium, which resides in the polar direction above
the obscuring medium, as indicated by polarization angle perpendicular
to the jet axis \citep{antonucci83}, allows to detect the hidden BLR
and continuum source through the scattered polarized light.

Spectropolarimetry of the continuum and broad lines in a variety of
type 1 Seyfert galaxies reveals a more elaborate picture, with a wide
range of polarization levels and polarization angles
\citep{smith05}. The polarization in some type 1 AGN is consistent
with a polar scattering medium, as observed in type 2 AGN. But, in the
majority of type 1 AGN spectropolarimetry suggests an equatorial
scattering medium located just outside the BLR, as indicated by the
polarization angle and its rotation across the broad line profiles. In
addition, some objects show an intermediate scattering medium, or show
no evidence for a scattering medium
\citep{goodrich88,goodrich89,goodrich94,smith02, smith04,smith05,
  afanasiev19}.

Spectropolarimetry allows us to constrain the properties and
kinematics of the scattering medium, and furthermore to view the AGN
from the direction of the scattering medium. The presence of two
lines-of-sight allows to constrain the inclination dependence of
various emission properties, and in particular the three dimensional
velocity field in the BLR. Since the BLR is very likely disk-like, as
indicated in radio-loud AGN (e.g., \citealt{wills86,fine11}), and
generally rotationally supported, the observed line-of-sight velocity
dispersion $W$ is inclination dependent. Since the virial black hole
mass estimate $M_{\rm BH}\propto W^2$, inclination effects can lead to
a significant bias. In particular, a face-on view and a low $M_{\rm
  BH}$ in objects with low $W$ are difficult to tell
apart. Spectropolarimetry allows to constrain the inclination angle,
and thus potentially reduce significantly the inclination bias in the
$M_{\rm BH}$ estimates. Furthermore, if the scattering region is
located in the AGN equatorial plane, it sees the full rotation
amplitude of the BLR, and the light reflected (and thence polarized)
will show the intrinsic value of the BLR line width.

The accuracy of $M_{\rm BH}$ estimates may be compromised by
additional systematic effects. First of all, the virial method
assumes that the BLR is dominated by rotation, while the presence of
outflowing material on different scales is a widespread characteristic
of AGN. An outflow will increase $W$ and induce a bias for the
$M_{\rm BH}$ estimates. The outflow intensity may be linked to the AGN
Eddington ratio, $\lambda = L/L_{\rm Edd.}$ (e.g., \citealt{king15}),
and then, ultimately, to $M_{\rm BH}^{-1}$. The effects of orientation
bias are even more complex in case of an outflowing component, as the
outflow occurs mainly along the polar direction, which will
produce an opposite bias compared to the rotational component.

We recently obtained spectropolarimetric observations with FORS2 at
the VLT of the narrow line Seyfert 1 PKS~2004-447 (NLSy1 are type 1
AGN in which the broad-line FWHM are $<$ 2000 km/s) (see
\citealt{baldi16}). We recovered all features typical of the
equatorial scattering model (ESM, \citealt{smith05}); in particular
its polarized spectrum shows two peaks separated by $\sim$9,100
km/s. The broadening with respect to the FWHM of 1447 km~s$^{-1}$
measured in direct-light \citep{oshlack01} is a dramatic indication that
in this source the $M_{\rm BH}$ value based on the virial method is
underestimated by a factor $\sim$40 (!). PKS~2004-447 is a very
challenging target for spectropolarimetric observations. Indeed, the
ESM predicts that the percentage of polarization decreases for objects
seen close-to face-on, a condition met by this source as witnessed by
its detection (and rapid variability) in $\gamma$-rays by Fermi
\citep{calderone11}, clear evidence for a relativistic jet oriented at
a small angle to the line-of-sight. Nonetheless, despite the extremely
low level of polarization $\sim$0.1\%, we were able to obtain a robust
$M_{\rm BH}$ estimate.

The purpose of this study is to follow up on the study of
\citet{baldi16} and derive systematic high signal-to-noise ratio
observations of the broad line and continuum spectropolarimetric
properties of a representative sample of 25 low redshift ($z<0.37$)
QSO. The QSO are selected to cover uniformly a wide range of
luminosities and BLR line widths. Specifically, the sample covers a
range in luminosities at 5100 \AA\ of $L_{5100}\sim 10^{44}-10^{46}
\ergs$, and FWHM$=1000-10,000$~\kms, which corresponds to black hole
masses (estimated with the virial method) of $M_{\rm BH} \sim 2\times
10^7$ to $5\times 10^9 M_\odot$.

We use the spectropolarimetry to explore the BLR structure and to
study the implications for the measurements of the black hole
masses. The plan of the paper is as follows: in Sect. \ref{sample} we
describe the sample selection; the observations and data analysis are
discussed in Sect. \ref{observations}; we then estimate the integrated
(Sect. \ref{intpol}) and spectrally resolved (Sect. \ref{respol})
polarization parameters of the BLR. In Sect. \ref{origin} we used our
observations to explore the polarization mechanism in QSOs. The
results are discussed in Sect. \ref{discussion}, while in
Sect. \ref{summary} we give a summary of our findings and draw our
conclusions. A cosmology with $H_0 = 72$ km s$^{-1}$ Mpc$^{-1}$,
$\Omega_m = 0.30$, and $\Omega_\Lambda = 0.70$ is assumed.

\begin{figure*}
\centering  
\psfig{figure=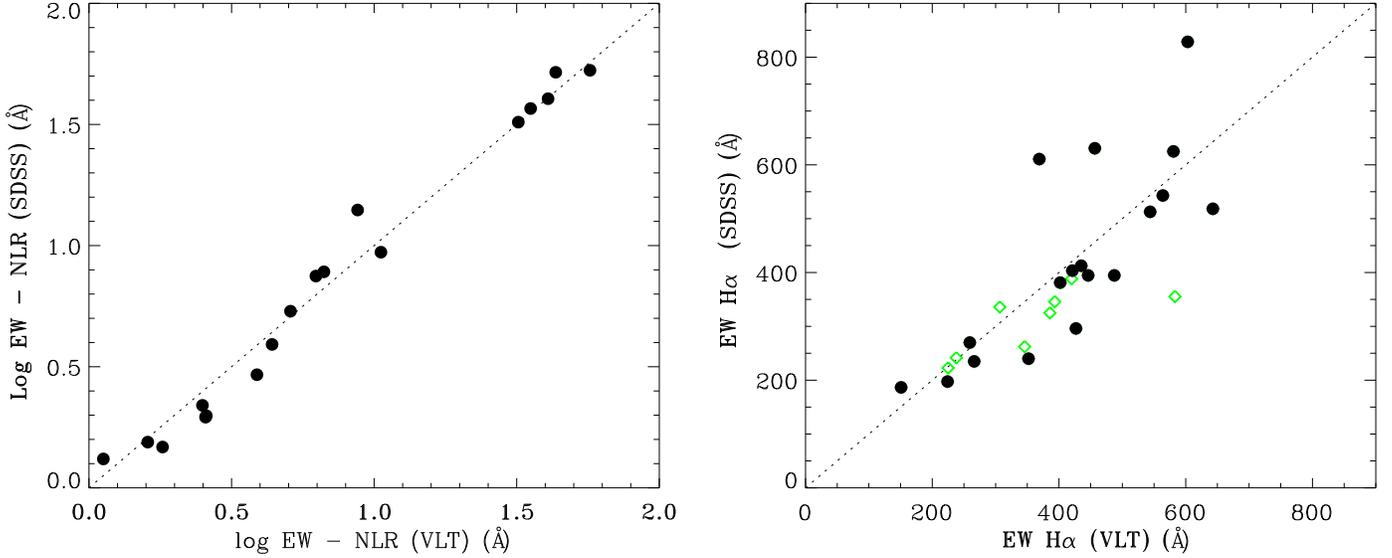,angle=90,width=0.99\linewidth}
\caption{Comparison of the SDSS and VLT measurements of the EW of the
  forbidden narrow lines (left) and the broad H$\alpha$ (right). The
  green diamonds in the right panel mark the sources where the narrow
  line fluxes could not be measured. These measurements were used to
  obtain an absolute flux calibration for the VLT spectra from the
  SDSS spectra. The left panel indicates generally small variability
  in both the continuum and BLR.}
\label{calib}
\end{figure*}

\section{Sample selection}
\label{sample}

The targets have been selected from the SDSS/DR7 catalog of QSOs
produced by \citet{schneider10b} and analyzed by \citet{shen11}. For
low redshift QSOs the best suited permitted line for an optical study
is H$\alpha$. The high accuracy needed for the spectropolarimetric
analysis requires to avoid the main telluric absorption features
(namely the O$_2$ bands at $\lambda \sim$ 7600 \AA\ and the H$_2$O
absorptions at $\lambda \gtrsim$ 9300 \AA) in the region of interest,
$\sim$10,000 km/s on each side of H$\alpha$, leading to the following
redshift constraints: $z < 0.12$ and $0.21 < z < 0.38$. The need for a
high S/N requires to focus on bright targets
and we adopted as an initial threshold $r < 17$. When combined with a
declination $\delta < 30^\circ$ to ensure VLT observability, these
constraints select 128 QSOs.

The final selection aims at exploring the properties of QSOs across
the widest possible range of luminosities and estimated black hole
masses. We located the 128 QSOs selected above in a diagram
comparing the $L_{5100}$ optical luminosity and the FWHM of the broad
H$\alpha$ line. We then divided this diagram with a (logarithmically)
uniform grid and within each rectangle we picked the source with the
highest optical flux to further improve the data quality (as a result
the median flux of the selected sources is $\sim 8 \times 10^{-16}$
erg s$^{-1}$ \AA$^{-1}$ cm$^{-2}$, about twice the selection
threshold).

Since the properties of radio loud and radio quiet QSOs (RLQs and
RQQs) might differ, we made separate selections of QSOs
belonging to the two sub-classes.  \citet{shen11} lists the
radio-loudness parameter $R$, i.e., the ratio of the flux density at
rest frame 6 cm and 2500 \AA, estimated following the method of
\citet{jiang07}: the rest frame 6 cm data are obtained by
extrapolating the Faint Images of the Radio Sky at Twenty centimeters
survey (FIRST, \citealt{becker95,helfand15}) 1.4 GHz measurements
assuming a power-law slope of 0.5, while the 2500 \AA\ flux density is
extrapolated from a power-law fit to the SDSS broad-band photometry. A
significant issue related to this definition is the presence, revealed
by the visual inspection of the FIRST images of the selected QSOs, of
several objects with radio structures extending well beyond the
30$\arcsec$ radius they adopted as the limit to include a radio component
into the total radio emission (see Sect. 4.3). For this reason we preferred to
define the radio-loudness parameter (see Table 1) by including the
radio flux densities of all the NVSS (National Radio Astronomy
Observatory Very Large Array Sky Survey, \citealt{condon98})
components (because this survey is more sensitive to diffuse radio
structure compared to FIRST) associated with each QSO. For the sources
not detected by the NVSS, we used either the FIRST detection or upper
limit.

The final sample consists of 16 RQQs and 9 RLQs (see Figure \ref{sel}),
respectively. RLQs only populate the upper part of the diagram because
there are no such objects, among the 128 selected initially, with a
FWHM smaller than 3000 \kms.

The main properties of the 25 QSOs of the sample, extracted from the
measurements obtained by \citet{shen11}, are given in Table
\ref{tab1}. As a result of our selection method we have objects
spanning a factor $\sim$100 in luminosity ($\log L_{5100}= 43.9-45.6$
in erg s$^{-1}$ units), and with a broad \Ha\ width from $\sim 1400$
to $\sim 11,000$ \kms, resulting in black hole masses (estimated with
the virial method) ranging from $2\times 10^7$ to $5\times 10^9
M_\odot$.

\section{Observations and data analysis}
\label{observations}

The selected 25 sources were observed in spectropolarimetric mode with
the FORS2 mounted at the UT1 telescope of the 8.2-m ESO Very Large
Telescope (VLT), program 098.B-0234(A), each source fitting into a
single 1 hour observing block. The log of the observations is
presented in Table \ref{tab2}.

The observations were obtained with the GRISM-300I, providing a
spectral resolution of $R\sim$660 at 8600 \AA, used in combination
with the blocking filter OG590. The usable wavelength range is 6100 -
9300 \AA. The 1$\arcsec$ wide slit was oriented along the parallactic
angle and the multi-object spectropolarimetry (PMOS) observation was
performed with a 2048 $\times$ 2048 pixel CCD with a spatial
resolution of 0.126 \arcsec/pixel.

The measurements at two retarder angles are sufficient to calculate
the linear Stokes parameters. However, in order to reduce instrumental
issues \citep{patat06} we used four half-wave plate angles (0, 22.5,
45 and 67.5 degrees) for our observations. For each source we then
obtained four spectra (split in two exposures for cosmic rays
removal), with the half-plate oriented at the four position
angles. The observing time per source is generally $\sim$41 minutes, with
a few exceptions due, e.g., to saturation limits or scheduling
problems, see Table \ref{tab2}. One source was observed twice.

The data were analyzed by using the standard VLT pipeline with the ESO
Reflex workflow \citep{hook08}. The frames were first bias-subtracted
and then flat-fielded using lamp flats. After removing cosmic ray
events, the different exposures for the four half-wave plate positions
were combined for each observing block. One-dimensional spectra, with
identical aperture widths for the ordinary (o) and extraordinary (e)
rays for each of the four half-wave plate orientations, were then
extracted (from 12 pixels corresponding to 1.5\arcsec). The wavelength
calibration was performed using standard arc lamps. The o- and e-rays
were resampled to the same linear spectral dispersion and combined to
determine the Stokes parameters I, Q, and U (as well as the associated
uncertainties) using the procedures described by \citet{cohen95} and
\citet{vernet01}. The S/N in the total spectrum in each
spectral element, estimated at 6750 \AA, ranges from $\sim 140$ to
$\sim$ 600 with a median value of $\sim$ 300 (see Table
\ref{tab2}). During the observations the seeing varied between
0\farcs36 and 1\farcs20, with a median value of 0\farcs60.

   \begin{table*}
\caption{Polarization measurements}
\begin{tabular}{l | c r c c r r r | r r r r r}
\hline
Name       & $P$ cont. (\%)   &PA cont.    & &  $P$ BLR (\%)     &aver. $P$ BLR (\%)&  PA BLR & PA swing & W (I)   & W (I$_P$)  &  $I_P$ - $I$ shift\\
(1)        &   (2)            &   (3)      & &            (4)    &              (5) &     (6) &     (7)  &  (8)    &       (9)  &       (10)   \\ 
\hline                                         
   J031027 &   0.18$\pm$ 0.06 &  68$\pm$ 9 & &    0.06$\pm$ 0.06 &  0.35   (0.15) &         --- &  78$\pm$11 &   1974   &     3890$_{- 650}^{+ 500}$    &    +670  $\pm$  510\\
   J074352 &   0.47$\pm$ 0.03 &  59$\pm$ 2 & &    0.24$\pm$ 0.03 &  0.27   (0.20) &   68$\pm$ 4 & ---        &   2148   &     5020$_{- 450}^{+ 320}$    &    +880  $\pm$  440\\       
   J083535 &   1.71$\pm$ 0.09 &  80$\pm$ 2 & &    0.05$\pm$ 0.07 &  0.26   (0.13) &         --- & ---        &   2239   &     4120$_{- 790}^{+ 720}$    &    +740  $\pm$  420\\       
   J084600 &   1.23$\pm$ 0.07 &  67$\pm$ 2 & &    0.36$\pm$ 0.05 &  0.32   (0.15) &   67$\pm$ 3 &  30$\pm$ 7 &   3186   &     3940$_{- 480}^{+ 550}$    &    -200  $\pm$  430\\
   J100402 &   0.36$\pm$ 0.05 &  18$\pm$ 4 & &    0.13$\pm$ 0.05 &  0.13   (0.07) &   23$\pm$10 & ---        &   1773   &     4220$_{- 730}^{+ 950}$    &    +890  $\pm$  580\\       
   J100447 &   0.36$\pm$ 0.06 &  35$\pm$ 5 & &    0.27$\pm$ 0.07 &  0.15   (0.17) &   83$\pm$ 7 & ---        &   3551   &     4110$_{- 870}^{+ 930}$    &    -290  $\pm$  520\\       
   J100726 &   0.54$\pm$ 0.06 &  75$\pm$ 3 & &    0.22$\pm$ 0.09 &  0.21   (0.11) &   73$\pm$11 & ---        &   4603   &     5860$_{-1040}^{+1670}$    &   +1680  $\pm$  810\\       
   J105151 &   0.87$\pm$ 0.07 & 128$\pm$ 2 & &    0.07$\pm$ 0.05 &  0.38   (0.14) &         --- &  79$\pm$ 9 &   3222   &     5230$_{- 610}^{+1040}$    &    -210  $\pm$  400\\
   J110205 &   0.14$\pm$ 0.07 &  50$\pm$14 & &    0.25$\pm$ 0.07 &  0.24   (0.12) &   29$\pm$ 7 & ---        &   1618   &     2960$_{- 590}^{+ 740}$    &    +300  $\pm$  260\\
   J110538 &   0.03$\pm$ 0.07 &   ---      & &    0.10$\pm$ 0.05 &  0.14   (0.07) &  111$\pm$14 & ---        &   2523   &     5170$_{-1010}^{+1360}$    &    -470  $\pm$  620\\
   J113422 &   0.59$\pm$ 0.05 &  72$\pm$ 2 & &    0.17$\pm$ 0.04 &  0.17   (0.09) &   56$\pm$ 7 & ---        &   2710   &     4460$_{- 770}^{+1610}$    &     +30  $\pm$  530\\
   J114306 &   0.21$\pm$ 0.10 &  28$\pm$14 & &    0.32$\pm$ 0.07 &  0.32   (0.27) &   45$\pm$ 6 & ---        &   2422   &     3460$_{- 610}^{+ 980}$    &    -500  $\pm$  350\\
   J140336 &   1.02$\pm$ 0.05 & 132$\pm$ 1 & &    0.76$\pm$ 0.06 &  0.36   (0.26) &   84$\pm$ 2 &  19$\pm$ 7 &   4808   &     4380$_{- 360}^{+ 420}$    &   +2050  $\pm$  310\\
   J140621 &   0.39$\pm$ 0.05 &  19$\pm$ 4 & &    0.35$\pm$ 0.05 &  0.40   (0.25) &    9$\pm$ 4 & ---        &   1225   &     1870$_{- 300}^{+ 500}$    &     +20  $\pm$   70\\
J140700(a) &   0.79$\pm$ 0.07 & 136$\pm$ 2 & &    0.48$\pm$ 0.05 &  0.56   (0.12) &   75$\pm$ 2 &  52$\pm$ 4 &   4123   &     5690$_{- 370}^{+ 380}$    &     -40  $\pm$  460\\
J140700(b) &   1.03$\pm$ 0.06 & 136$\pm$ 2 & &    0.56$\pm$ 0.04 &  0.50   (0.14) &   64$\pm$ 2 &  41$\pm$ 5 &   4255   &     5700$_{- 480}^{+ 430}$    &    -570  $\pm$  220\\
   J142613 &   0.18$\pm$ 0.07 & 125$\pm$10 & &    0.19$\pm$ 0.06 &  0.29   (0.15) &   20$\pm$ 8 & ---        &   2020   &     4740$_{- 770}^{+ 570}$    &    -200  $\pm$  450\\
   J142725 &   0.84$\pm$ 0.05 &  70$\pm$ 2 & &    0.52$\pm$ 0.04 &  0.36   (0.13) &   77$\pm$ 2 &  22$\pm$ 4 &   3515   &     4750$_{- 490}^{+1010}$    &    +550  $\pm$  260\\
   J142735 &   1.86$\pm$ 0.10 &  67$\pm$ 2 & &    0.37$\pm$ 0.05 &  0.37   (0.29) &   44$\pm$ 3 &  14$\pm$ 9 &   4982   &     5410$_{- 450}^{+ 500}$    &   +1820  $\pm$  620\\
   J145108 &   0.23$\pm$ 0.03 &  73$\pm$ 3 & &         ---       &  0.40   (0.20) &   ---       & ---        &   1261   &     ---                    &    ---             \\
   J154007 &   0.72$\pm$ 0.09 &  48$\pm$ 4 & &    0.43$\pm$ 0.06 &  0.35   (0.09) &   56$\pm$ 3 &  13$\pm$ 7 &   3003   &     3870$_{- 610}^{+ 940}$    &   +1030  $\pm$  360\\
   J154019 &   2.03$\pm$ 0.26 &  89$\pm$ 4 &*&    1.30$\pm$ 0.08 &  0.77   (0.25) &   80$\pm$ 1 &  12$\pm$ 4 &   4978   &     5220$_{- 510}^{+ 330}$    &    -500  $\pm$  300\\
   J154743 &   1.29$\pm$ 0.06 &  18$\pm$ 1 & &    0.28$\pm$ 0.04 &  0.20   (0.11) &   66$\pm$ 3 &   5$\pm$ 8 &   3186   &     3730$_{- 670}^{+ 650}$    &    -160  $\pm$  350\\
   J155444 &   1.23$\pm$ 0.09 &  86$\pm$ 2 &*&    1.14$\pm$ 0.06 &  0.82   (0.22) &   71$\pm$ 1 &   4$\pm$ 3 &   1252   &     1480$_{- 160}^{+ 230}$    &    +100  $\pm$   70\\
   J214054 &   0.67$\pm$ 0.12 & 131$\pm$ 5 &* &    0.27$\pm$ 0.06 &  0.36   (0.14) &  137$\pm$ 6 & ---        &   1325   &     2990$_{- 590}^{+1430}$    &     +20  $\pm$  280\\
   J222024 &   0.72$\pm$ 0.07 &  21$\pm$ 3 & &    0.36$\pm$ 0.06 &  0.58   (0.30) &   30$\pm$ 4 &  58$\pm$ 5 &   2719   &     4250$_{- 400}^{+ 540}$    &    +230  $\pm$  550\\
\hline                                                                          
\end{tabular}
\label{tab3}

\medskip
Column description: 1) source name, 2) and 3) percentage of
polarization and polarization position angle of the continuum at 6563
\AA. For the three sources marked with an * we corrected the continuum
for the interstella polarization. 4) and 5) integrated $P$ and averaged
$P$ (see Section 5 for its definition) of the BLR over twice the width
of the broad \Ha\ line after continuum subtraction, 6) PA of the BLR
7) PA swing between the blue and red wing of the BLR, 8) $W (I)$:
interpercentile width (between 25\% and 75\%) of the BLR in total
intensity in \kms\ and 9) in polarized light ($W (I_P)$), 10) shift
between the line centroid in $I$ and $I_P$ in \kms.
\end{table*}

\begin{figure}
\psfig{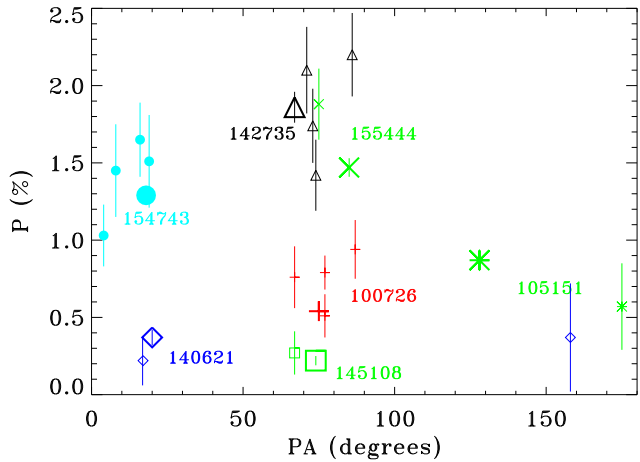}
\caption{Comparison of $P(\%)$ and PA$({\rm deg.})$ measured here with
  the white light measurements of \citet{berriman90} for the seven
  sources in common. Each object is marked with a different
  symbol. The large symbol is the VLT measurement. For clarity we
  reproduce only the errors for $P$. The differences are generally
  consistent with the measurement errors. The polarization appears to
  be stable over $\sim$30-40 years.}
\label{berriman}
\end{figure}

\begin{figure*}
\psfig{figure=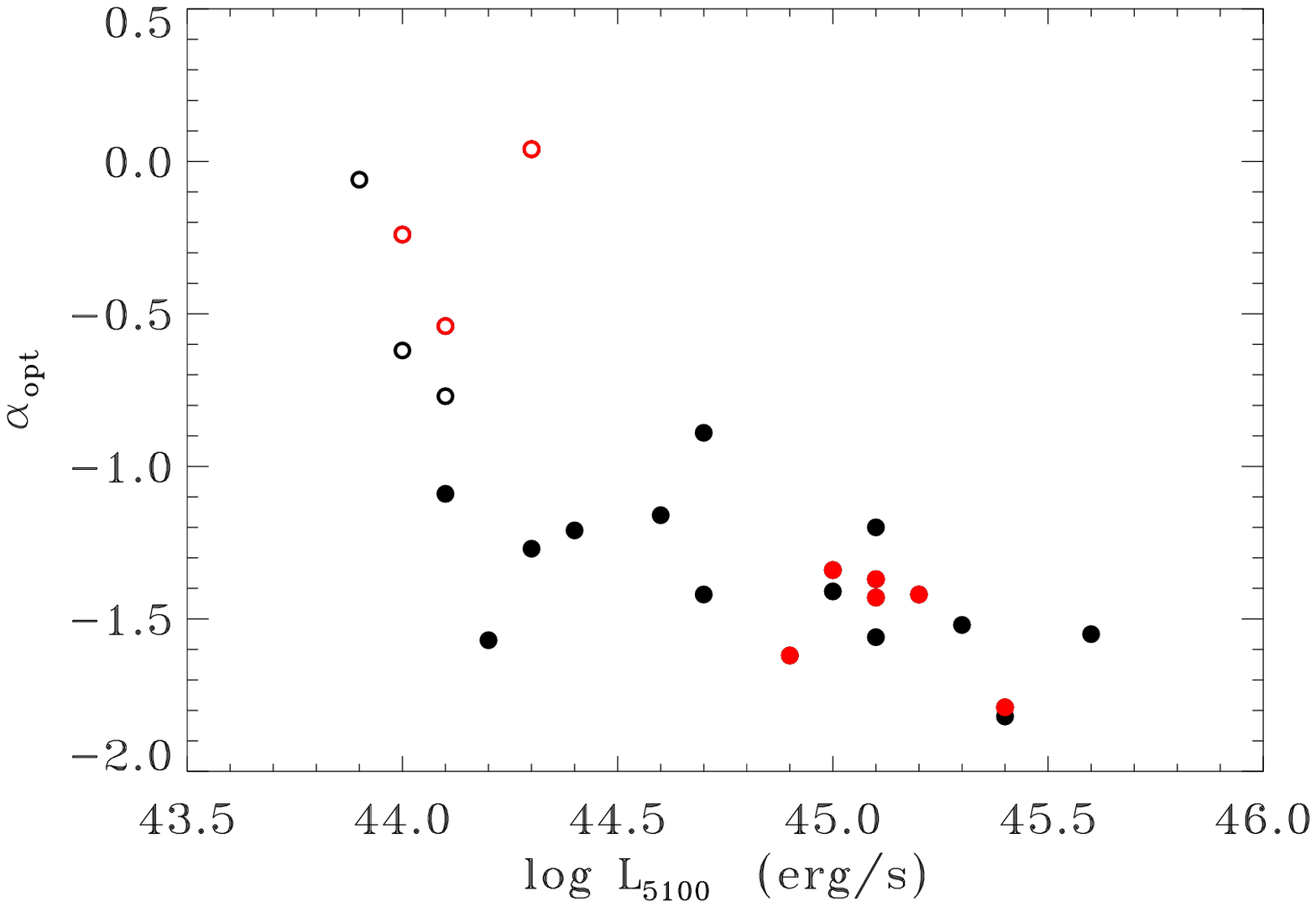,width=0.49\linewidth}
\psfig{figure=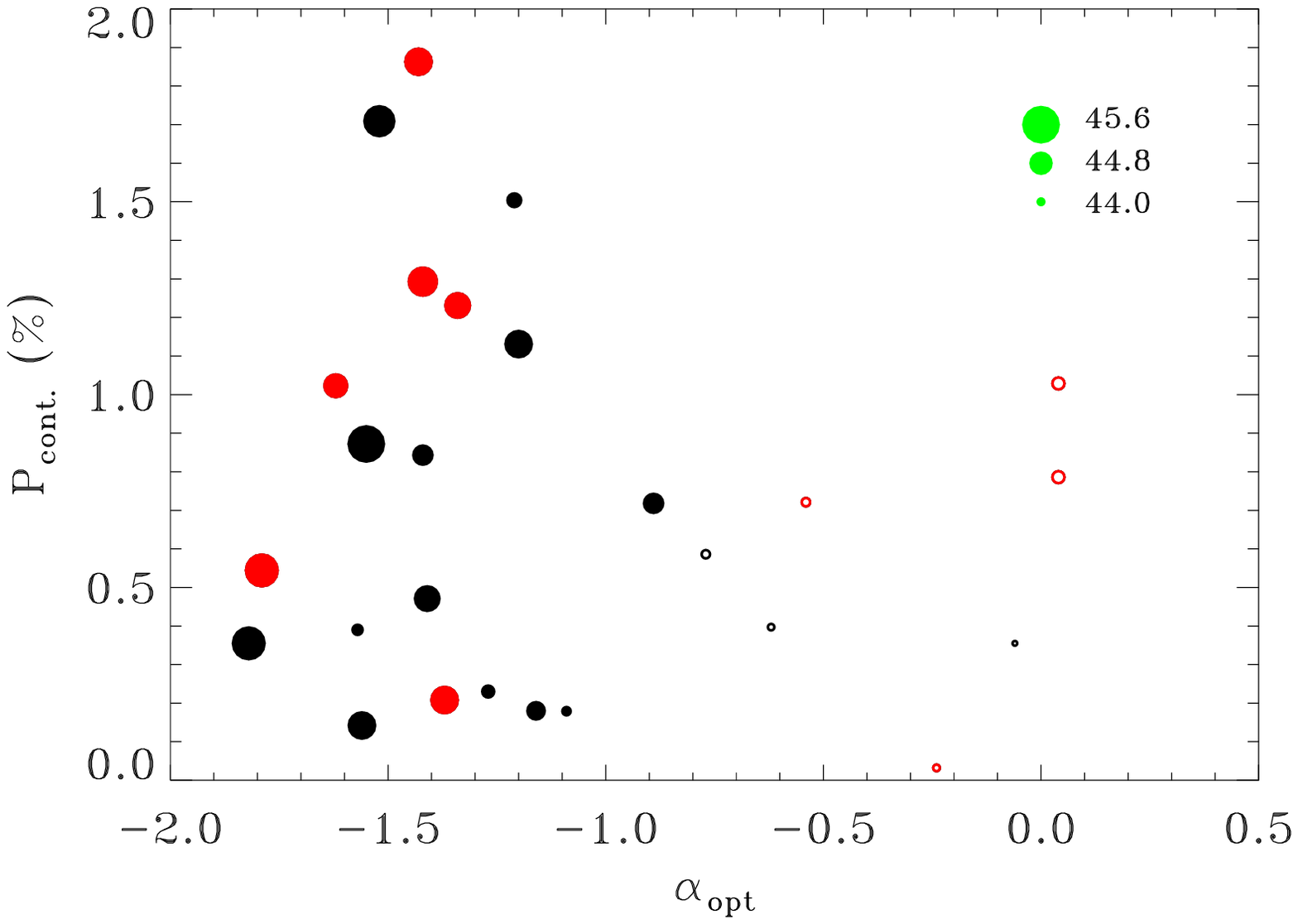,width=0.49\linewidth}
  \caption{Left: The optical slope $\alpha_{\rm opt}$ from SDSS versus
    $\log L_{5100}$. Red dots mark RLQs. All luminous QSOs are blue,
    and all red QSOs are low luminosity, indicating substantial host
    contribution. We therefore designate here and below all objects at
    $\log L_{5100}<44.5$ and $\alpha_{\rm opt}>-1$ with empty circles.
    Right: The value of $P_{\rm cont}(\%)$ versus $\alpha_{\rm opt}$;
    the symbol sizes increases with the value of $L_{5100}$ (see
    legend in the top right corner). The red low luminosity QSOs
    extend to a lower maximal polarization than the blue high
    luminosity ones, probably due to the host dilution. Note that the
    two data points at $\alpha_{\rm opt}>0$ refer to the two
    observations of the same object, J140700 (see Table 3).}
\label{slopes}
\end{figure*}

\begin{figure*}
\psfig{figure=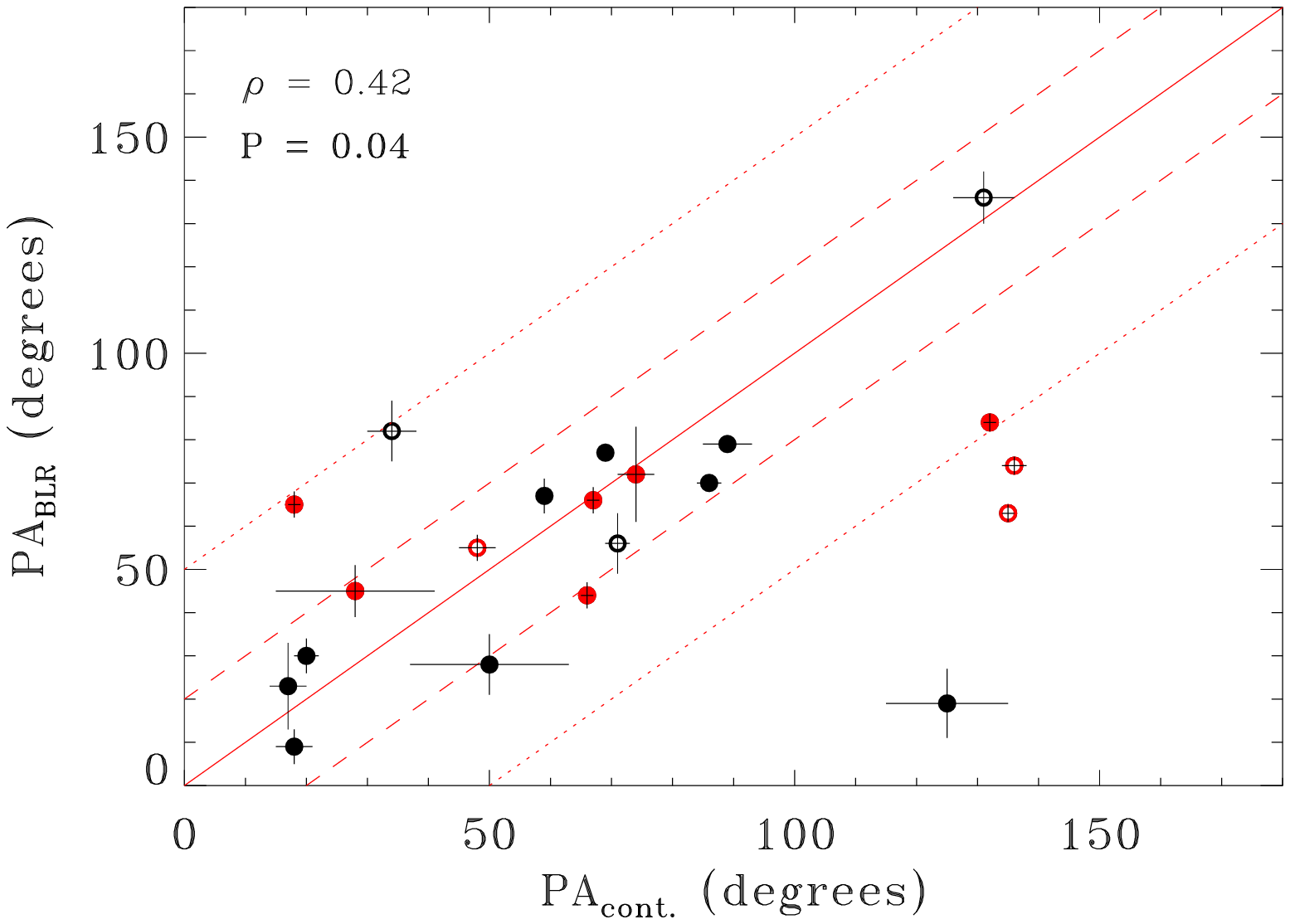,width=0.49\linewidth}
\psfig{figure=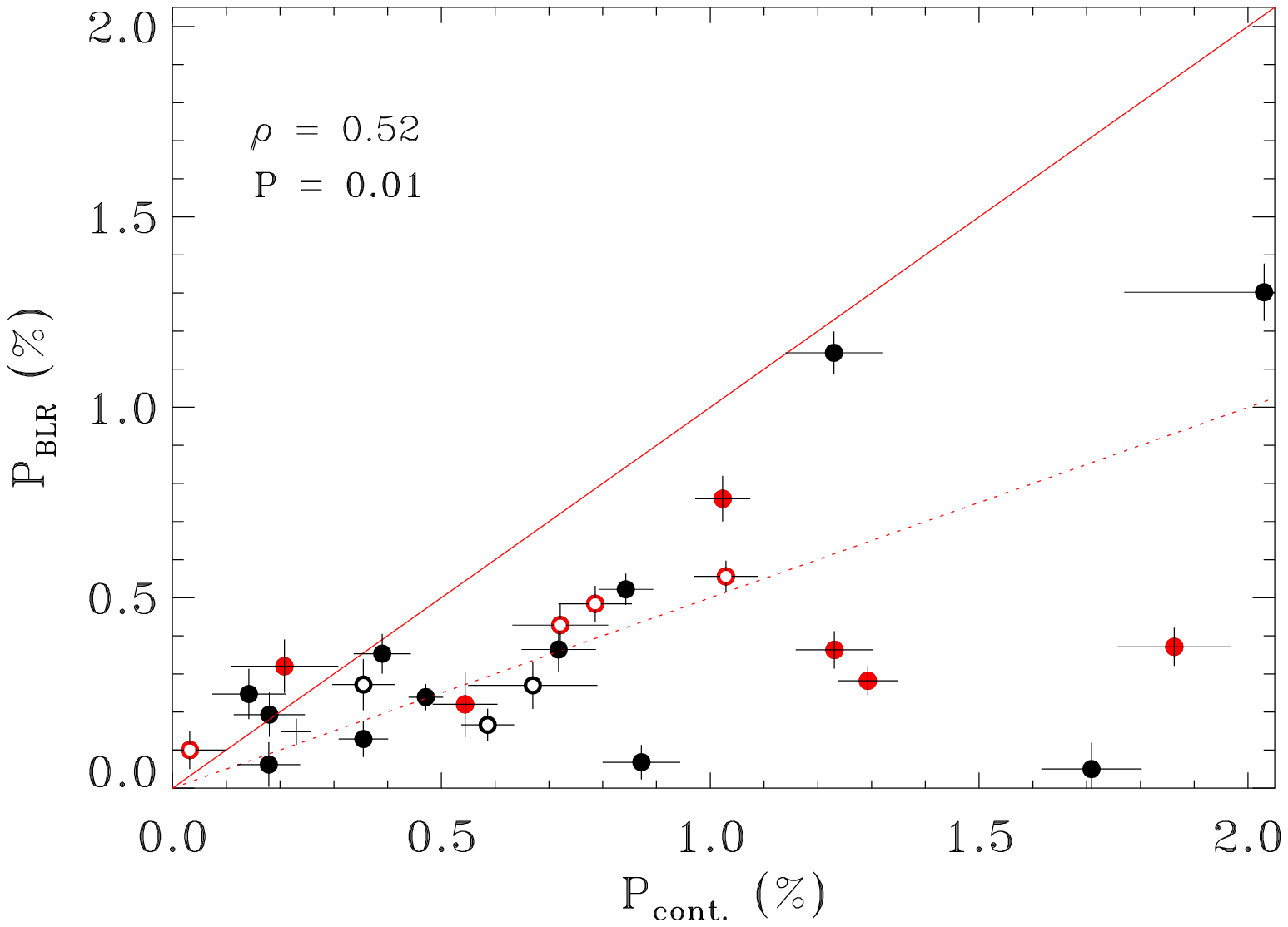,width=0.49\linewidth}
  \caption{Left: a comparison of the polarization PA of the continuum
    and of the BLR. The dashed (dotted) lines correspond to a
    difference of $\pm 20^{\circ}$ ($\pm 50^{\circ}$). The two angles
    differ by less than 20$^\circ$ in 15/20 objects, and only in 2/20
    by more than 50$^\circ$. This result indicates that in most
    objects the polarization of the continuum and the BLR are both
    produced by a single scattering medium. The Spearman rank test
    indicates a positive correlation between these two quantities with
    $\rho =0.42$ and a chance probability of 0.04. Right: a
    comparison of the percentage of polarization of the continuum and
    of the BLR. The continuum polarization is on average 1.7 times
    larger than the BLR polarization. The rank test returns $\rho
    =0.52$ and $P$=0.01. The low luminosity red QSOs, with
    $\alpha_{\rm opt} > -1$ and $\log L_{5100} < 44.5$, are
    represented by empty circles (see Figure 4). The red symbols are the
    RLQs. In essentially all objects $P_{\rm BLR}\le P_{\rm
      cont}$. This result suggests that the scattering medium is
    physically close to the BLR, leading to a significant geometrical
    cancellation.}
\label{pablrcont}
\end{figure*}

To maximize the efficiency of the observing program we did not obtain
observations of standard stars. Nonetheless, a comparison of the
SDSS and VLT spectra can be used to derive a relative flux
calibration. We estimated the ratio of the raw VLT and the calibrated
SDSS spectra for each source. The shape of this ratio is very similar
for all sources, reproducing mainly the changes of efficiency with
wavelength of the VLT spectropolarimetric observations. The curve
representing the median ratio versus wavelength can be well reproduced
with a quadratic curve over the whole spectral range, with residuals
never exceeding 3\%. We will use this curve to correct the VLT
spectra and to obtain their relative flux calibration.

We obtained also an absolute calibration for our spectra by measuring
the equivalent width (EW) of the forbidden narrow lines. By assuming
that their fluxes do not change between the epoch of the SDSS (2001 -
2008) and VLT spectra, any variation of their EW can be interpreted as
the result of continuum flux variability. The EW comparison is
possible for the 17 sources (with two measurements for J140700) for
which we could obtain an accurate (with an error smaller than $\sim$
20\%) EW measurement of either the [O~III]$\lambda$5007 line or
[S~II]$\lambda$6716,6731 doublet, depending on the target
redshift. The result is shown in Figure \ref{calib} (left panel). The
SDSS and VLT values for the EW agree, in all but one case, to within
30\%. The resulting variation of the continuum flux (ranging from 0.76
to 1.60) from the SDSS to the VLT data is reported in Table
\ref{tab2}.

The EW of the broad H$\alpha$ line show changes of similar amplitude,
see Figure \ref{calib} (right panel). We can extend this comparison also
to the objects lacking measurable narrow lines (green diamonds in the
figure). For these 8 objects the EW changes are of comparable (or even
smaller) amplitude. We conclude that in all of the 25 QSOs observed
there were no large changes in the last 10-15 years in the relative
fluxes of the continuum and the broad \Ha\ line, and in at least 17 of
the QSOs the absolute continuum and BLR fluxes also did not show large
changes.

\section{Integrated polarization measurements}
\label{intpol}

From the $I$, $Q$, and $U$ Stokes parameters we computed the polarized
spectrum $I_P = (Q^2 + U^2)^{1/2}$, the corresponding degree of
polarization $P$ = $I_P/I$, and the polarization position angle, PA
(measured counterclockwise starting from North). We started the
analysis by deriving integrated polarization measurements for the BLR
and the continuum emission. The continuum polarization was estimated
by considering two spectral regions between 150 and 300 \AA\ wide
flanking the broad \Ha\ line. The width and location of the region of
integration was selected, on an object-by-object basis (see the
spectra presented in the Supplementary Material), in order to avoid telluric
absorption bands and, in some cases, the narrow emission lines
[O~I]$\lambda\lambda6300,6363$ or [S~II]$\lambda\lambda6716,6731$. The
Stokes parameters measured on the blue and red side were linearly
interpolated to obtain their values at the rest frame
\Ha\ position. All measurements of the polarized flux are at a
significance higher that 2$\sigma$, except for J110538 (see Table
\ref{tab3}).

Seven of the sources of our sample are in common with those studied by
\citet{berriman90}: they obtained white light polarization of the 114
PG QSOs between 1977 and 1986. In Figure \ref{berriman} we compare
their results with the continuum polarization from the VLT data. There
is close agreement between these measurements, with changes overall
consistent within the errors. Thus, the polarization of these QSOs
appears to be stable over $\sim$30-40 years.

We measured the BLR polarization integrating over a spectral range
spanning twice the \Ha\ FWHM of each given source. In order to isolate
the intrinsic BLR polarization we removed the underlying continuum
polarization: the linear fit to the continuum $I$, $Q$, and $U$
derived above was subtracted from the data before deriving the
polarization parameters. The measurements of the BLR polarized flux
are at a significance higher that 2$\sigma$, in all but two sources.
In addition, in J145108 the BLR polarization measurements are
compromised by the presence of a cosmic ray that is not effectively
removed by the reduction pipeline.

\subsection{Dilution from unpolarized emission}
The presence of unpolarized emission might affect the measurements of
the intrinsic nuclear polarization. In particular, the host emission
dilutes the continuum polarization, while the narrow lines (i.e., the
[N II] doublet and the narrow \Ha) might contaminate the BLR
polarization. We stress, however, that the presence of unpolarized
components only affects the percentage of polarization, while the
polarized flux and the polarization angle are not affected.

\subsubsection{Polarization dilution by the host galaxies}

Analysis of type I AGN SDSS spectra by \citet{stern12a} yields that at
the SDSS aperture the host dominates the AGN flux at 7000\AA, on
average, for luminosities $\log L_{\rm bol} \le 45 $, or equivalently
$\log L_{5100} \le 44 $, which holds for 3 of the objects in our
sample. However, there is still a significant contribution
(host/AGN$>30$\%) at $\log L_{5100} < 45 $, which holds for all of the
11 low redshift ($z <0.12$) objects in our sample.

In order to explore this issue more quantitatively we measured the
power-law slope $\alpha_{\rm opt}$ ($=d\ln I(\lambda)/d\ln \lambda$)
of the continuum in the SDSS spectra. While the higher luminosity
sources have generally blue spectra $\alpha_{\rm opt} \sim -1.5$,
several low luminosity QSOs are quite red (see Figure \ref{slopes}, left
panel), suggestive of a significant host contribution.

\begin{figure*}
\centering  
\includegraphics[width=3.3cm]{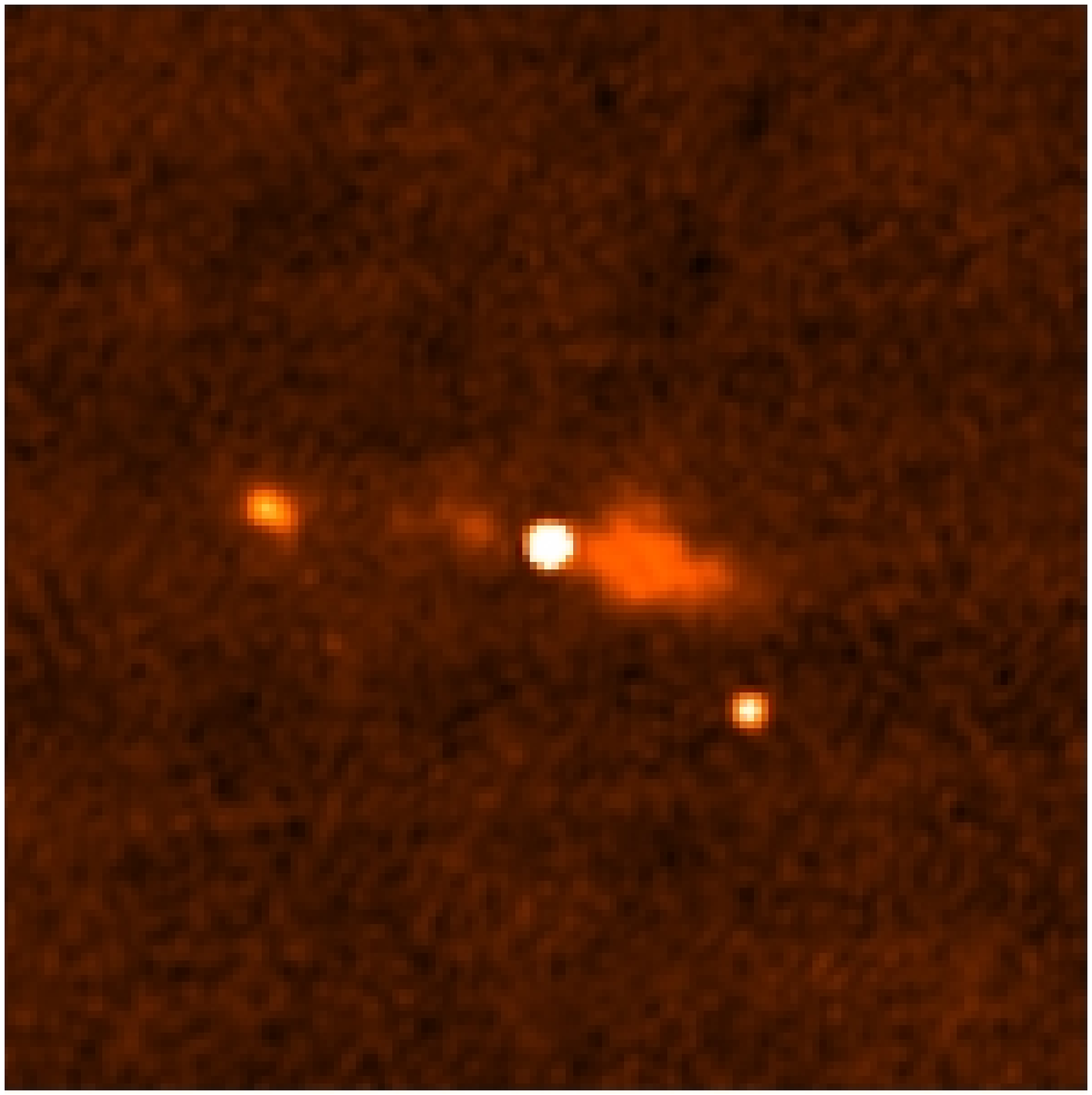}
\includegraphics[width=3.3cm]{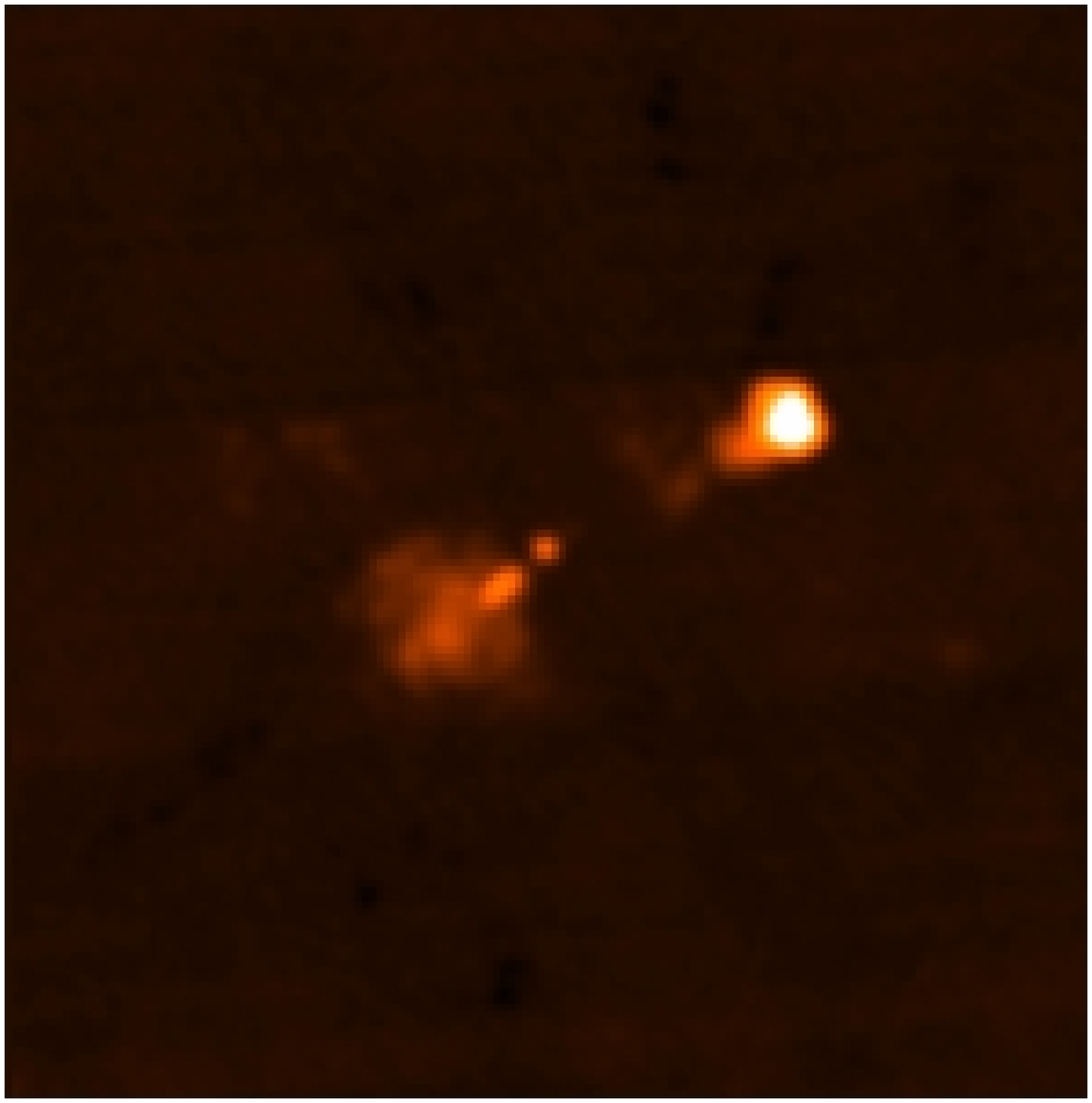}
\includegraphics[width=3.28cm]{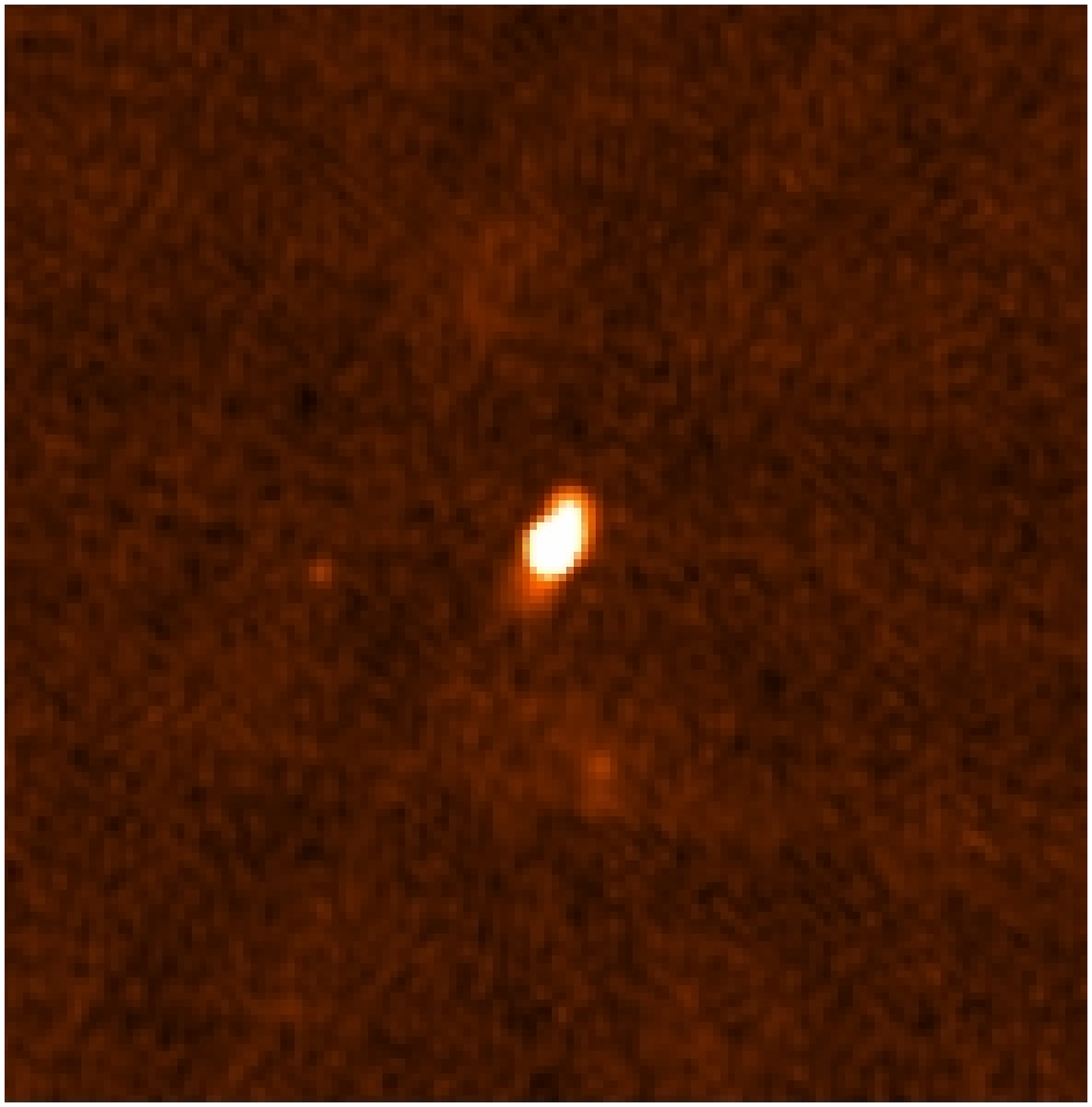}
\includegraphics[width=3.32cm]{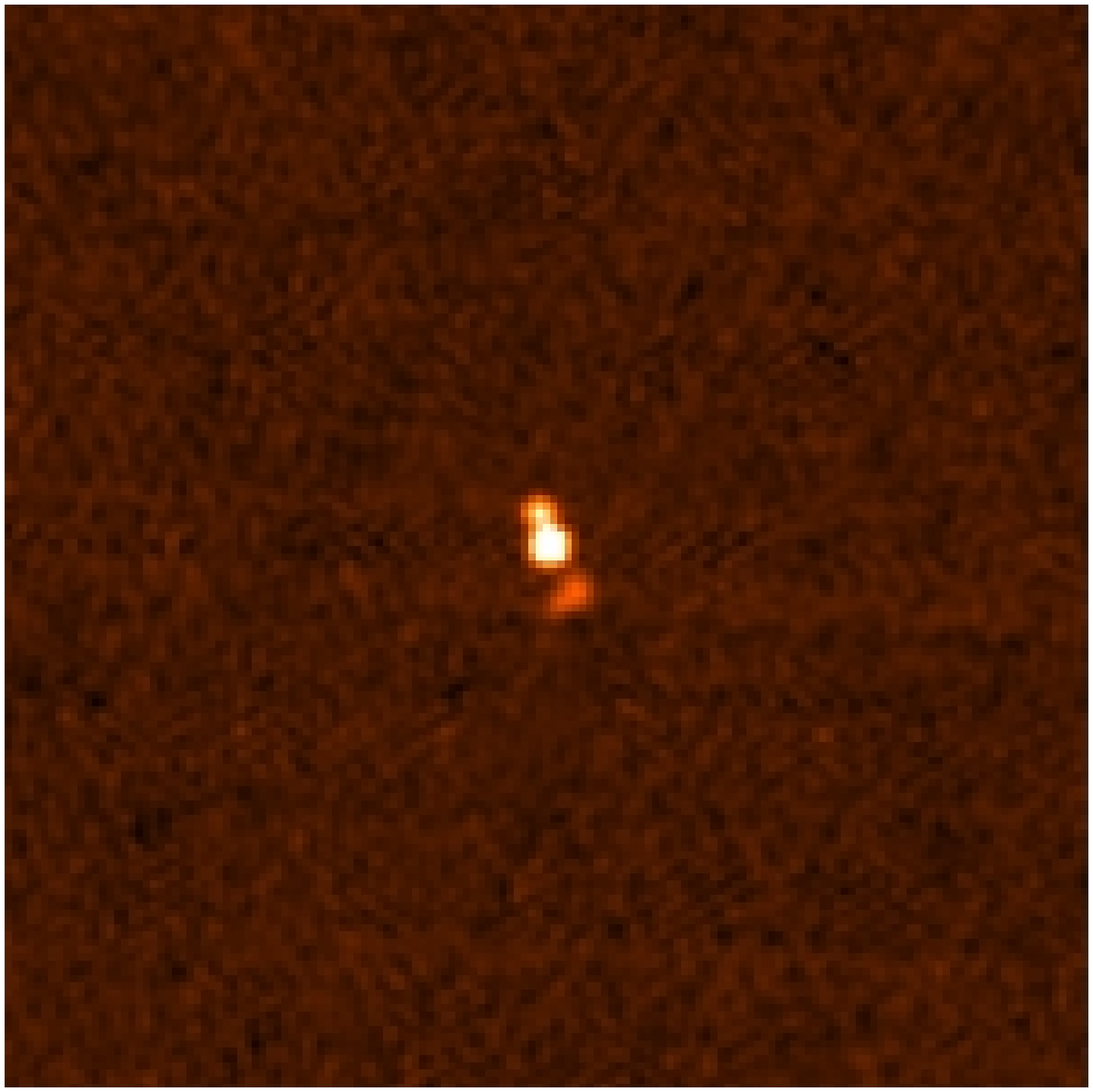}
\includegraphics[width=3.28cm]{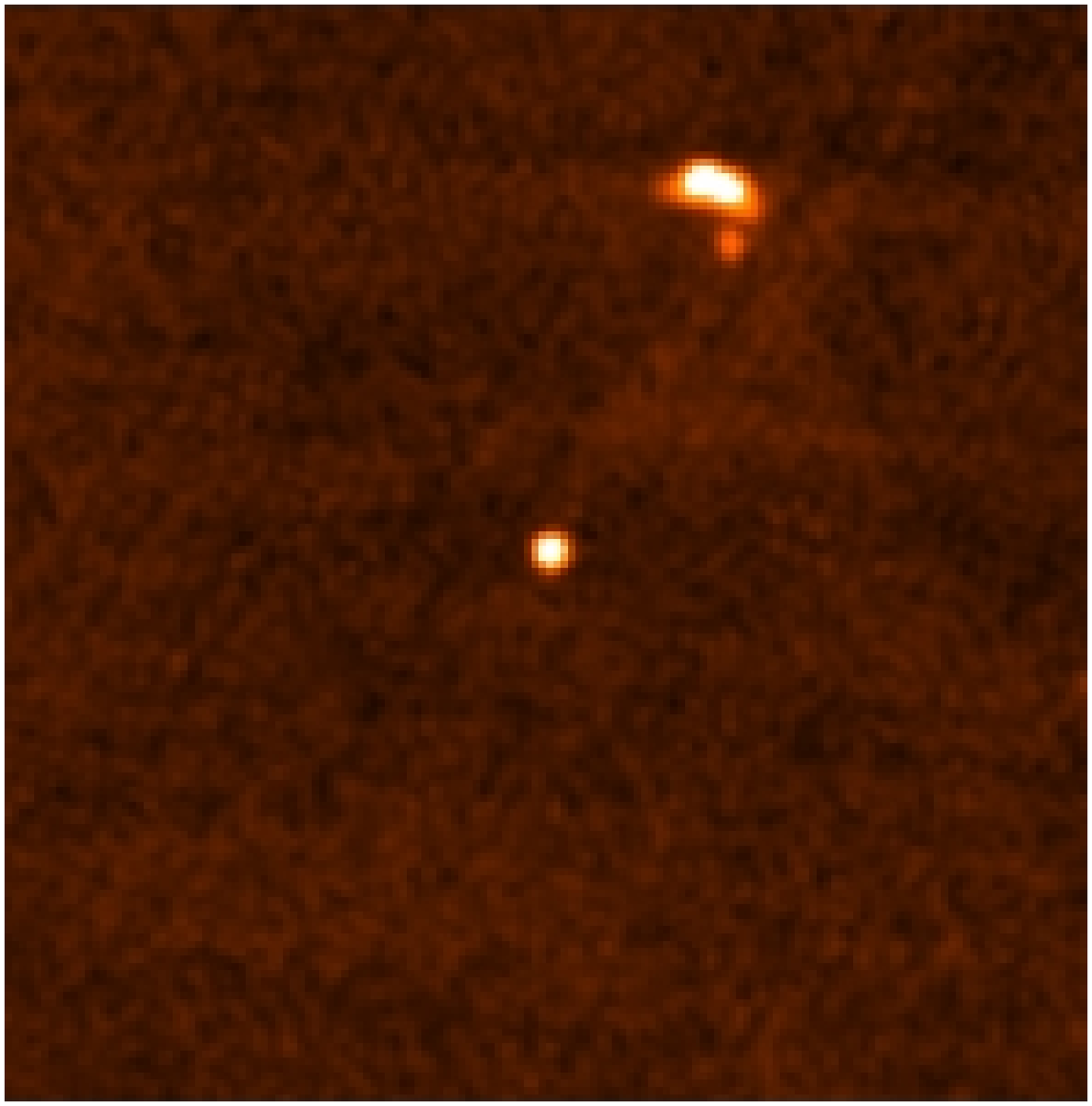}
\includegraphics[width=3.28cm]{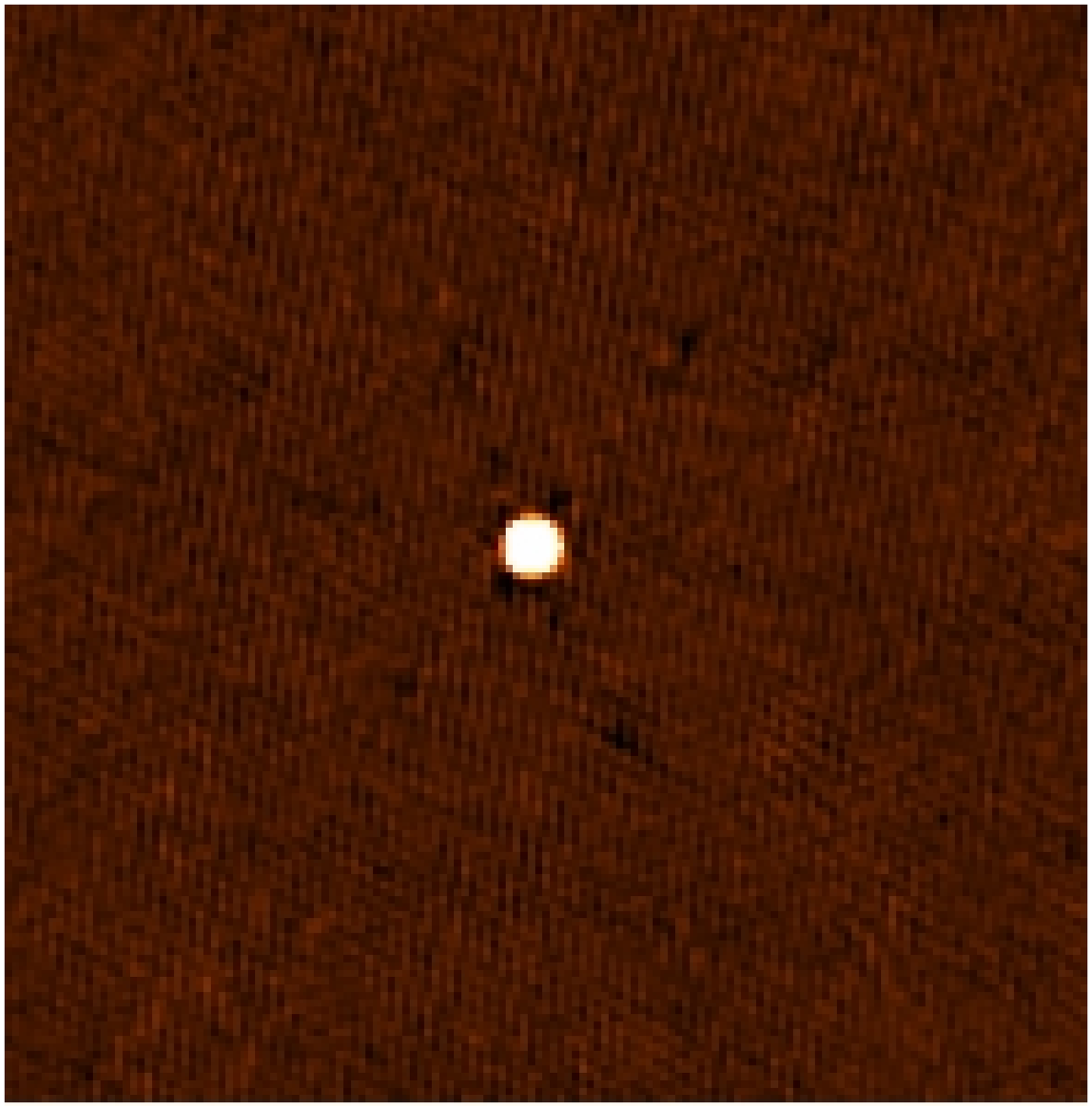}
\includegraphics[width=3.3cm]{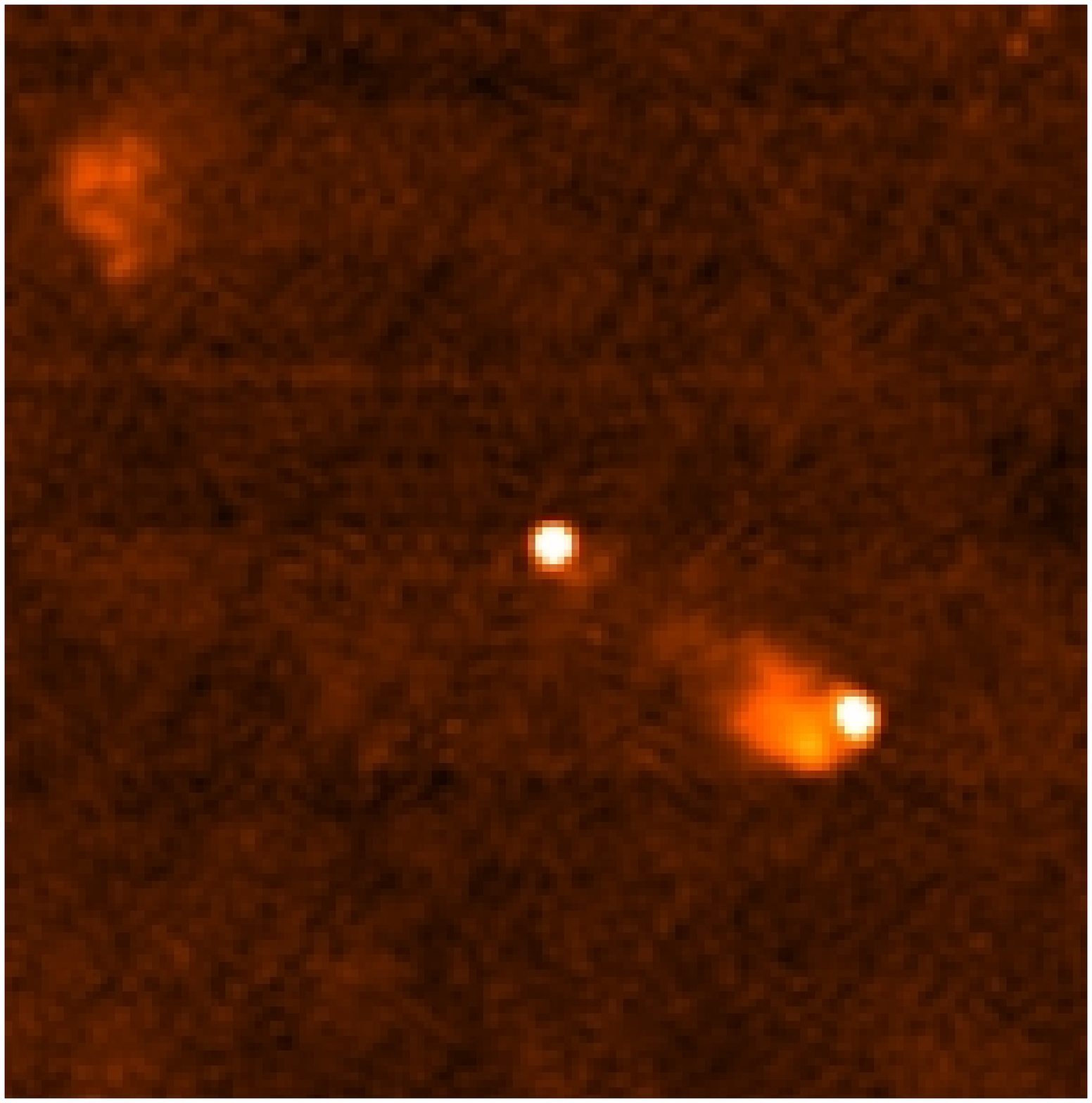}
\includegraphics[width=3.32cm]{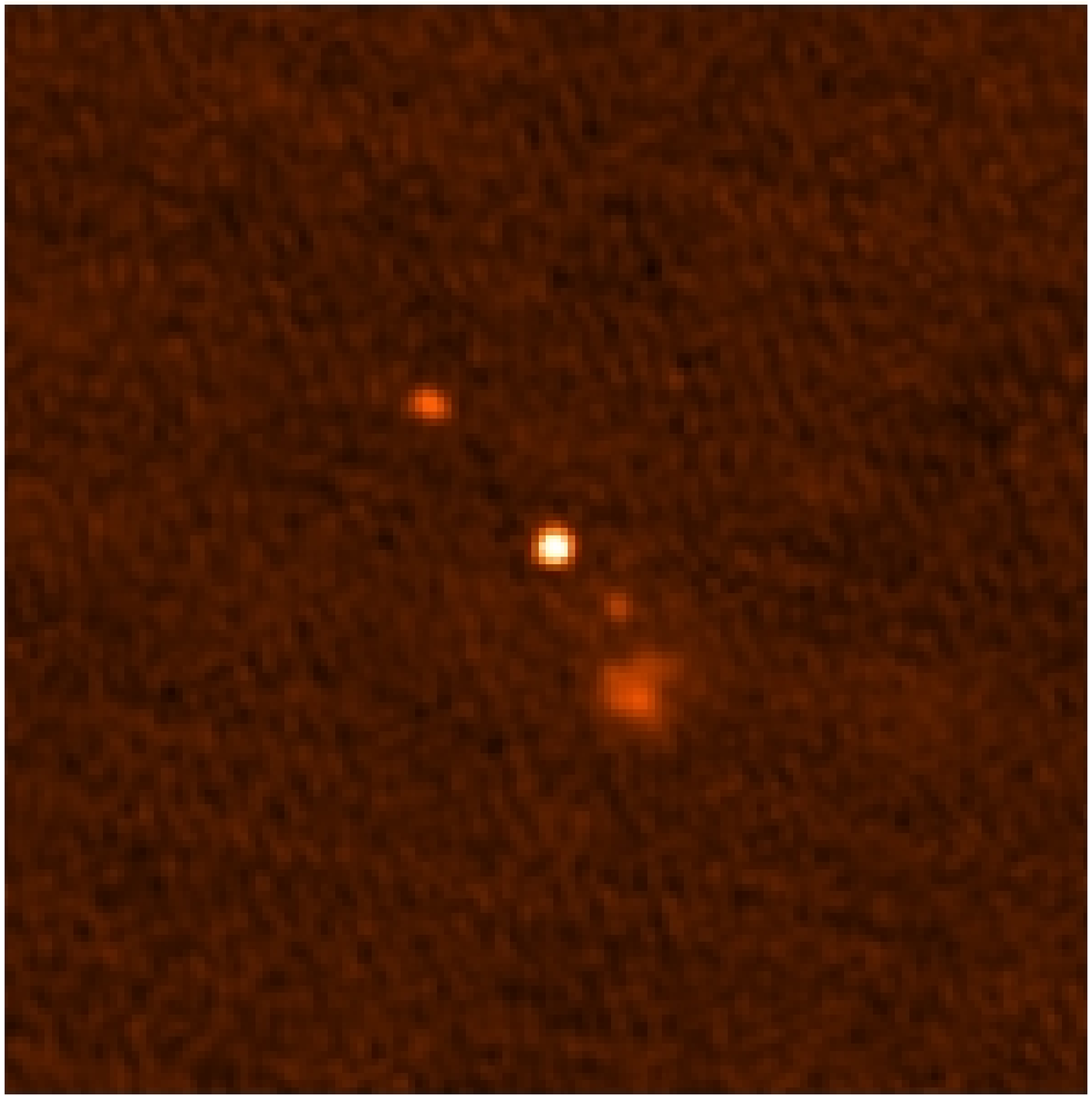}
\includegraphics[width=3.3cm]{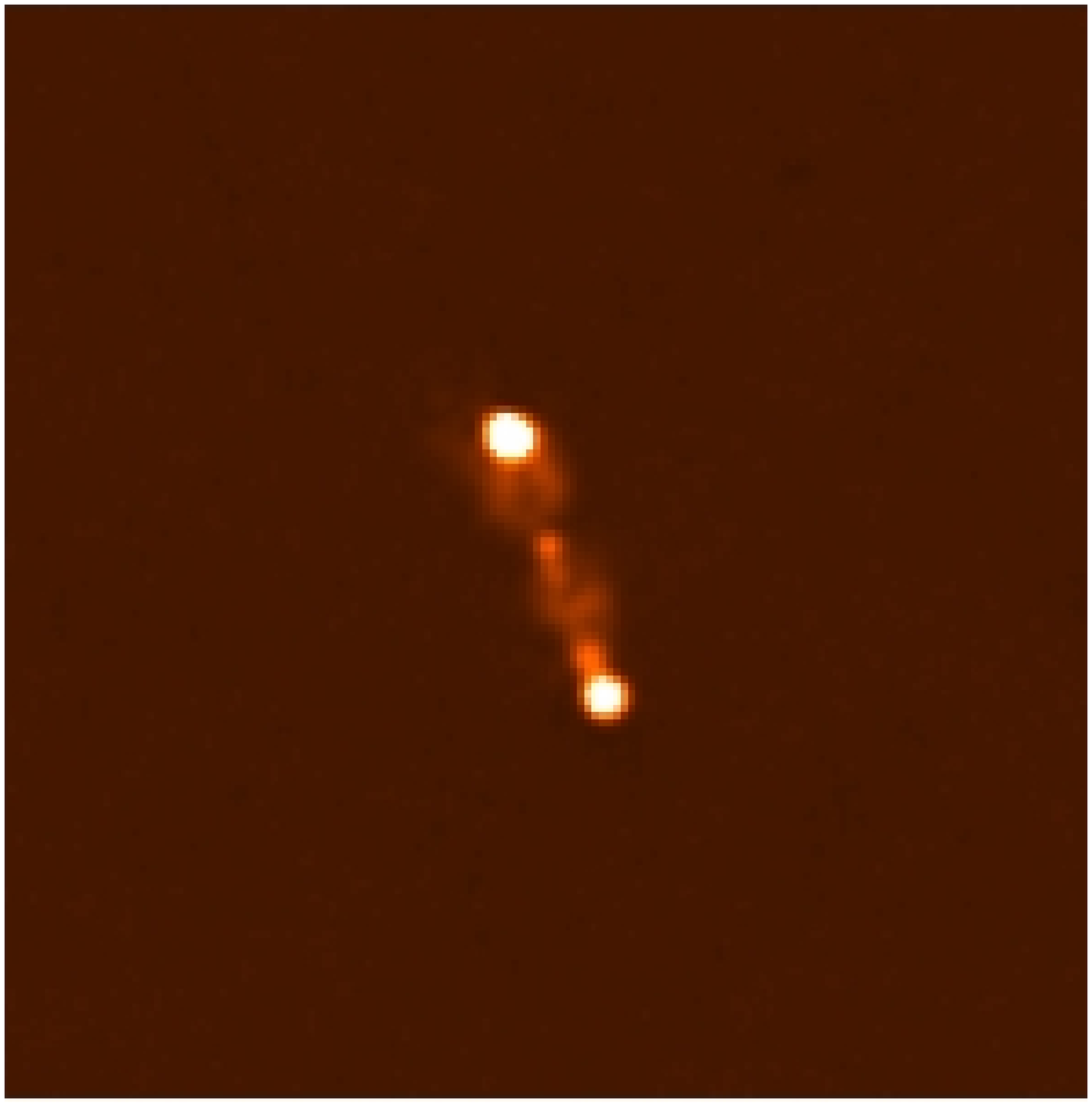}
\caption{FIRST images of the nine radio-loud QSO, all but one are
  extended radio sources, reaching sizes of $\sim$ 1 Mpc. Top, from
  left to right: J084600, J100726, J110538, J114306, and J140336.
  Bottom: J140700, J142735, J154007, and J154743. The fields of view
  are $4.5\arcmin \times 4.5\arcmin$. All objects, except J140700, are
  resolved, which allows to measure the radio axis and the core
  dominance. J140700 remains unresolved on VLBA scale, and is a
  GHz-peaked source \citep{devries97}, which indicates it is a
  physically compact source, rather than highly beamed.}
\label{radio}
\end{figure*}

The presence of the host manifests its effects also on the continuum
polarization (see Figure \ref{slopes}, right panel): while the QSO
with bluer spectra (and of higher luminosity) show a polarization up
to $\sim$ 2\%, the reddest (and least luminous) ones only reach $\sim$
1\%. This effect is likely caused by the dilution of the unpolarized
starlight. Due to the relatively large spread of the distribution of
the optical slopes, a correction to derive the genuine nuclear
polarization on an object-by-object basis is not deemed to be
sufficiently accurate. Furthermore, the extraction region used for the
VLT spectra (1$\arcsec \times 1\farcs5$) is significantly smaller than
the SDSS fiber (3$\arcsec$ in diameter) and thence the host
contribution in the spectropolarimetry data is reduced with respect to
the SDSS spectra.  In order to highlight possible dilution effects we
mark below the six red, low luminosity QSOs ($\alpha_{\rm opt} > -1$,
$\log L_{5100} < 44.5$) when relevant for our analysis.

\subsubsection{Polarization dilution from the narrow lines}
The presence of narrow emission lines within the spectral region of
integration might reduce the BLR polarization with
respect to the continuum. The narrow line region (NLR) extends to a
much larger scale than the continuum and the BLR and its emission
generally shows a very low level of polarization (e.g.,
\citealt{smith02}). As already mentioned, the resolution of the VLT
data we are using ($\sim$ 450 km s$^{-1}$) is not sufficient to
properly measure the NLR polarization. We can nonetheless address this
issue by considering the SDSS spectra which we used to decompose the
different emission lines and to estimate their EWs. The EW values span
a large range, but they are all generally smaller than $\sim$ 10\AA.

\begin{table}
\caption{Optical polarization and radio axis of the RL QSOs.}
\begin{tabular}{|l r r r r r r r}
\hline
Name  & $PA$ cont. & $PA$ BLR  & Radio &Core dominance\\
\hline
   J084600 &     67$\pm$ 2 &  67$\pm$ 3   &    82 & 0.35\\
   J100726 &     75$\pm$ 3 &  73$\pm$11   &   111 & 0.01\\
   J110538 &      ---      & 111$\pm$14   &   150 & 0.64\\ 
   J114306 &     28$\pm$14 &  45$\pm$ 6   &    24 & 0.70\\
   J140336 &    132$\pm$ 1 &  84$\pm$ 2   &   156 & 0.20\\
   J140700 &    136$\pm$ 2 &  70$\pm$ 2   & unres.& --- \\
   J142735 &     67$\pm$ 2 &  44$\pm$ 3   &    56 & 0.12\\
   J154007 &     48$\pm$ 4 &  56$\pm$ 3   &    38 & 0.38\\
   J154743 &     18$\pm$ 1 &  66$\pm$ 3   &    20 & 0.01\\
\hline                                                                          
\end{tabular}
\label{tabrl}
\end{table}

By assuming that the NLR emission is unpolarized the percentage of
polarization of the BLR is reduced with respect to its intrinsic value
by a factor $1+EW_{\rm NLR}/EW_{\rm BLR}$.  For the sources considered
the $EW_{\rm BLR}$ ranges from 300 to 900 \AA\ and thence the NLR
effects on the integrated BLR polarization measurements are
negligible.

Nonetheless, the NLR might affect the polarization within individual
bins in the spectra shown in the Supplementary Material, particularly
those corresponding to the \Ha\ and [N~II] doublet. However, as
already noted above, the only effect is a reduced value of the
percentage of polarization.

\subsection{The effects of interstellar polarization}

Dust grains aligned by magnetic fields in our Galaxy are known to
polarize optical light (e.g., \citealt{mathewson70}) and it is
important to assess the effects of interstellar polarization (ISP) on
the QSO properties, in particular on the continuum polarization. In
fact, our estimates of the BLR polarization are not affected by the
ISP because, by subtracting the continuum polarization, we effectively
removed also the ISP contribution.

In the Appendix A, we estimate the ISP for our sources based on the
polarization maps produced by the {\it Planck} satellite. We conclude
that the effects of polarization induced by the interstellar medium
have generally a negligible effect on the QSOs polarization
properties. We applied the ISP correction only to three sources
(namely J154019, J155444, and J214054) where the observed and
corrected percentage of polarization differ by more than
2$\sigma$.

\subsection{Continuum vs. BLR polarization}
In Figure \ref{pablrcont}, left panel, we compare the continuum
polarization percentage and position angle, interpolated to 6563 \AA,
with the BLR polarization. Recall that, in order to isolate the net
BLR polarization, we derived its polarization parameters after
subtraction of the underlying continuum polarization in the Stokes
parameters space.

The polarization PA of the continuum and of the BLR differ, in most
sources, by less than 30$^\circ$. More quantitatively, the
r.m.s. difference between these two values is $\sim 27^\circ$.

The percentage of polarization of the BLR is generally smaller than
that measured in the continuum (Figure \ref{pablrcont}, right
panel). The median ratio $P_{\rm{cont}}/P_{\rm{BLR}}$ is
1.7. Furthermore, as discussed above, the dilution from the host
galaxy in the red, low luminosity QSOs (the empty symbols) can lead to
an underestimate of $P_{\rm{cont}}$, further increasing the difference
between continuum and BLR polarization.

\subsection{The radio-loud QSO}
\begin{figure}
\psfig{figure=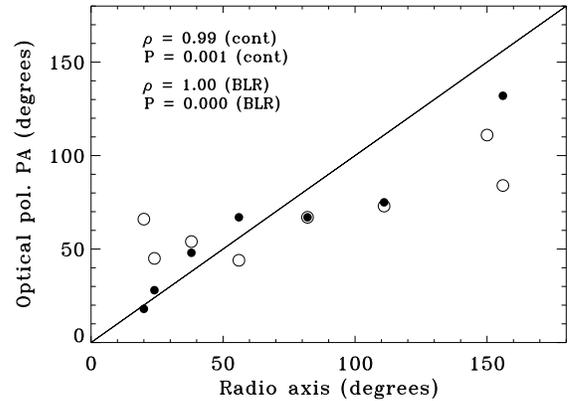,width=0.99\linewidth}
\caption{Comparison of the PA of the radio emission and the optical
  continuum polarization (filled dots) and of the BLR (empty dots) for
  the resolved radio loud QSO. The PA of the Radio and the optical
  continuum polarization differ by 18$^\circ$ on average. The parallel
  position angles indicate the scattering is by an equatorial medium,
  in a plane perpendicular to the jet axis.}
\label{ropt}
\end{figure}

\begin{figure}[h]
\psfig{figure=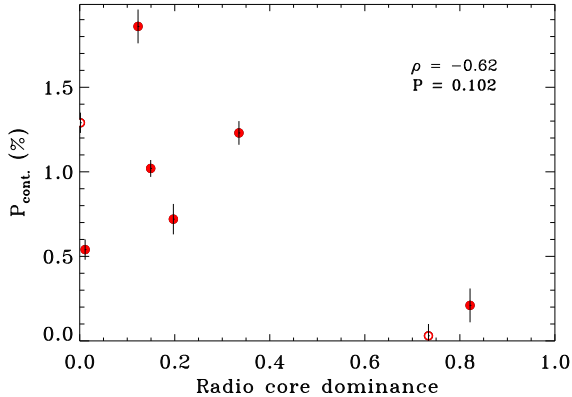,width=0.99\linewidth}
\caption{Comparison of the radio core dominance (a proxy for the jets
  orientation) and the optical continuum polarization for the radio
  loud QSO. A moderate negative correlation is present ($\rho =-0.62$,
  P = 0.10). The empty dots mark the two red low luminosity RLQs.
  J140700, which is a young GHz-peaked spectrum source
  \citep{devries97}, is omitted from this plot, as its compactness is
  physical and not a projection effect. The low optical polarization
  of the high core dominance objects suggest the polarization is
  induced by scattering from a medium with an axial symmetry,
  presumably due to cancellation in a face-on geometry.}
\label{coredom}
\end{figure}

There are 9 radio-loud QSO in our sample. Their radio images from the
FIRST are presented in Figure \ref{radio}. They are all associated
with extended radio sources, with the only exception being J140700, a
young GHz-peaked spectrum source \citep{devries97} that, nonetheless,
has one of the largest value radio-loudness ($R = 580$).

\begin{figure*}
\psfig{figure=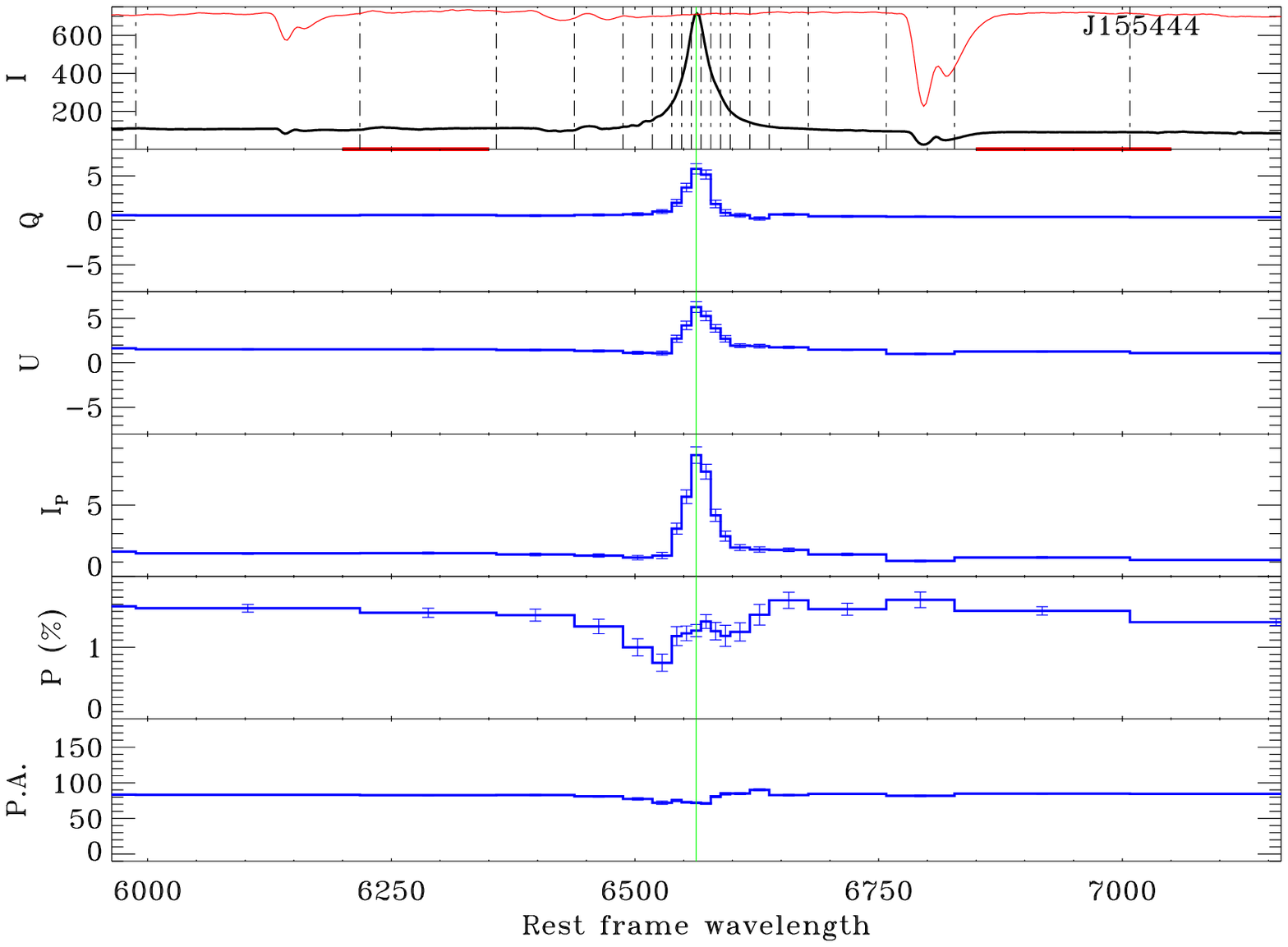,width=0.495\linewidth}
\psfig{figure=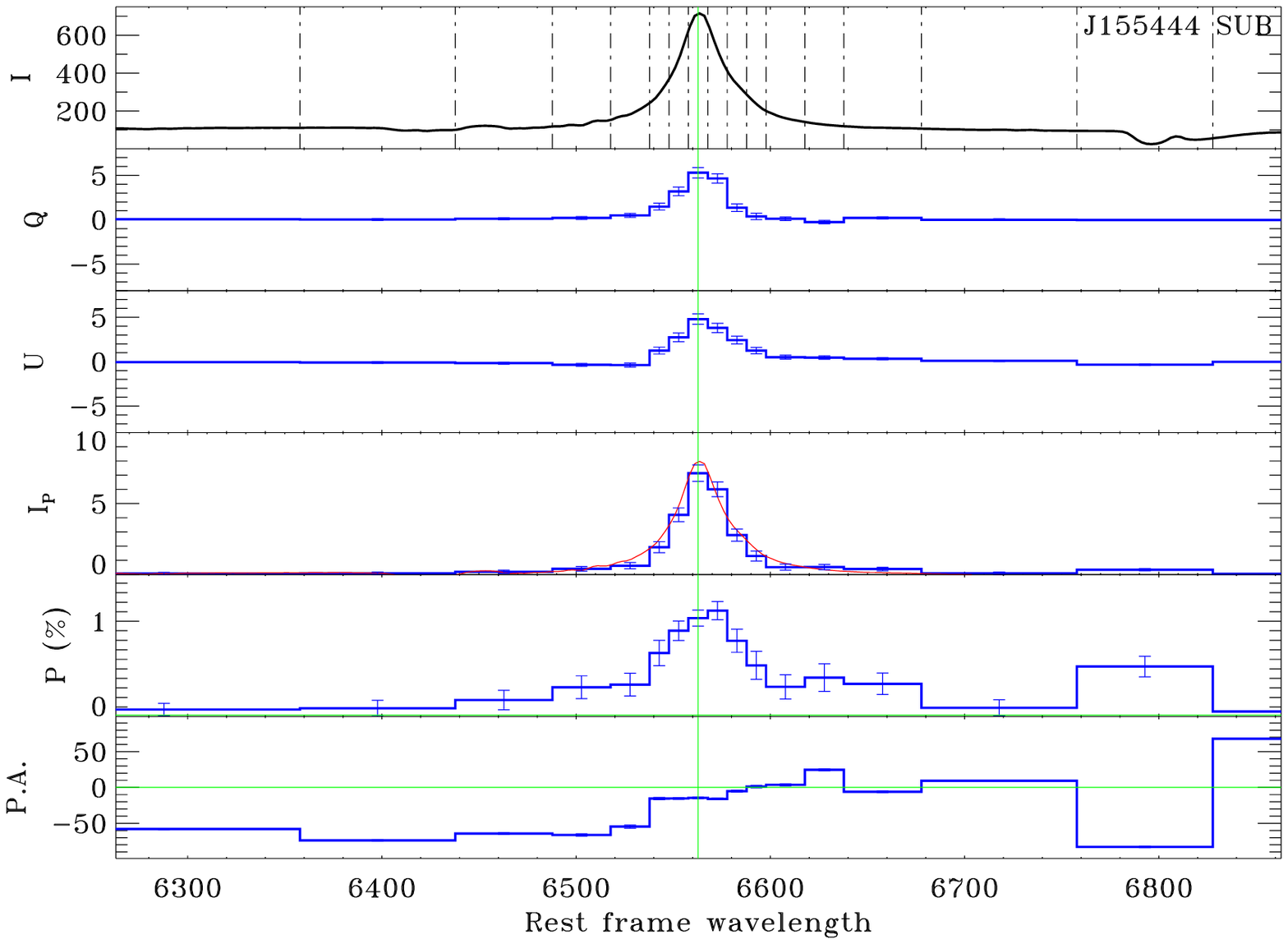,width=0.495\linewidth}
\caption{Example of the spectropolarimetric parameters obtained for
  each source of the sample, here we show the results for J155444.
  Left: top three panels: I, Q, and U Stokes parameters shown over a
  spectral region 1200 \AA\ wide. I is shown at full spectral
  resolution, the vertical dashed lines mark the boundaries of the
  regions used for the rebinning. Wavelengths are rest frame in \AA,
  fluxes are in arbitrary units. The green vertical line marks the
  location of the \Ha\ line. The red curve in the top panel represents
  the ratio between the SDSS and the corrected VLT spectra in which
  can be located the telluric absorption bands. The two horizontal red
  lines locate the spectral regions used to estimate the continuum
  polarization. Bottom three panels: polarized flux $I_P$, percentage
  of polarization $P$, and polarization position angle PA. Right:
  same as the left panels, but after subtraction of the continuum
  polarization and with a reduced spectral region (600 \AA\ wide).
  The total flux is plotted in red over the polarized flux. The PA is
  now measured with respect to the continuum PA.}
\label{spectra1}
\end{figure*}

Figure \ref{ropt} compares the radio axis and optical continuum polarization
 PA for the RLQs: their difference is always smaller than $\sim
35^\circ$, with a r.m.s. of $\sim 18^\circ$. The BLR polarization PA
(including J110538 for which PA$_{\rm cont}$ cannot be measured) is also
aligned with the radio axis, with a larger r.m.s. difference of $\sim
31^\circ$.

A further connection emerges when comparing the polarization and radio
properties of the RLQs. A suggestive trend emerges between their radio
core dominance, defined as the ratio between the nuclear core
component in the FIRST images and the total NVSS flux (both at 1.4
GHz) and the continuum polarization (see Figure \ref{coredom}), which
decreases for larger values of core dominance. The Spearman rank test
indicates a moderate negative correlation between these two quantities
with $\rho =-0.62$ and a chance probability of 0.10. The low
optical polarization of the high core dominance objects suggest the
polarization is induced by scattering from a medium with an axial
symmetry, presumably due to cancellation in a face-on geometry.

\section{Spectrally resolved polarization measurements}
\label{respol}

\begin{figure*}
\psfig{figure=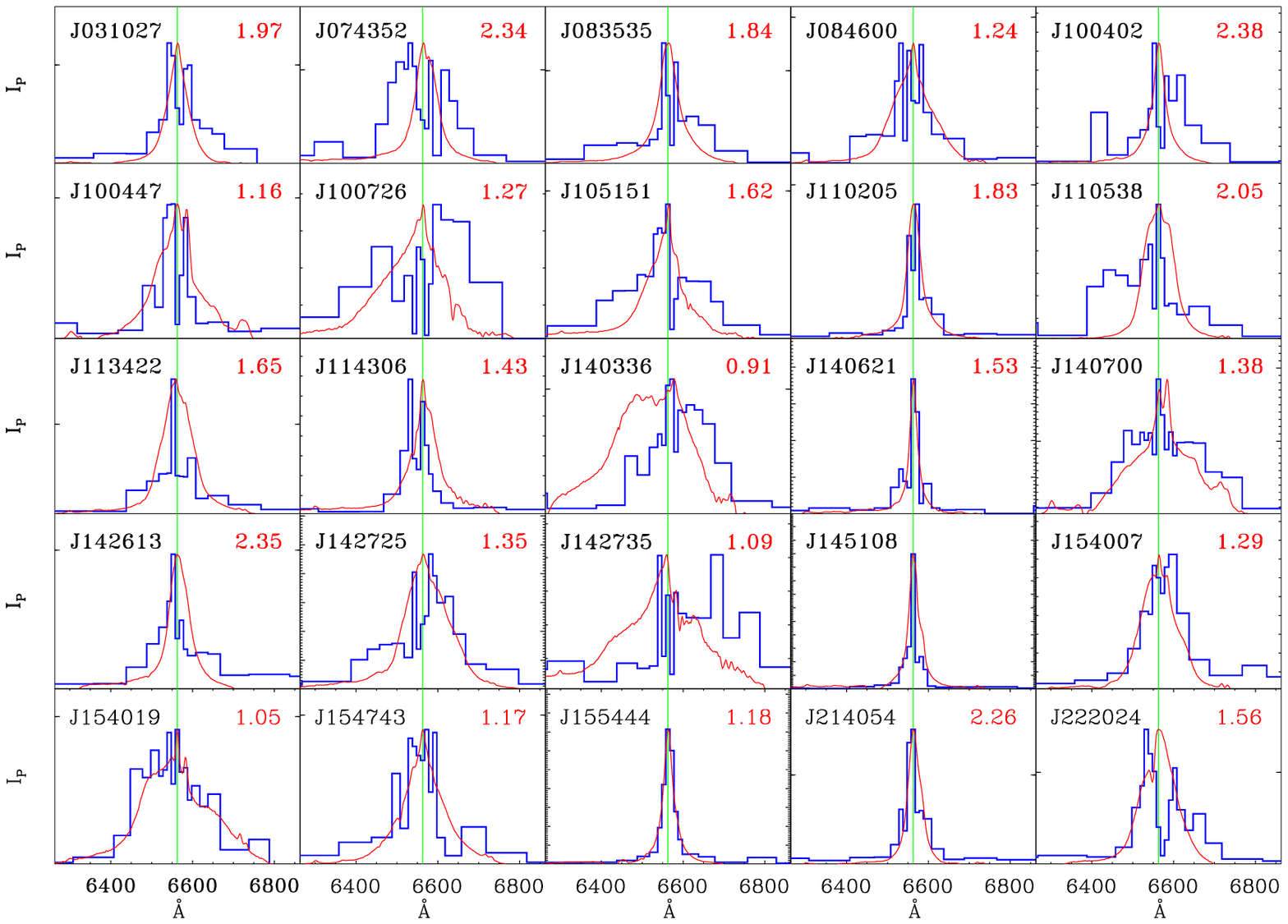,width=0.99\linewidth}
  \caption{Summary of the spectropolarimetric results for the 25
    QSOs. The red curve shows the total flux, the blue histogram
    the polarized flux after continuum subtraction. The green
    vertical line marks the location of the narrow \Ha\ line. The red
    number on the top right of each panel shows the ratio between the
    width in polarized and in total flux ($W(I_P)$ and $W(I)$,
    respectively; see below for details). Note that this ratio is
    generally above unity. This reflects the higher velocity
    dispersion from the equatorial vantage point of the scattering
    medium, compared to the direct view. The broadest line objects
    (J100447, J140336, J142735, J154019) show a ratio close to unity,
    which suggest the direct view is close-to equatorial. The narrower
    line objects generally show a large ratio (e.g. J100402, J214054),
    indicating a close-to face-on view, but some have a low ratio
    (e.g. J155444), indicating a high inclination view. The scattered
    line is significantly shifted to the red in some objects (J140336,
    and possibly J100726 and J142735), which may indicate scattering
    from an outflow.  The full spectropolarimetry results are
    presented in the Supplementary Material}
\label{panels}
\end{figure*}

Due to the general low level of polarization of the sources,
spectrally resolved polarization measurements must be obtained by
decreasing the resolution of the data. The spectra were rebinned in
order to obtain, where possible, statistically significant
measurements. The bin sizes used are smaller across the BLR and they
increase far from the spectral regions of interest. Initially, the bin
size was set to 10\AA\ (rest frame, corresponding to $\sim$450 \kms)
the original instrumental resolution of the spectra. Five narrow bins
across the core of the BLR enable us also to separate the possible
polarization contribution of the \Ha\ and \nii\ narrow lines. At
larger spectral distances from the BLR core, the 10\AA\ grid is
maintained when the continuum subtracted polarized flux has a
significance $> 3 \sigma$. When this requirement is not met, the bin
size is increased until 1) the required significance is reached or 2)
when the bin width reaches half its velocity difference from the BLR
core: i.e., a bin whose center is located 100\AA\ from the BLR core
cannot be broader than 50\AA. This strategy maintains a fixed minimum
relative velocity resolution across the binned spectra. The procedure
continues at larger velocity offsets on both sides of the spectrum
independently.

In Figure \ref{spectra1} we present one example of the results
obtained, i.e., the Stokes parameters, the polarized flux $I_P$, the
degree of polarization $P$, and the polarization angle over a spectral
region extending $\pm 600$ \AA\ (rest frame, $\sim$27,000 \kms) on
each side of the \Ha\ line center. The continuum subtracted spectra
are shown in the right panels. Here we use the continuum PA as
reference for the angles. The results for all sources are shown in the
Supplementary Material. In Figure \ref{panels} we show a sub-set of the spectra
obtained, comparing the polarized flux after continuum
subtraction with the total flux for all sources.

\begin{figure*}
\psfig{figure=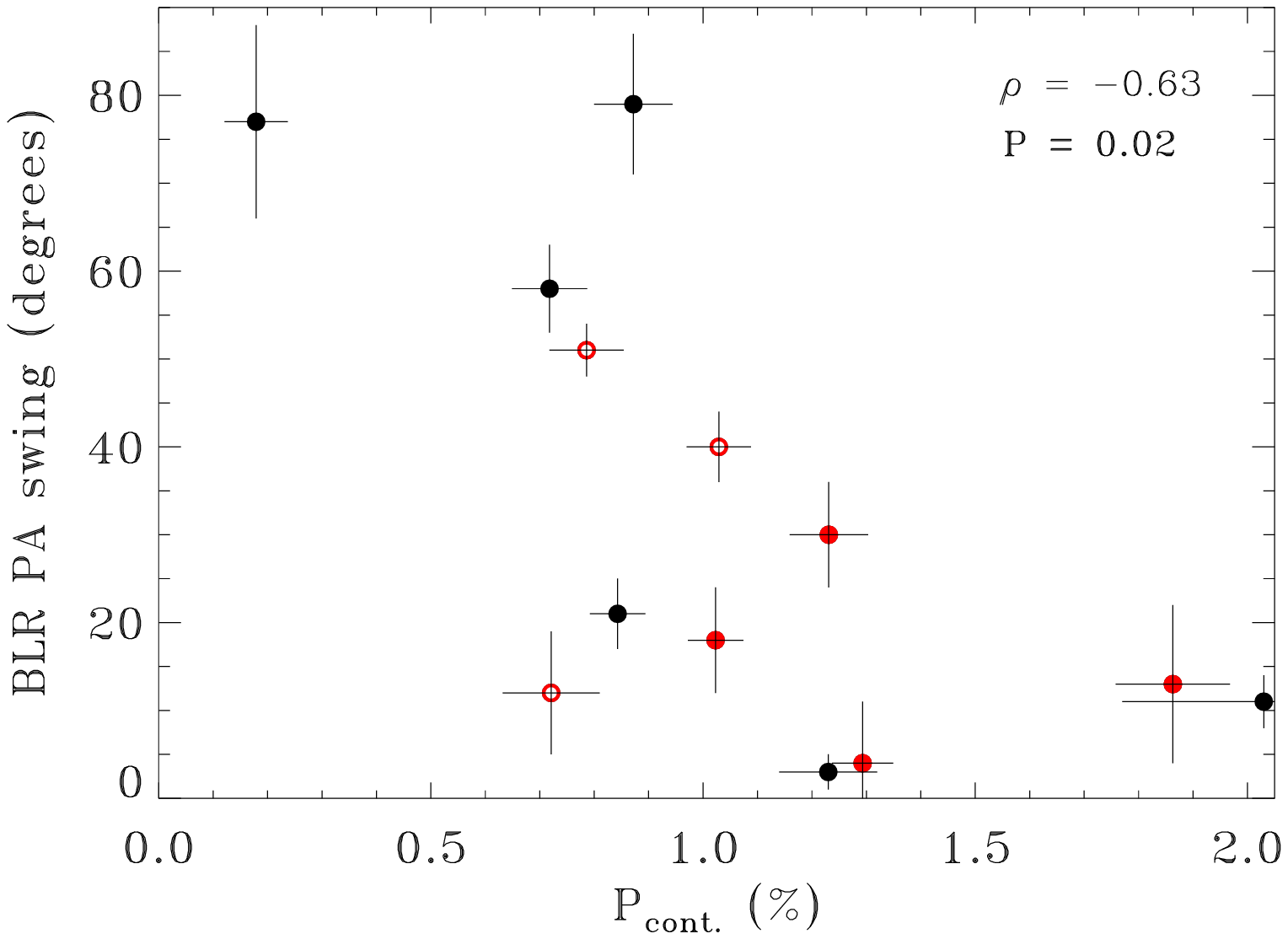,width=0.49\linewidth}
\psfig{figure=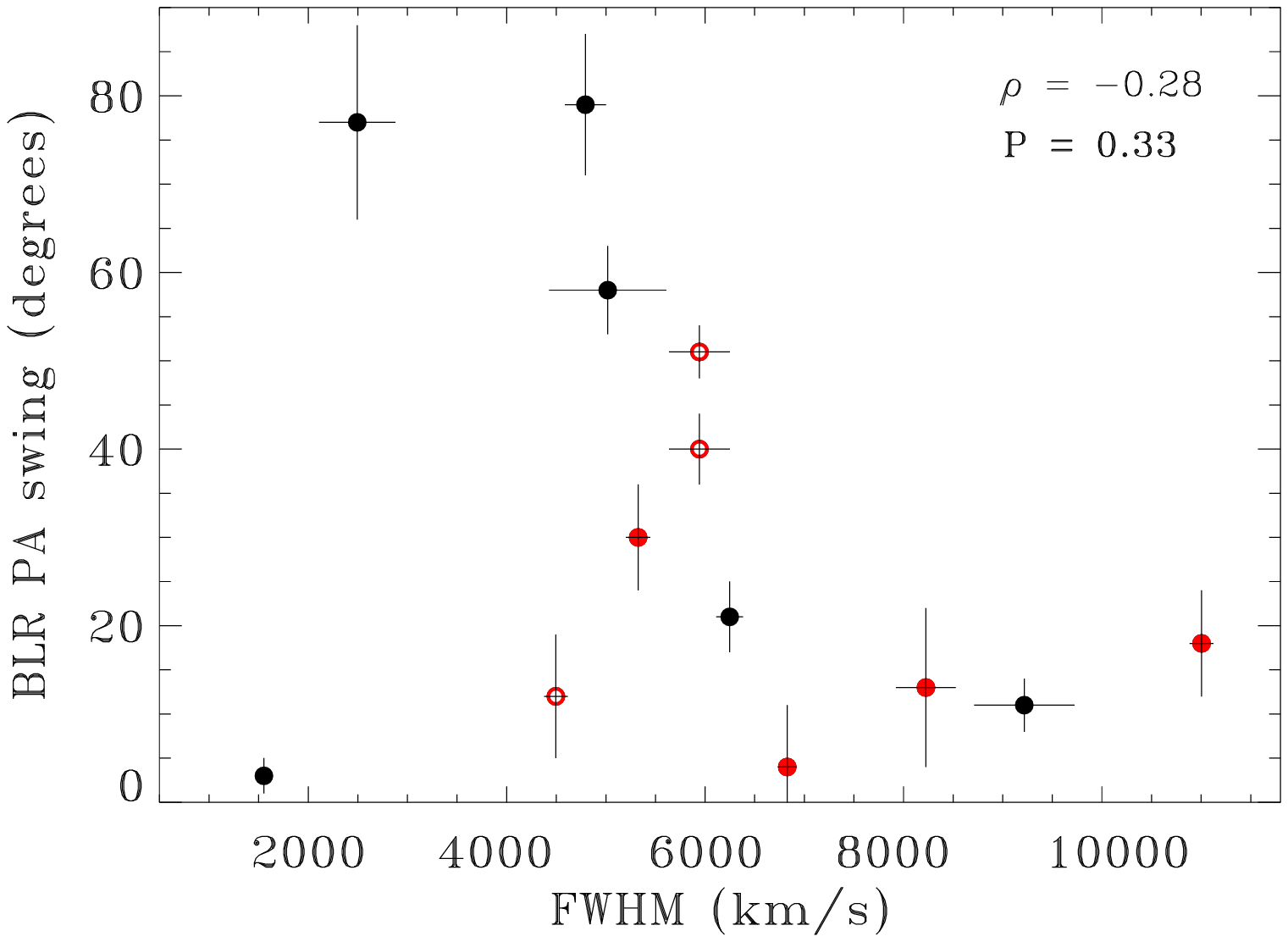,width=0.49\linewidth}
  \caption{The dependence of the swing in the polarization PA between
    the red and the blue wing of \Ha\ on: (left) the continuum
    polarization and (right) the \Ha\ line width. There is a highly
    significant negative correlation between PA swing and continuum
    polarization, as expected for an equatorial scattering medium.
    The relation of the PA swing and line width is weaker, as the line
    width is affected by both the inclination and the black hole
    mass. The largest swings are seen in the sources with FWHM
    $\lesssim 6000$ \kms, which suggests they are all high inclination
    objects (see also Figure 16). The low luminosity red QSOs, with
    $\alpha_{\rm opt} > -1$ and $\log L_{5100} < 44.5$, are
    represented by empty circles (see Figure 4). The red symbols are the
    RLQs. }
\label{swing}
\end{figure*}

\begin{figure}
\psfig{figure=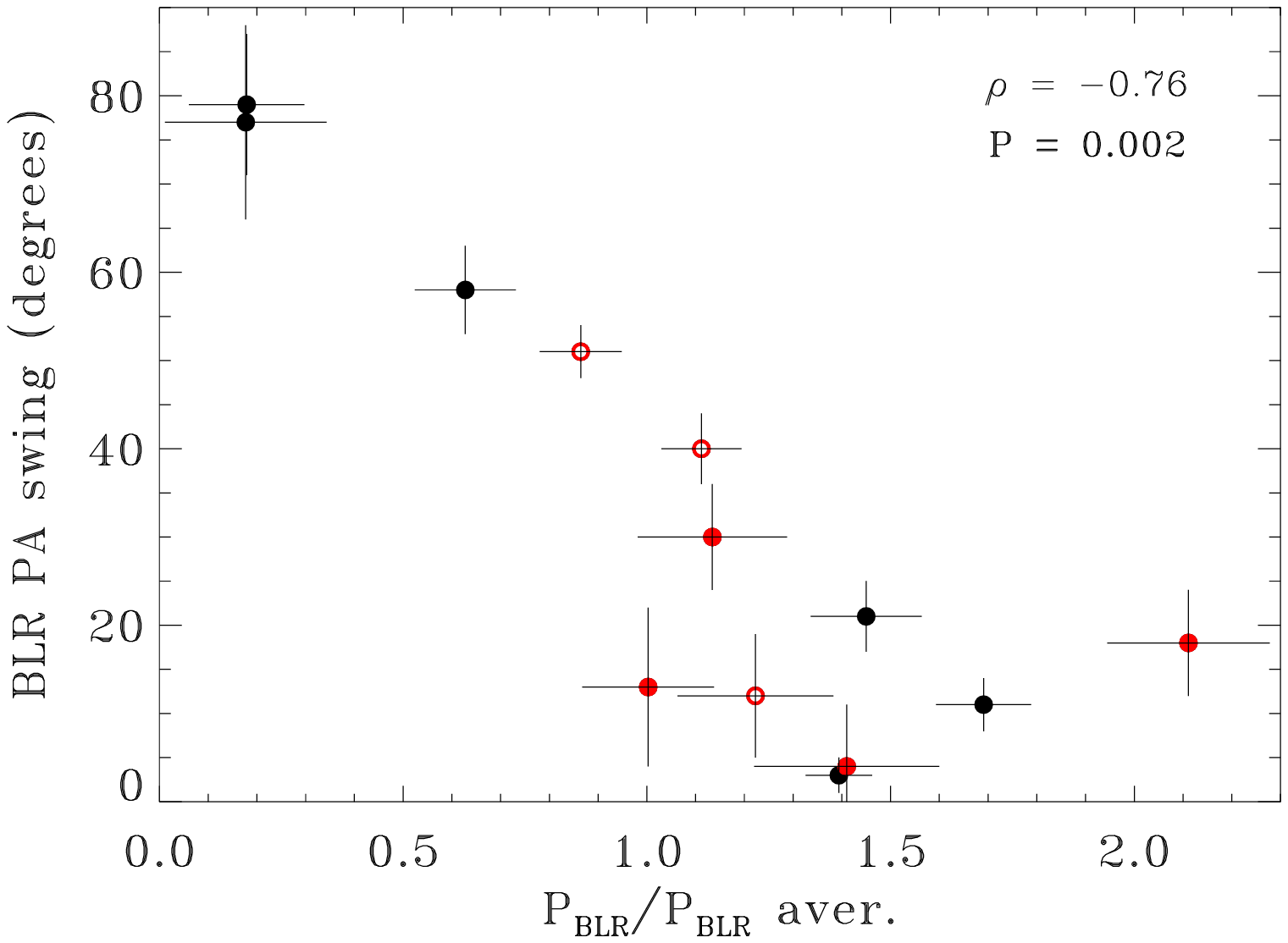,width=0.99\linewidth}
\caption{The ratio between the \Ha\ polarization, which is measured from
the integrated line flux, and the averaged wavelength dependent \Ha\ 
polarization, versus the PA swing. The tight correlation
indicates that the integrated \Ha\ polarization is strongly affected 
by the PA swing cancellation (i.e., summing polarization vectors oriented 
at different angles), which explains some of the reduced \Ha\ polarization
compared to the continuum polarization.   }
\label{cancellation}
\end{figure}

A common feature in the spectra is a sign change of $Q$ and $U$ across
the broad \Ha\ line profile, corresponding to a rotation of the PA. In
order to estimate the PA swing across the BLR we integrated the Stokes
parameters on the red and blue sides of the broad \Ha\ line. We
experimented with various integration windows. The results reported
refer to the case in which we split the BLR into a blue and a red
region, each of them with a width equal to the \Ha\ FWHM. This
measurement is possible for 13 out of 25 sources in which the
polarization is defined on both regions with a significance larger
than 2$\sigma$.

Note that in two sources (J031027 and J105151) the BLR polarization is
very low and we cannot measure the integrated BLR PA. Nonetheless, we
were able to estimate the PA swing. In these sources the very large
differences in PA causes polarization cancellation when integrating
the whole BLR, but the individual wings are significantly polarized.

Figure\ref{swing} compares the amplitude of the PA swings, $\Delta
PA_{\rm BLR}$, and the continuum polarization. The largest swings are
seen in the least polarized sources, while sources with $P_{\rm cont}
\gtrsim 1\%$ have swings smaller than $\lesssim 30^\circ$. The
Spearman rank test indicates a significant negative correlation
between these two quantities: for the 13 QSOs for which the PA BLR
swing can be measured, we obtain $\rho =-0.63$ and a chance
probability of P=0.02. If the red low-luminosity QSOs are excluded,
the correlation values are $\rho =-0.76$ and P=0.011 for 10 objects.
$\Delta PA_{\rm BLR}$ is not significantly correlated with the
\Ha\ width ($\rho$ = -0.28, P=0.33), but, nonetheless, the largest
swings are seen in the sources with FWHM $\lesssim 6000$ \kms.

The PA swings across the BLR suggest that the lower integrated BLR
polarization with respect to the continuum discussed in Sect. 4.2
might be due to cancellation, resulting from the presence of
polarization vectors oriented at different angles. To test this idea
we introduce a different estimate of the BLR polarization: instead of
integrating the Q and U Stokes parameters over the BLR spectral
region, we measured the ``average'' percentage of BLR polarization
from the $P$ values obtained from the rebinned spectra.  These
measurements are reported in Table 3, where we also give in
parenthesis the observed rms of $P$ across the BLR. While these values
are affected by the polarization positive bias, they are less
influenced by the PA changes. We then tested the importance of
cancellation by comparing the ratio between integrated and averaged
BLR polarization and the PA swing across the BLR (see
Fig. \ref{cancellation}). Apparently, the sources with the largest
angle swings are those where the BLR polarization is more reduced
respect to the averaged value. Indeed, a strong significant negative
($\rho=-0.64$, P=0.018 for 13 objects) trend is present.

\subsection{A comparison between the BLR width in total flux and polarized light}

\begin{figure}
\psfig{figure=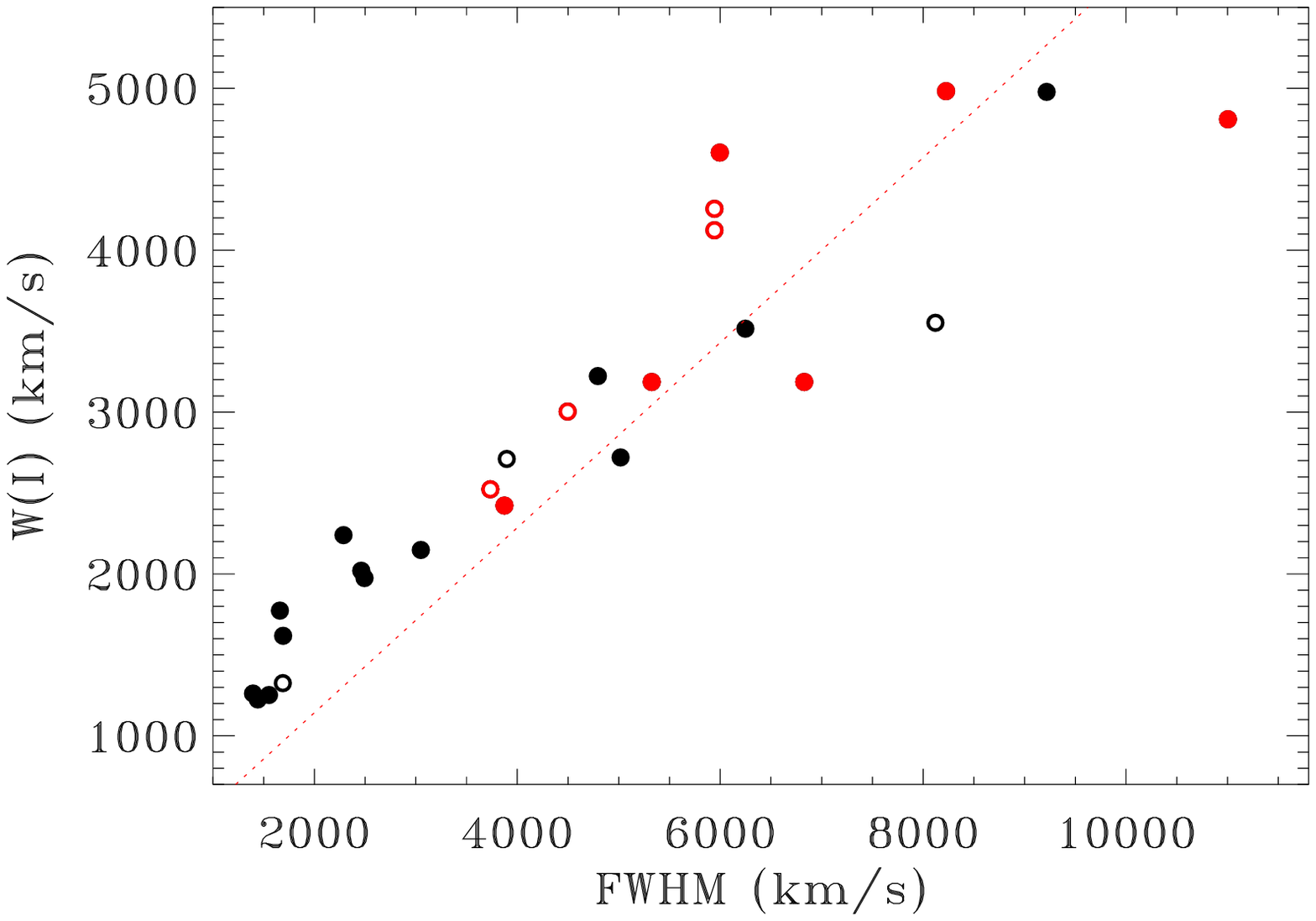,width=0.99\linewidth}
\caption{Comparison of the broad \Ha\ FWHM and $W(I)$ - the IPV width,
  which is the separation between the 25$^{\rm th}$ and 75$^{\rm th}$
  percentiles of the integrated line flux. The two parameters follow
  roughly a ratio of 1.75 (dotted line). The ratio is closer to unity
  at the lowest FWHM objects, due to the more prominent line wings in
  these sources. Thus, although $W(I)$ provides a good measure of the
  line width, it can deviate systematically from the FWHM in very
  narrow or very broad line objects.}
\label{ipvi-fwhm}
\end{figure}

\begin{figure*}
\psfig{figure=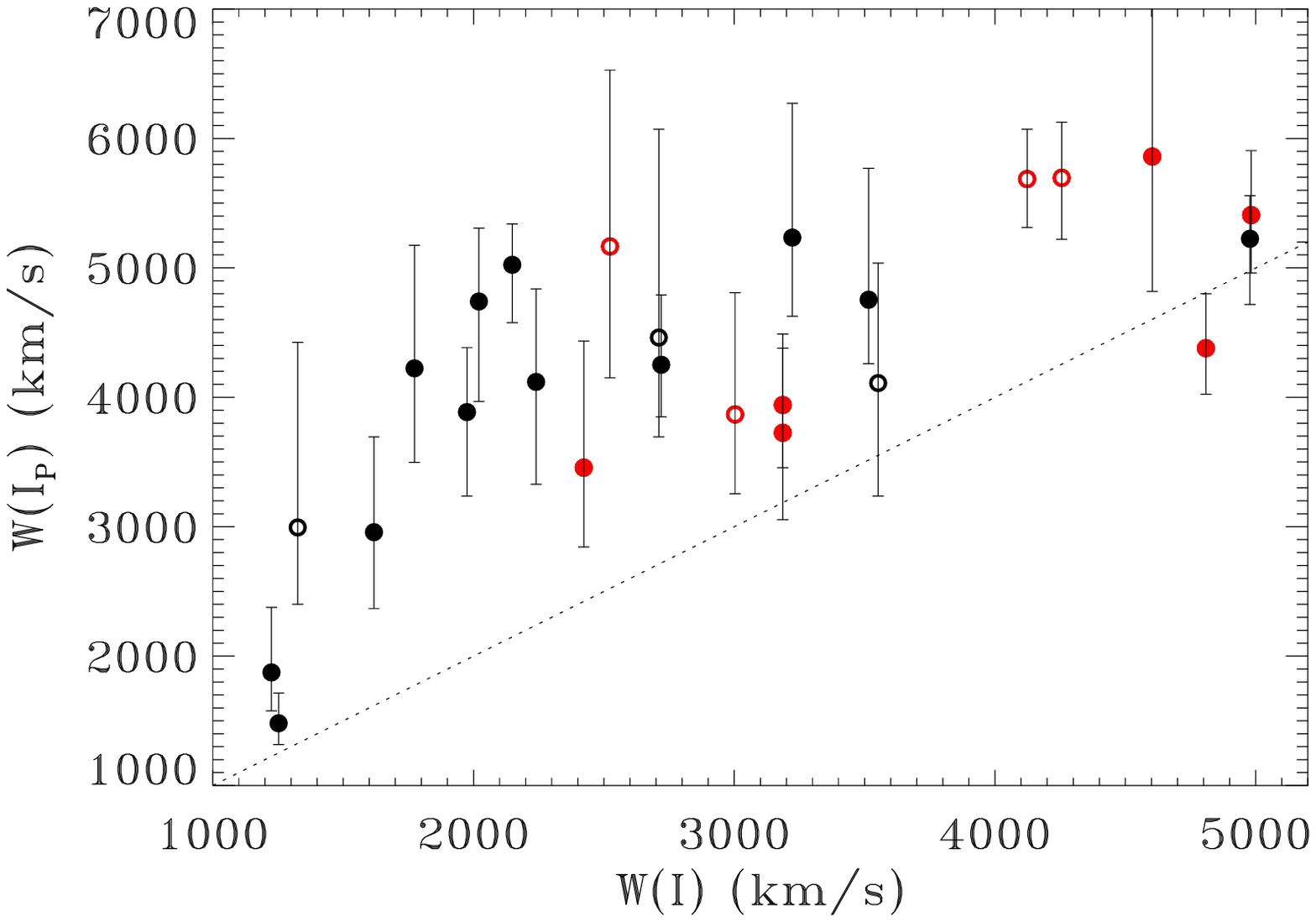,width=0.49\linewidth}
\psfig{figure=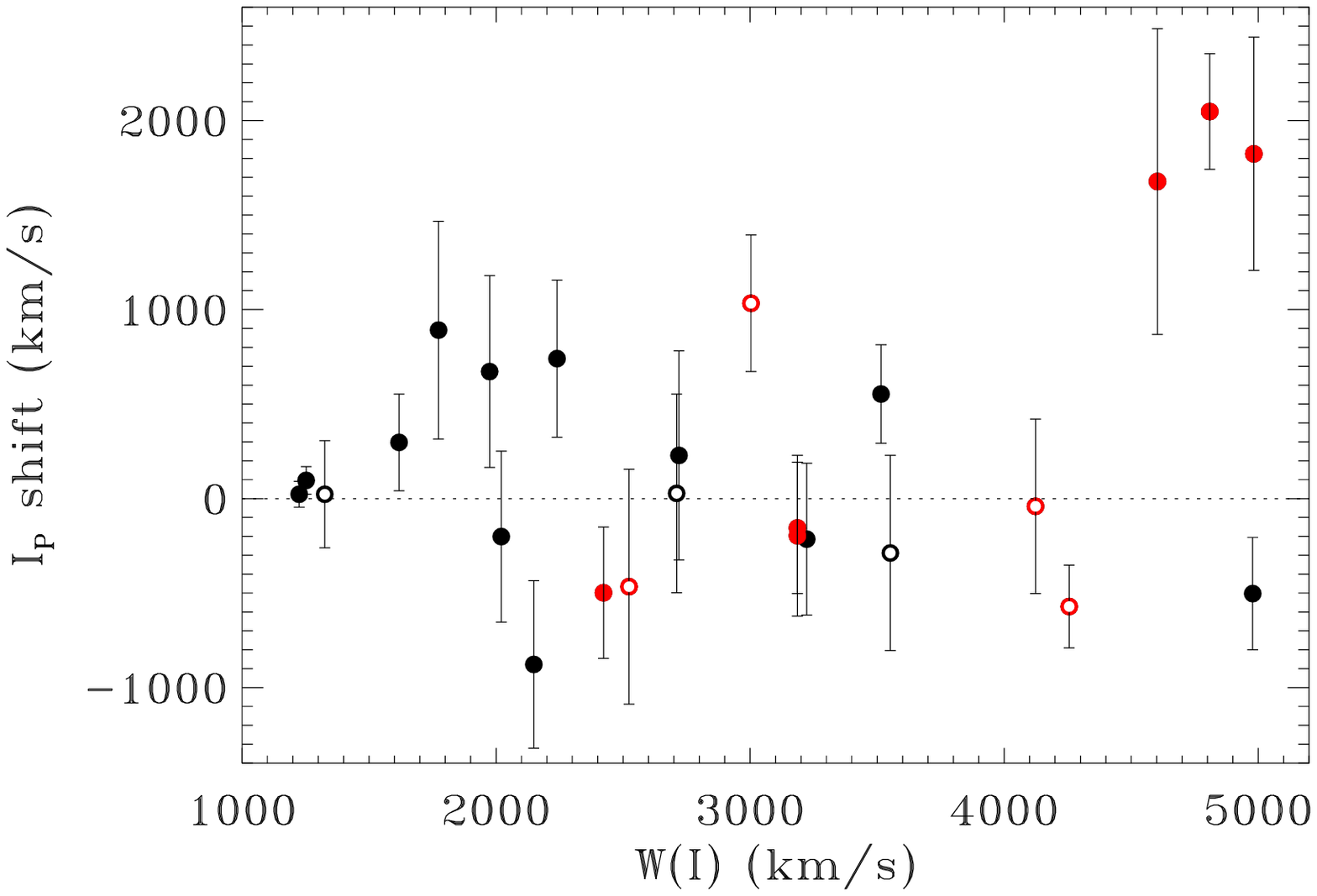,width=0.49\linewidth}
\caption{Left: comparison of the IPV widths measured in total
  flux and polarization, $W(I)$ and $W(I_P)$, respectively. There
  is a general broadening of the BLR in polarized light with respect
  to what is seen in total flux. Right: velocity shift, in \kms,
  between the BLR centroid in $I_p$ with respect to the BLR in total
  flux.  Polarized line shifts are generally small ($\lesssim
  1000$ \kms) and rarely significantly different from a null value,
  but there is a group of three RLQs with a more significant shift of
  $\sim 2000$ \kms.}
\label{shift}
\end{figure*}

\begin{figure*}
\psfig{figure=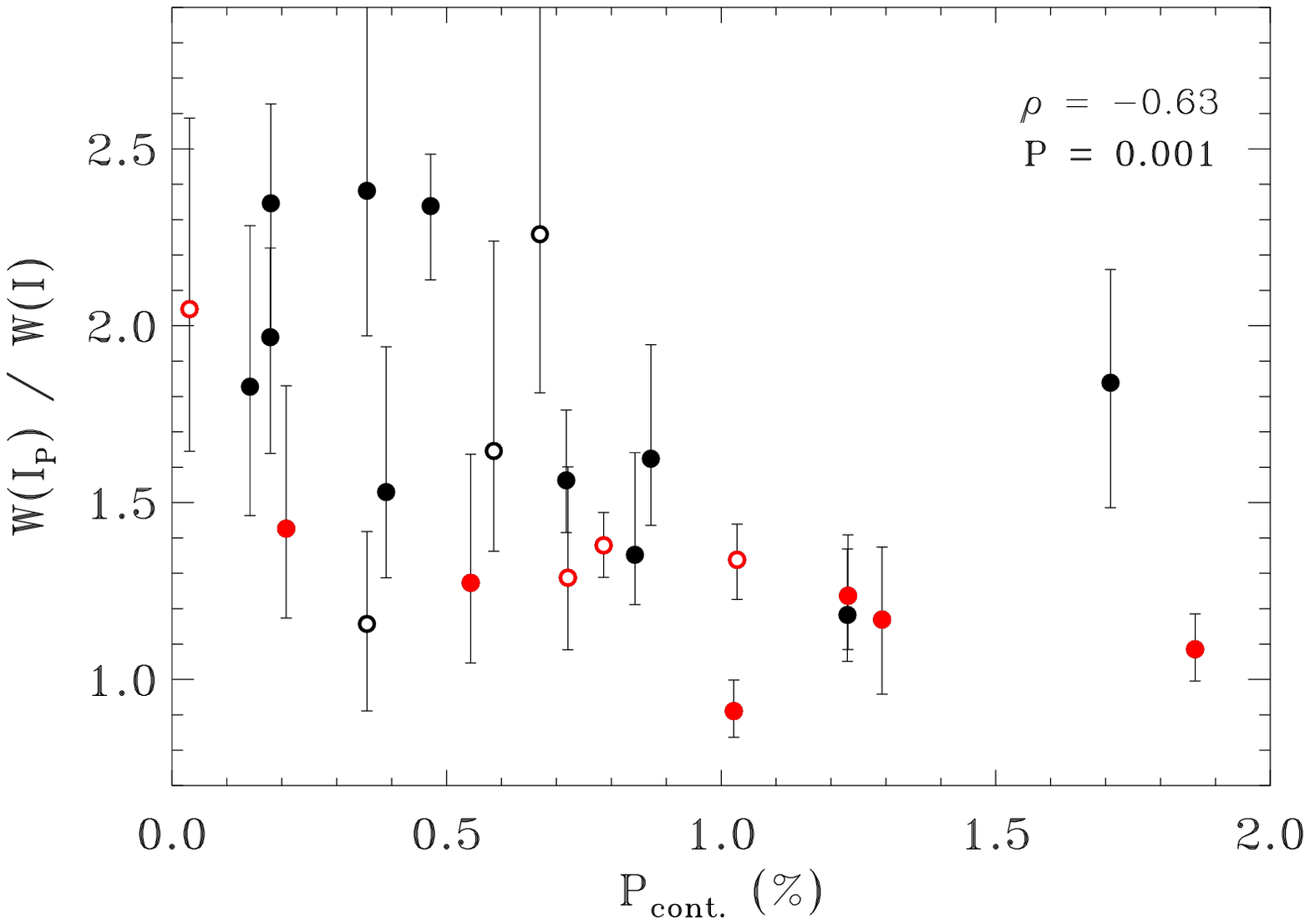,width=0.49\linewidth}
\psfig{figure=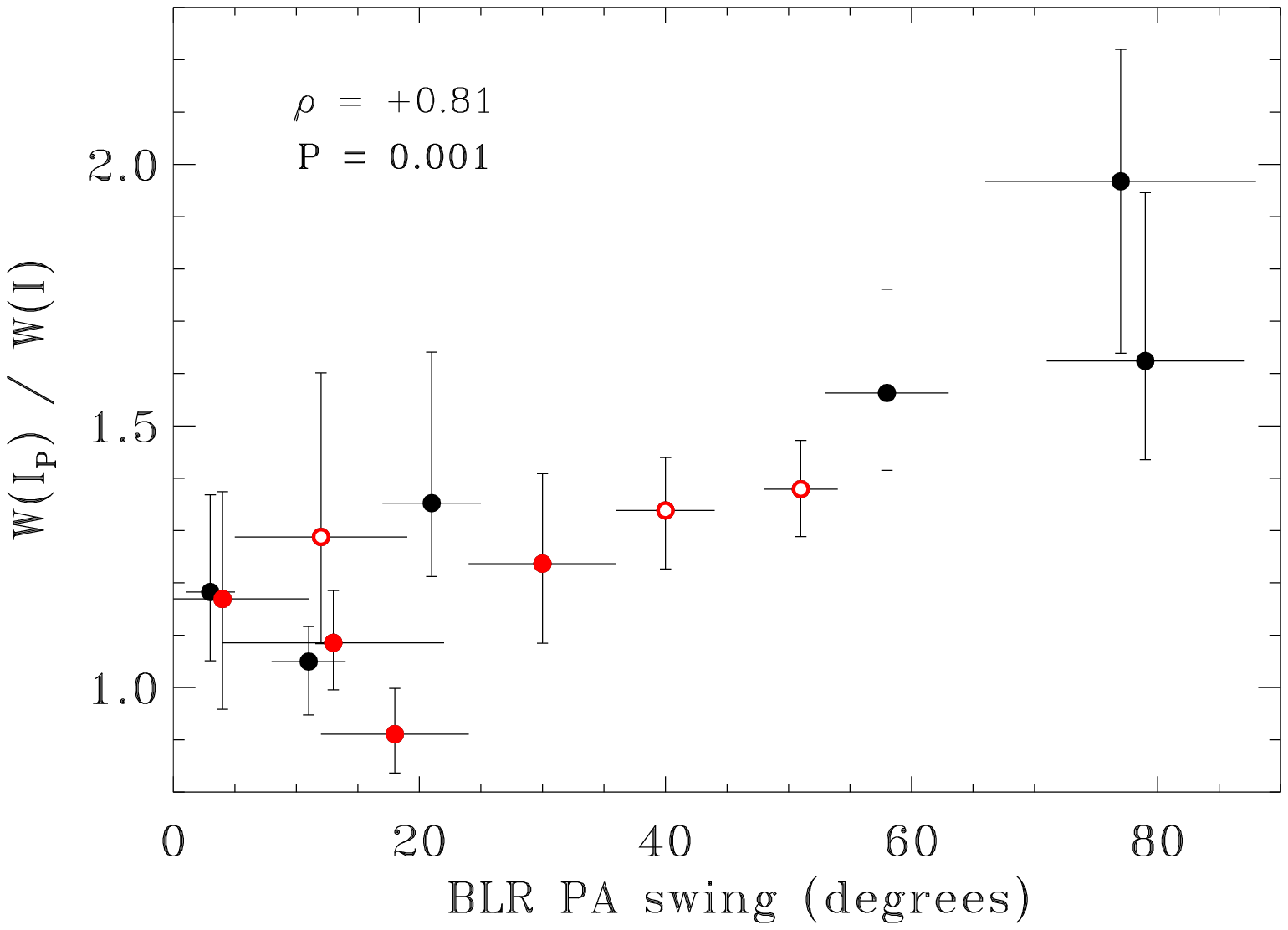,width=0.49\linewidth}
\caption{The polarization broadening of the BLR, i.e., the ratio
  between the line width in polarized and total flux versus:
  (left) the percentage of polarization of the continuum, and (right)
  the PA swing across the BLR. The sources with the largest broadening
  are the least polarized and with the largest PA swing across the
  BLR.  Both trends are expected for equatorial scattering observed at
  a range of inclinations. With decreasing inclination, $W(I)$ becomes
  smaller, while $W(I_p)$ is unchanged, so $W(I_p)/W(I)$ increases,
  the PA swing is also expected to rise, while the polarization
  strength decreases.  }
\label{broadening}
\end{figure*}

\begin{figure}
\psfig{figure=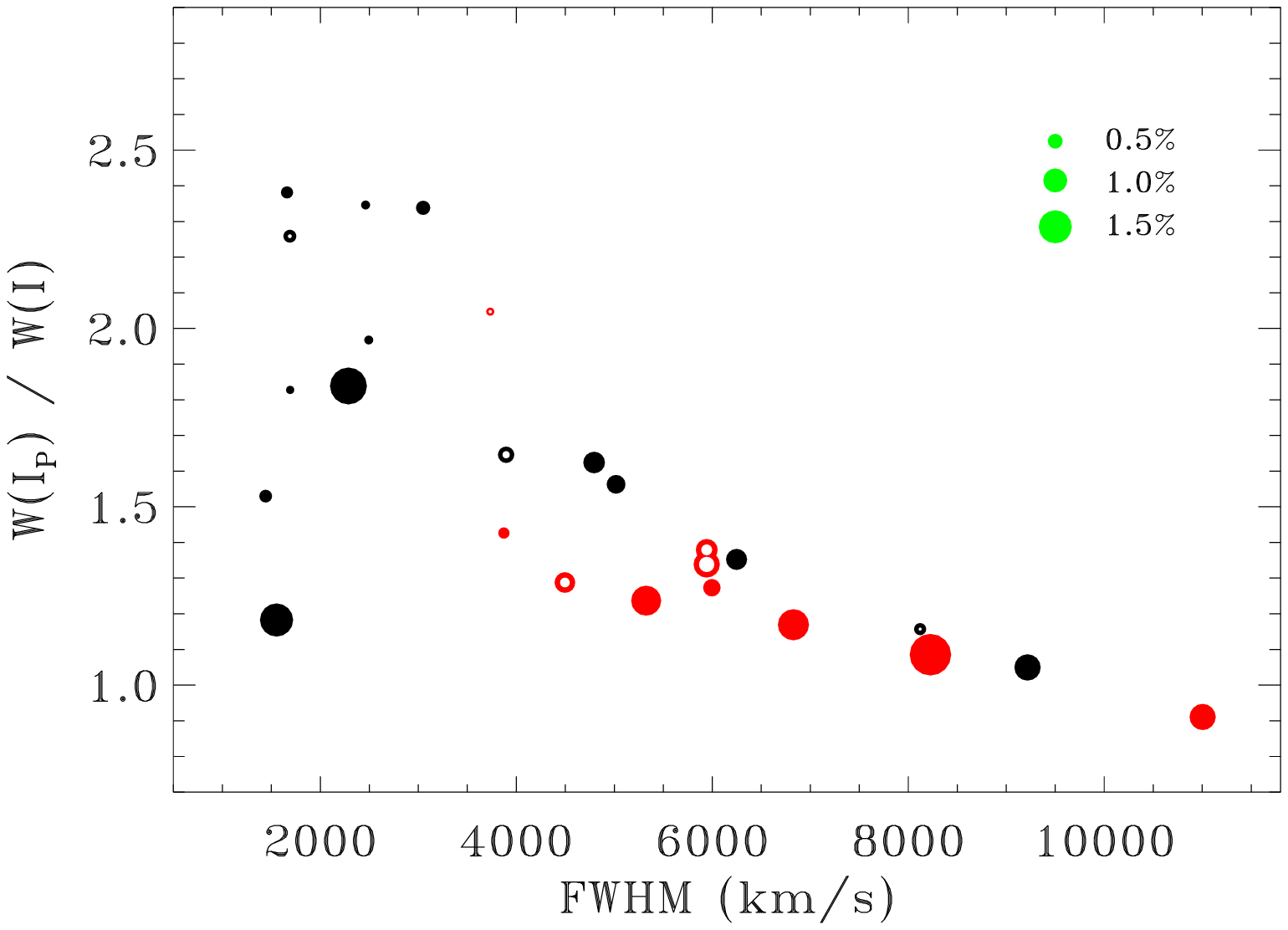,width=1.05\linewidth}
\caption{The polarization broadening of the BLR, i.e. $W(I_p)/W(I)$,
  as a function of the \Ha\ FWHM. Symbols sizes are proportional to
  the continuum polarization. The very broad line objects
  (FWHM$>6500$~km~s$^{-1}$) have small broadening ($<1.2$) which
  indicates the objects are observed at a high inclination, close to
  the inclination over which the equatorial scattering medium
  extends. These objects also have a high continuum \%P, as expected
  for their high inclination (the one deviant object with a low \%P is
  host diluted). At low FWHM values, there is a large scatter in
  $W(I_p)/W(I)$, which indicates it is a mix of objects observed over
  a wide range of inclinations.  The objects with highest
  $W(I_p)/W(I)$ at low FWHM are expected to be closer to face-on, as
  indeed indicated by their low \%P. The highest \%P objects at low
  FWHM have lower $W(I_p)/W(I)$, which both indicate a higher
  inclination.}
\label{fw-broad}
\end{figure}

We now estimate the width of the broad \Ha\ line in total flux
and in polarized light. The methods commonly used to measure the BLR
width are the estimates of its FWHM or dispersion $\sigma_{\rm
  BLR}$. Both have disadvantages, in particular when dealing with
noisy spectra. For example, the measure of $\sigma_{\rm BLR}$ gives a
strong weighting to the values of pixels in the wings of the lines,
making it an unreliable estimator of line-width in low S/N spectra. In
the case of spectropolarimetric measurements, we are dealing not only
with low S/N data, but they are also affected by the polarization bias
\citep{wardle74,vaillancourt06}: the polarization is the length of a
pseudo-vector, i.e., a definite positive quantity. As a consequence,
the polarization measurements follow a Rice (and not a Gaussian)
distribution. A direct measurement of the width of a line in polarized
light, both the FWHM or $\sigma_{\rm BLR}$, is generally unreliable.

Following \citet{whittle85}, we used instead the inter-percentile
values (IPV) to characterize line widths (see also
\citealt{fine08}). The dependence of IPV widths on the cumulative flux
distribution rather than on the flux density at a given point makes
the IPV measurements more robust. IPV widths, $W$, can then be directly
compared to the values derived from any BLR model or a different width
measurement. For example, for a Gaussian profile, FWHM = 1.75 $\times
W$.

For each of the 25 QSO we built the cumulative distribution of the
total flux $I$ and of the polarized flux $I_P$ and measured the
wavelengths of the 25$^{\rm th}$, 50$^{\rm th}$, and 75$^{\rm th}$
percentiles, $\lambda_{25}$, $\lambda_{50}$ and $\lambda_{75}$. In
order to reduce the effects of noise and of the polarization bias it
is necessary to consider the rebinned spectra. As a consequence the
cumulative distribution is coarsely defined. The percentile values are
estimated by performing linear interpolations between the adjacent
measurements.  The width between $\lambda_{25}$ and $\lambda_{75}$ is
$W$ while $\lambda_{50}$ corresponds to the line centroid. For the
sake of simplicity we refer to the IPV width measurements in direct
and polarized light as $W(I)$ and $W(I_P)$, respectively.

IPV widths are also affected by the choice of the integration region
which is not known a priori. For each source we considered, as first
guess, a wavelength range of three times the FWHM of the broad
\Ha\ line in total flux and measured the line width in polarized
light. In a second iteration, we used as wavelength range three times
the value of $W(I_P)$ estimated in the first iteration. We proceed
until convergence, which required at most three iterations.

We estimated the error on $W(I_P)$ with a Monte Carlo
simulation. While the polarization measurements follow a Rice
distribution, the individual Stokes parameters obey Gaussian
statistics. We perturbed $Q$ and $U$ independently  (the errors in $I$
are negligible) in each pixel of the original spectra by adding to the
data a gaussian noise with amplitude given by the error at each given
wavelength, producing 100 different realizations. The resulting
spectra are then treated with the same strategy described in Sect. 2
and 3, including the adaptive rebinning and the subtraction of the
continuum polarization. The $W(I_P)$ width in polarized light is
measured for each realization and from their distributions we
estimated the $1 \sigma$ error. The relative errors are between
$\sim$10\% and $\sim$25\%.

The resulting values of $W(I)$ and $W(I_P)$ are tabulated in Table
\ref{tab3} where we also give the shift between the centroid of the
line in total and in polarized light.  In Figure \ref{ipvi-fwhm} we
compare the broad \Ha\ FWHM to the separation between the 25$^{\rm
  th}$ and 75$^{\rm th}$ percentiles, W(I). Overall, they show a
ratio of 1.75, but the objects with FWHM $\lesssim 4,000$
\kms\ show smaller values, with FWHM/W(I) between 1.0 and 1.3. This
effect is due to the well known connection between the line width and
its profile, where the broad-line objects have more flat-topped
profiles, while the narrow-line objects have more extended wings
(e.g., \citealt{collin06}).

In Figure \ref{shift}, left panel, we compare the width measured in
polarized light $W(I_P)$ and total flux $W(I)$: the line width in
polarized light is, in all but four QSOs, significantly broader than
in total flux with a broadening reaching a factor $\sim 2.5$. In
many cases, a large line broadening is associated with a double-humped
profile in polarized light. The median value of the broadening is
$\sim 1.5$.

In the right panel we instead compare the shift between the BLR
centroid in total and polarized light with the total flux
width. Most QSOs do not show a statistically significant shift, with
values of shift smaller than $\sim 1000$ \kms\ and no relation with
$W(I)$. However, there are three sources, among those with the largest
line widths ($W(I) \sim 5000$ \kms), showing large shifts. In
particular, a highly significant redshift of 2,050$\pm$350 \kms\ is
measured in J140336. These three QSOs are all radio-loud and do not
show a significant line broadening, i.e., $W(I_P)/W(I) \sim 1$.

In Figure \ref{broadening} we compare the broadening, i.e. the ratio
between $W(I_P)$ and $W(I)$, with the percentage of polarization of
the continuum and the PA swing across the BLR. In both cases the two
quantities are strongly correlated: the sources with the largest
broadening are the least polarized
($\rho=-0.63$ and P=0.001 for 25
sources) and with the largest PA swing across the BLR ($\rho=0.81$ and
P=0.0008 for 13 sources).

A slightly different view of the same effect is obtained by looking at
the dependence of the line broadening with the FWHM, taking into
account also the continuum polarization (see Fig. \ref{fw-broad}). The
very broad line objects have small broadening which indicates the
objects are observed at a high inclination. These objects also have a
high continuum \%P, as expected for their high inclination. At low
FWHM values, there is a large scatter in $W(I_p)/W(I)$, which
indicates it is a mix of objects observed over a wide range of
inclinations. The objects with highest $W(I_p)/W(I)$ at low FWHM are
expected to be closer to face-on, as indeed indicated by their low \%P.

\section{The Origin of the polarization in QSOs}
\label{origin}

We here summarize the main results obtained in the previous sections.

$\bullet$ The continuum \%P and PA appears to remain steady over
timescales of years, as indicated by a comparison to the values
measured $\sim 30$-40 years ago for seven sources from our sample.

$\bullet$ The median continuum polarization is 0.59\%.  The QSOs with
the lowest luminosities are redder and less polarized ($\overline P =
0.39\%)$ compared the more luminous sources ($\overline P =
0.72\%$). The difference is most likely due to the significant
contribution of the unpolarized host galaxy light, which reddens the
observed continuum, and lowers its polarization.

$\bullet$ When integrated over the broad \Ha\ profile, the BLR PA is
closely aligned with the continuum PA. But, $P_{\rm cont}$ is higher
than the average $P_{\rm BLR}$ by a factor of $\sim 2$.

$\bullet$ RLQs show a close alignment between their radio axis and
both the continuum and BLR polarization.

$\bullet$ RLQs show an inverse relation between $P_{\rm cont}$ and the
radio core dominance, where $P_{\rm cont}$ drops with increasing core
dominance.

$\bullet$ A significant swing of the polarization PA vector is commonly
observed across the BLR profile. The largest swings are seen at the
lowest $P_{\rm cont}$ and the lowest \Ha\ FWHM.

$\bullet$ The broad \Ha\ line width in polarized light is generally
broader than in direct-light, by a median factor of $\sim 1.5$.  The
largest BLR broadening in polarized light is observed in the objects
with the lowest \Ha\ FWHM, the lowest $P_{\rm cont}$, and the largest
PA swing.

$\bullet$ The polarized broad \Ha\ line profile of J140336 is
redshifted by $\sim$ 2,000 \kms\ with respect to the total flux
profile, but the line width is essentially unchanged, in the highest
\Ha\ FWHM object.  A similar effect may be present in two other high
\Ha\ FWHM objects.

$\bullet$ In most cases, however, the shift between the broad
\Ha\ line in total and polarized light is not statistically
significant, with values of shift smaller than $\sim 1000$ \kms\ and
no relation with $W(I)$.

$\bullet$ The 16 RQQs and 9 RLQs follow in a similar manner the
various relations described above.

$\bullet$ The interstellar polarization has, in general, a negligible
effect on the polarization properties of the QSOs we observed.

\medskip

There is a host of mechanisms which can induce line and continuum
polarization in AGN.  The continuum polarization may be produced by
synchrotron emission. The thermal disk emission may
be polarized by electron scattering within the disk atmosphere (e.g.,
\citealt{laor90, agol96}).  The BLR emission lines may be resonantly,
Rayleigh, or Ramann scattered and thus polarized by ambient gas clouds
\citep{lee97, lee98, korista98, chang15}. Transmission of the quasar
emission through aligned dust grains along our line-of-sight can
polarize all the quasar emission \citep{serkowski75}. In addition,
continuum and BLR polarization can be induced through electrons or
dust grain scattering, as commonly observed in type 2 AGN, the
phenomena which underlies the AGN unification scheme
\citep{antonucci93}.

The similarity of the PA of the continuum and the BLR suggests that
the same mechanism is responsible for the polarization of both
components. This excludes the separate line and continuum polarization
mechanisms, as mentioned above. Despite the similarity in PA, the mean
amplitude of $P_{\rm cont}$ is different from the amplitude of $P_{\rm
  BLR}$, which excludes dust transmission, which would produce the
same polarization of all the transmitted radiation.

The only remaining viable mechanism is scattering, either by free
electrons or dust. Scattering induced polarization depends on the
geometry of the scattering configuration.  Since AGN are characterized
by axial symmetry, the geometry can be ideally separated into two
situations, polar and equatorial scattering (spherically symmetric
scattering produces no polarization).  The alignment of the
polarization PA and the radio axis PA indicates equatorial scattering.
Since the RLQs and RQQs follow the various relations found here in a
similar manner, the polarization mechanism for both types of AGN is
likely similar.  We therefore conclude that equatorial scattering
applies for both types of quasars.

Furthermore, scattering can be divided into a near-field and a
far-field regime. The near-field regime occurs when the emission
region and the scattering regions overlap or are next to each
other. The detection of a swing of the PA across the BLR in most QSO
indicates that we are generally observing near-field equatorial
scattering.

Further support for the near field equatorial scattering, is the
additional observation that the average $P_{\rm BLR} \sim 0.5 P_{\rm
  cont}$, which is naturally interpreted as a geometrical cancellation
effect.  Different emitting regions in the BLR are scattered at
different angles from a given region in the extended scattering region
(see \citealt{smith05}), which lowers the polarization of the
integrated flux scattered. In contrast, the continuum source is far
smaller than the scattering region, and all parts of the continuum
emission region are scattered by the same set of scattering angles in
the scattering medium, which is less affected by the geometrical
cancellation effect.

A key additional observable, in the case of near field scattering, is
the PA swing across the BLR, $\Delta$PA$_{\rm BLR}$. The models
indicate that this is related to the orientation of the BLR with
respect to our line-of-sight: at smaller inclinations we expect larger
$\Delta PA_{\rm BLR}$ and smaller $P_{\rm cont}$. This relation is
indeed found here, where Fig. \ref{swing} demonstrates a highly
significant negative correlation between $P_{\rm cont}$ and
$\Delta$PA$_{\rm BLR}$.  The specific velocity dependence of PA$_{\rm
  BLR}$ is consistent with the predicted PA velocity dependence for a
rotating disk emitter, scattered by an outer rotating disk medium
\citep{smith05, Savic18, Lira20}. The maximal observed PA swing of
$\sim 80^\circ$ found here, which is expected for a close-to face-on
view, suggests the scattering medium radius is $\sim 1.5\times$ the
size of the emitting region (see Fig.6 in \citealt{smith05}), so the
scattering medium may be effectively considered as the outer BLR.

Polarization angle swings in the \Ha\ profile might also cause
cancellation because polarization vectors of different orientation are
added up when integrating over a large velocity range. Indeed we found
that sources with the largest ratio between the average and the
integrated $P_{\rm BLR}$ are those with the largest PA swings: the
reduced level of polarization integrated over the whole BLR is indeed
due to this effect.

The importance of orientation on the polarization properties of QSOs
is also suggested by the connection between $\Delta$PA$_{\rm BLR}$ and
the BLR width. Larger $\Delta$PA$_{\rm BLR}$ are generally found in
QSOs with a smaller FWHM, indicating that orientation affects the
observed line width. However, this is not a one-to-one correspondence,
as there are sources with small $\Delta$PA$_{\rm BLR}$ associated with
a smaller FWHM. This is expected because the BLR width distribution is
produced by the folding of intrinsic widths with orientational
effects.

Further support for the idea of equatorial scattering and, at the same
time, another signature of the importance of orientation is found when
considering the broadening of the BLR in polarized flux compared to
total flux. The polarized broad lines result from scattering by
material distributed mainly in the AGN equatorial plane. From this
vantage point, the rotational component of the velocity is maximized
and it is then higher than along our line-of-sight. The broadening
reaches values as large as $\sim$2.5, with a median value of
$\sim$1.5. A significant negative trend links the broadening with the
\Ha\ FWHM, and with the continuum polarization, and a strong positive
trend is seen between the broadening and PA swing across the BLR.  The
three effects are all expected for equatorial scattering: at smaller
inclination the polarization and the typical FWHM decreases, while the
BLR PA swing and the line width ratio in total to polarized light,
both increase.

The median broadening expected for a random orientation of the QSOs is
equal to \, 1/sin$(60^\circ) \sim 1.15$, smaller than the observed
value of $\sim$1.5 that corresponds instead to a median viewing angle
of $\sim 42^\circ$. This effect is likely due to the presence of
circum-nuclear selective obscuration, as predicted by the unified
model of active nuclei. Another indication of the role of orientation,
already discussed in Sect. 4.3, is the negative correlation between
degree of polarization and radio core dominance in the RLQs: the
former increases at larger viewing angles while the latter decreases.

Thus, there is a single scenario which explains the polarization
properties of this representative sample of QSO. The polarization of both
the continuum and the BLR are produced by an equatorial scattering
medium which resides close, or possibly at, the BLR.

While the polarization properties of most of the QSO in this sample
are well described by near-field equatorial disk scattering, there
are some deviants. For example, in J142613 the continuum PA and the
integrated \Ha\ flux PA are orthogonal. The very low \%P($<0.2$\%) of
both the continuum and the BLR, and the non physical erratic structure
in the wavelength dependence of the BLR PA (see Supplementary Material), suggest
possible systematic effects. However, some variations in the
equatorial scattering scheme are also possible.

Another variation is observed in J140336. Its polarized BLR is
redshifted by $\sim$ 2,050$\pm$310 \kms\ with respect to the total
flux BLR. The BLR profile, leaving aside its redshift, is quite
similar in both polarized and direct-light and, in particular it does
not show a significant broadening ($W(I_P) / W(I) = 0.91 \pm
0.07$). There are two further sources with apparently similar
properties, but with poorer data quality: J142735 where the polarized
BLR is redshifted by 1,820 $\pm$ 620 \kms, and J100726 where the
redshift is 1,680 $\pm$ 810 \kms. Their line broadenings are
1.09$\pm$0.10 and 1.27$\pm$0.29, respectively. These three sources
have several characteristics in common: they are all radio-loud QSO
and are the sources with the broadest BLR in this group, with a FWHM
between 6,000 and 11,000 \kms, among the QSO with the highest
luminosity, and are likely to be observed closer-to edge-on (high \%P
and small core dominance).

\section{Discussion}
\label{discussion}

\subsection{Comparison with earlier studies}

Some of the relations found here were already noted in earlier
studies.  Specifically, the typical optical continuum polarization
amplitude of QSOs, which is 0-2\% \citep{stockman84,berriman90}. The
parallel alignment of the radio axis PA and the polarization PA in
RL quasars \citep{stockman79,antonucci83, berriman90}.

In contrast with our conclusion that the continuum and BLR
polarization are produced by a single scattering medium,
\citet{kishimoto04} suggest that the optical polarization in quasars
is produced by electron scattering within the accretion disk
atmosphere. The conclusion is based on the detection of a drop in the
polarization amplitude below 4000\AA. This drop is interpreted as the
Balmer edge opacity jump which may make the absorption opacity larger
than the scattering opacity, and thus reduce the polarization induced
by scattering (see a similar effect predicted at the Lyman edge,
\citealt{laor90}). However, this interpretation is inconsistent with
the PA of a disk atmosphere scattering, which is predicted to be
perpendicular to the radio axis, rather than parallel, as observed.
Furthermore, scattering from dust is expected to induce a drop in the
polarization amplitude in the UV due to the wavelength dependence of
the scattering cross section \citep{zubko00}. The exact wavelength
dependence of the polarization amplitude depends on the dielectric
function, i.e. the grain composition, and to some extent also on the
grain size distribution.

A polarization PA swing across the \Ha\ line in a fraction of type 1
AGN was noted in various studies \citep{goodrich94, corbett00,
  smith02, smith04, afanasiev19}. A PA swing is also observed in the
rest frame UV broad lines \citep{Alexandroff18}.  As noted above, the
polarization PA swing is consistent with near field equatorial
scattering \citep{smith05, Savic18, Lira20}.

A redshifted polarized line profile, suggesting scattering from a fast
nuclear outflow, was also found in \citet{robinson99} and
\citet{young07} from spectropolarimetric observations in two QSOs,
4C~74.26 and PG~1700+518, as found here for the redshifted polarized
line objects.

Despite the similarity to earlier studies, a major difference is that
earlier high S/N spectropolarimetric studies (e.g.,
\citealt{goodrich94}) tend to find a complex situation with a variety
of polarization properties, suggesting a mix of polarization
mechanisms. This is in contrast with the more coherent picture derived
here. This likely results from the nature of the earlier samples, which
include a greater variety of objects, including significantly host
dominated AGN, reddened AGN, and very low $L/L_{\rm Edd}$ AGN. The new
aspect in this study is that it is based on a uniformly selected and
complete sample of quasars, which extends evenly over a wide range of
luminosity and line width, and is thus likely more representative of
the more common quasars.

\subsection{Further support for near field equatorial scattering}

The various correlations found between the continuum \%P, the BLR PA
swing, and the line width ratio $W(I_p)/W(I)$, provide further
independent support for the equatorial scattering scenario. The
continuum \%P is expected to increase with inclination, the PA swing
is expected to decrease with increasing inclination \citep{smith05},
and the line width ratio $W(I_p)/W(I)$ is expected to decrease with
inclination. All expected relations are indeed observed.

The relations of the above three quantities with the BLR FWHM are
generally weaker, showing a growing dispersion with decreasing
FWHM. This trend is indeed expected, as the FWHM is affected by both
the inclination and the absolute gas velocity at the BLR, which is set
by both the size of the BLR and the black hole mass. The highest FWHM
values are expected to come from a combination of both a high velocity
BLR, and a high inclination view. Indeed these objects all show high
\%P, low BLR PA swing, and $W(I_p)/W(I)$ close to unity, as
expected. In contrast, the lower FWHM values includes both low
velocity BLR observed at a high inclination, and higher velocity BLR
observed at low inclination. Indeed, the low FWHM QSOs show a large
range in the continuum \%P, BLR PA swing, and in
$W(I_p)/W(I)$. However, as expected the low FWHM objects with a high
$P_{\rm cont.}$ show a smaller $W(I_p)/W(I)$, and the reverse, which
indicates the scatter at low FWHM is an inclination effect.

The large swing inevitably leads to a drop in the integrated BLR flux
polarization, compared to the mean value of the wavelength dependent
polarization (Fig. \ref{cancellation}). Thus, the BLR needs to be
adequately spectrally sampled in order to resolve its true
polarization level.

Another more minor piece of supporting evidence for equatorial scattering is
that the large redshift offset of the polarized profile is observed in
objects with a high inclination (high continuum \%P, small BLR PA
swing, $W(I_p)/W(I)\sim 1$), as expected for a closer-to edge-on view
of an equatorial wind.  

Additional support for near field equatorial scattering in type 1 AGN
is provided by the nature of the type 2 AGN polarization, which is
produced by polar scattering \citep{antonucci83, antonucci85}. The
absence of the equatorial scattering polarization signature in type 2
AGN indicates this polarized light is obscured together with the
continuum and BLR emission. Thus, the scattering region in type 1 AGN
must be physically close to the BLR. In addition, only a small
minority of type 1 AGN appear to be dominated by polar scattering
\citep{smith04}, despite the fact that the polar scattering medium
must be directly observed in type 1 AGN, together with the equatorial
scattering region. This indicates that the equatorial scattered flux
generally dominates the polar scattered flux.  The polar scattering is
consistent with far field scattering, while the equatorial scattering
is near field. Thus, one naturally expects the covering factor and
thus the scattered flux of the equatorial scattering medium to
dominate.

Reverberation mapping of the polarized versus the direct-light
continuum in NGC 4151 by \citet{gaskell12} reveals that the scattering
medium has the size comparable to the BLR. Clearly, additional
reverberation mappings of the continuum and of the BLR polarized
fluxes can measure directly the size and the spatial distribution of
the scattering medium in AGN (e.g., \citealt{rojas20}).

We note in passing that intensive spectropolarimetric studies of broad
absorption line quasars \citep{cohen95, goodrich94, Ogle99,Schmidt99},
lead to the same picture of an equatorial scattering medium.

\subsection{Implications for black hole mass estimates}

The relations found in this study show that the continuum \%P, the BLR
PA swing, and the line width ratio $W(I_p)/W(I)$, can all be used as
independent inclination indicators in AGN. We also find evidence that
the BLR FWHM is indeed affected by both inclination and the absolute
gas velocity at the BLR, as expected for an equatorial BLR structure.

In principle one can use the three quantities, the continuum \%P, the
BLR PA swing, and the line width ratio $W(I_p)/W(I)$, to derive the
inclination, and use that to correct the observed FWHM to the absolute
gas velocity. This will allow to mitigate the inclination bias in the
BLR based black hole mass estimate. However, the BLR is likely not a
thin disk, as it needs to subtend over a significant solid angle
($\sim 0.3$, see \citealt{baskin18}), which means it forms a torus
structure. This structure inevitably implies a random velocity
component, which is a fraction of the planar Keplerian velocity.
Since this random component may be object dependent, it is not
straightforward to derive the absolute gas velocity from the observed
FWHM. The simplest and most straightforward approach is just to use
the line width ratio $W(I_p)/W(I)$ as the FWHM correction factor, as
this directly measures the increase in the velocity spread from our
vantage point, to a close-to equatorial point of view.

A practical procedure is to first use broad band polarimetry at a
number of wavelengths, in order to estimate the continuum \%P
level. If say P$<1$\% than the black hole mass correction factor
$(W(I_p)/W(I))^2$ may become comparable or larger than the black hole
mass uncertainty (factor $\sim 3$). This can then justify investing
the significant telescope time required to get a high quality
spectropolarimetry of the object, in order to measure $W(I_p)/W(I)$ as
accurately as possible.

\subsection{Further questions}

Equatorial scattering at, or just outside, the BLR is the
major polarization mechanism in quasars. However, this is clearly just
the first order level approximation.

The various relations found here often show significantly offset
objects, which indicates second order effects must be present. For
example, compact RLQ which are beamed, like 3C273, are observed very
close-to face-on, but still show optical polarization of a few tenths
of percent, which cannot be due to scattering, so there may be some
synchrotron contribution to the polarization.  Of the six quasars with
P$_{\rm cont}>1$\% and P$_{\rm BLR}<0.5\%$, five are RL
(Fig. \ref{pablrcont}). Is their continuum polarization boosted by
contribution from a synchrotron component?  If yes, this synchrotron
source should not be beamed, as these objects are generally lobe
dominated (Fig. \ref{radio} and \ref{coredom}).
 
Is the scattering produced by dust or by free electrons?  The
available wavelength baseline is too small to test the two options,
and one clearly needs to extends this study to the UV.  However, since
the scattering occurs on the BLR scale, or just outside, dust can
already survive in the gas. In dusty gas dust inevitably dominates the
scattering, given its optical opacity which is likely a factor $\sim$
100 larger than the electron scattering opacity ($\sim 10^{-22}$
vs. $6.56\times 10^{-25}$~cm$^2$ per H atom, e.g. \citealt{baskin18}).

Are there additional “fake” NLS1 objects, like PKS~2004-447
\citep{baldi16}? In PKS~2004-447 $W(I_P)/W(I)\sim 6$, whereas here the
maximal ratio found is $\sim 2.4$. Clearly, objects like PKS~2004-447
are not common in type 1 AGN.  Since the typical covering factor in
AGN is $\sim 0.3$ \citep{baskin18}, these objects are inevitably
obscured in a close-to edge-on view.  Also the random velocity
component at the BLR will be of the order of $\sim 1/2$ of the
Keplerian speed, which again precludes a very high $W(I_P)/W(I)$.  The
high $W(I_P)/W(I)$ value observed in PKS~2004-447 requires a thin
scattering medium, and also a thin disk BLR, which are not common.
But, may be more common in RLQ, as these can be found at relatively
low $L/L_{\rm Edd}$, which may be required for these properties
\citep{baskin18}.

Inclination is clearly a fundamental property which controls the
observed properties of AGN. Polarization is sensitive to the geometry
of the system, and can thus serve as a sensitive inclination
indicator. Various other methods were suggested to derive the
inclination of individual objects (e.g., \citealt{fischer13,
  marin14}). AGN show a set of emission line and continuum
correlation, which are generally termed the EV1 set of correlations
\citep{boroson92}. It is of great interest to establish whether and
how the polarization properties change along the EV1
sequence. Answering this question can provide important hints on the
physical mechanisms which produce the EV1 sequence.  We plan to
compare the indications on orientation provided by polarization with
other suggested indicators in a forthcoming paper.

Finally, it will be interesting to explore the spectropolarimetric
properties of higher redshift, z $\sim$ 2-3. The primary line used for
such studies is C IV at 1549\AA, that, however, appears to provide a
significantly less accurate, and possibly biased, estimate of $M_{\rm
  BH}$ \citep{baskin05}. In fact, the C IV line is known to be
blue-shifted with respect to the lower-ionization lines, and often
shows strong asymmetries, suggesting that non-gravitational effects
affect its profile. Furthermore, the line properties appear to be
connected with the object's Eddington ratio. Spectropolarimetry of a
sizeable sample of high-z QSOs can be used to explore their geometry,
to isolate outflow dominated sources, and, eventually, to obtain more
accurate estimates of their black hole masses.

\section{Summary and conclusions}
\label{summary}

We obtained spectropolarimetric observations of a sample of 25 QSOs
extracted from the SDSS/DR7 catalog with $z<0.37$, the redshift limit
at which the broad \Ha\ line is visible in the SDSS spectra. We
focused on the objects sufficiently bright to obtain good data quality
with a single one hour observing block with FORS2 at the VLT: the
median S/N of the total flux spectra in each spectral
element is $\sim$300. The objects were chosen to explore the QSO
properties across the widest possible range of luminosities and BLR
width (and consequently estimated black hole masses) by uniformly
sampling the $L_{5100} - {\rm FWHM}$ plane. We made separate
selections of RL (9) and RQ (16) QSOs. As a result we obtained a
representative sample of QSO with BLR widths spanning between 1,500
and 11,000 \kms\ and about two orders of magnitude in optical
luminosity. There were no large changes in the spectra of these
sources compared to the epoch in which SDSS observations were obtained, 10-15
years ago.

The median continuum polarization of sample is 0.59\%; the less
luminous QSO (log L$_{5100} < 44.5$) are redder and less polarized
($\overline P=0.39\%$) than the more luminous ones ($\overline
P=0.72\%$), most likely due to the dilution from the unpolarized
starlight. Conversely, the presence of the narrow lines does not
affect significantly the polarization properties. The comparison with
previous observations of some sources of the sample indicate that the
polarization is stable over a timescale of 30-40 years.

The integrated PA of the BLR is well aligned with the continuum PA,
but the BLR is less polarized than the continuum: this is due to the
PA swing across the BLR that produces significant cancellation. We
also found that the BLR width in polarized light, W(I$_P$), exceeds
the line width in total flux, W(I), by a median factor
$\sim$1.5. 

Several connections between the various polarization parameters
emerge, e.g., between the PA swing across the BLR, the continuum
polarization fraction and the increase of the line width in
polarization with respect to its value in direct-light. All these
results point to a simple explanation in which 1) the dominant
polarization mechanism in QSO is due to scattering from material
located in the equatorial plane, 2) the QSO polarization properties
depend on their orientation. In particular a more face-on view results
in low polarization, large PA swing across the BLR, and large
polarized line broadening. Both the RQQs and RLQs exhibit similar
relationship. RLQs show 1) a close alignment between their radio axis
and both the continuum and BLR polarization and 2) a connection
between radio core dominance and polarization, both indications of the
role of orientation and equatorial scattering.

Nonetheless, there are three
sources, all RL QSOs, in which the behavior is different: their
polarized is redshifted by $\sim$ 2,000 \kms\ with respect to the
total flux BLR, but with a BLR width essentially unchanged. 

The information derived from spectropolarimetry can be used to
mitigate the inclination bias in the virial estimates of the black
hole masses. One can use the W(I$_P$)/W(I) ratio as correction factor
for the measured FWHM, as this directly measures the increase in the
velocity spread from the vantage (equatorial) point of view of the
scatterers. In the sample of QSOs studied here the correction factor
can be as large as $\sim 6$. The dependence of W(I$_P$)/W(I) on the
continuum polarization indicates that the correction is stronger for
the least polarized sources. Significant care is needed to obtain a
robust measurement of this broadening effect: we adopted as
measurement of the line width the separation of the 25$^{\rm th}$ and
75$^{\rm th}$ percentiles of the line profile, a definition that is
robust against the uncertainties due to the low S/N and of the
positive bias of the polarization data.

\begin{figure}
\psfig{figure=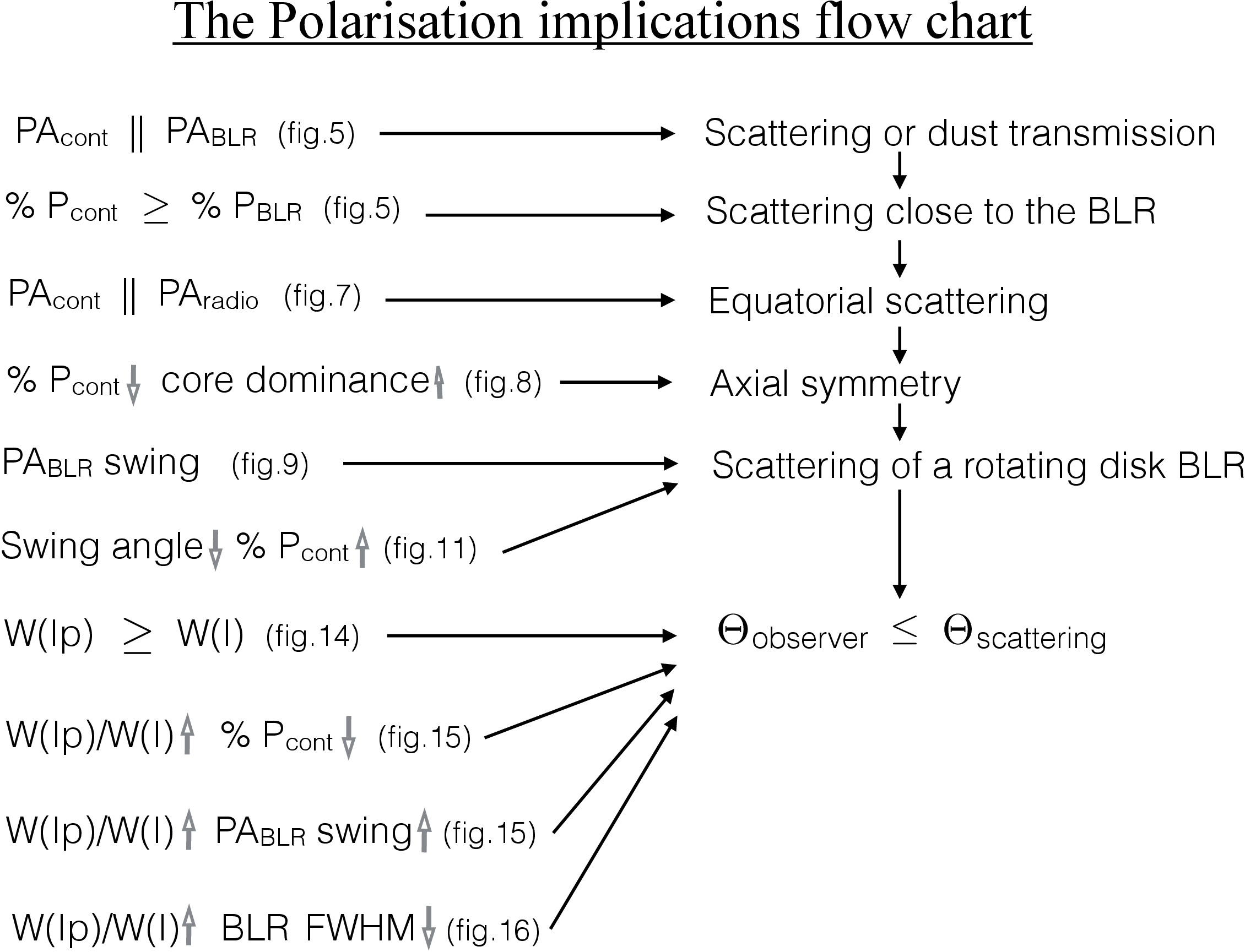,width=1.\linewidth}
\caption{Summary of the the observed polarization properties, which
  all lead to a consistent interpretation. The polarization of
  unobscured quasars is produced by scattering, the scattering medium
  is axially symmetric, equatorial, and resides at the outskirts of
  the BLR. }
\label{flow}
\end{figure}

Fig. \ref{flow} shows a short summary, in the form of a flow chart,
which relates the various observed polarization properties, and their
physical implications, which lead to the implied polarization
scenario. It appears that the polarization of both the continuum and
the BLR are produced by an equatorial scattering medium which resides
close to, or possibly at, the BLR. This scattering geometry explains
the polarization properties of this representative sample of QSO.  The
polarization properties can be used to explore the nature and the
geometry of the scattering medium in type 1 AGN, and eventually
understand its origin. In addition, spectropolarimetry may provide a
useful tool to improve the accuracy of the black hole mass estimates.

\section*{Acknowledgments}
This research was supported by the Israel Science Foundation (grant
no. 1008/18).

\section*{Data availability}
The data underlying this article will be shared on reasonable request
to the corresponding author.

\appendix

\section{The interstellar polarization}
\label{isp}

Thermal emission from the dust grains is polarized
\citep{hildebrand99} and \citet{planckXXI} have shown that the
polarization at 353 GHz measured by the {\it Planck} satellite is
closely connected with the optical polarization of stars. More
specifically they found that 1) the sub-millimeter polarization is
perpendicular to that measured in the optical and 2) the ratio
$R_{S/V}= p_S/(p_V/\tau_V)$ (where $p_s$ and $p_V$ are the degree of
polarization in the sub-mm and V band, respectively, and $\tau_V$ is
the absorption optical depth) is fairly constant across different
lines of sight in the diffuse interstellar medium with a measured
value of $R_{S/V} = 4.2 \pm 0.2 {\rm(stat.)} \pm 0.3$ (syst.).

We estimated the contamination level of the quasar polarization from
the ISP following the analysis of \citet{pelgrims19}. At the location
of each of the 25 QSOs we measured the Stokes parameters of the 353
GHz emission after smoothing the images with a Gaussian with a FWHM of
15$^\prime$. The expected contribution to the optical polarization is
estimated as $q_V = -(q_S \tau_V) / R_{S/V}$ and $u_V = -(u_S \tau_V)
/ R_{S/V}$ and it can be subtracted in the Stokes parameters space
from the observed QSO polarization.

The actual effect on our data must be estimated by considering the
wavelength dependence of the ISP to transform the ISP polarization in
the V band to the wavelength of the \Ha\ line in each individual
QSO. We used the empirical law from \citet{serkowski75} adopting as
wavelength at which the ISP reaches its maximum $\lambda_{\rm max} =
5500$ \AA.  This correction is fairly small, with a ratio between
$p_{{\rm H\alpha}}$ and $p_V$ ranging from 0.76 to 0.94.

\begin{figure}
\psfig{figure=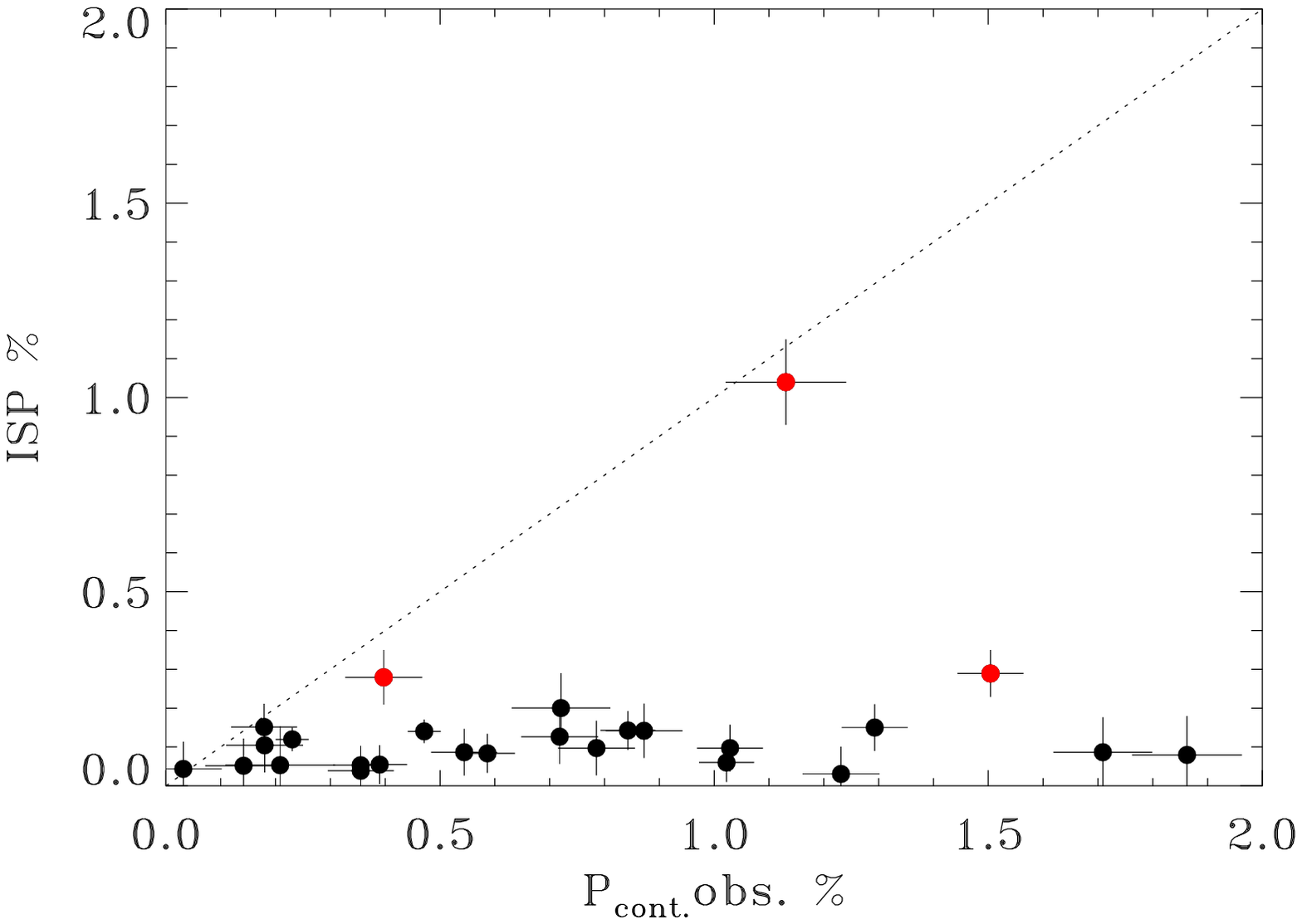,width=1.0\linewidth}
\psfig{figure=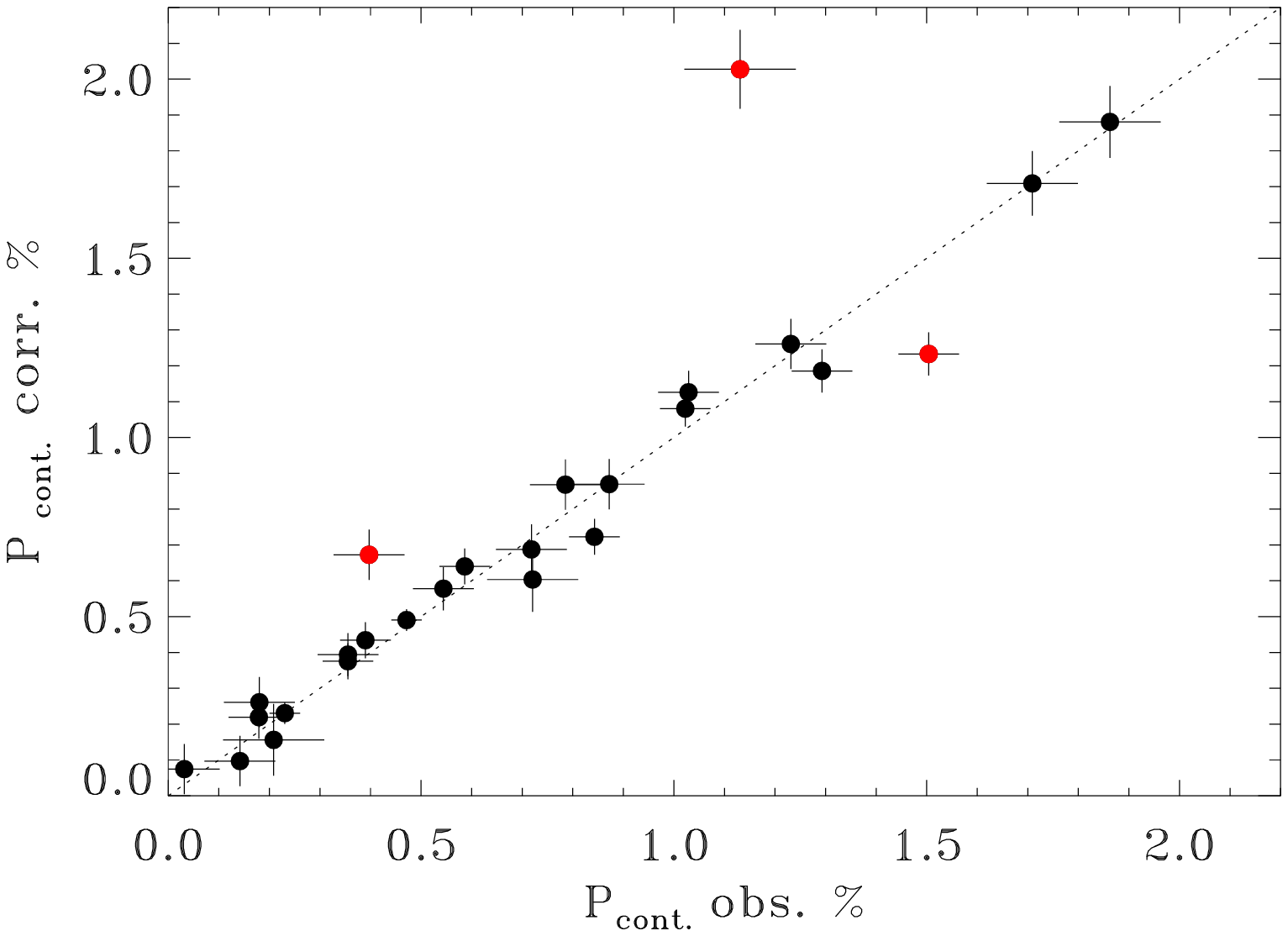,width=1.0\linewidth}
\psfig{figure=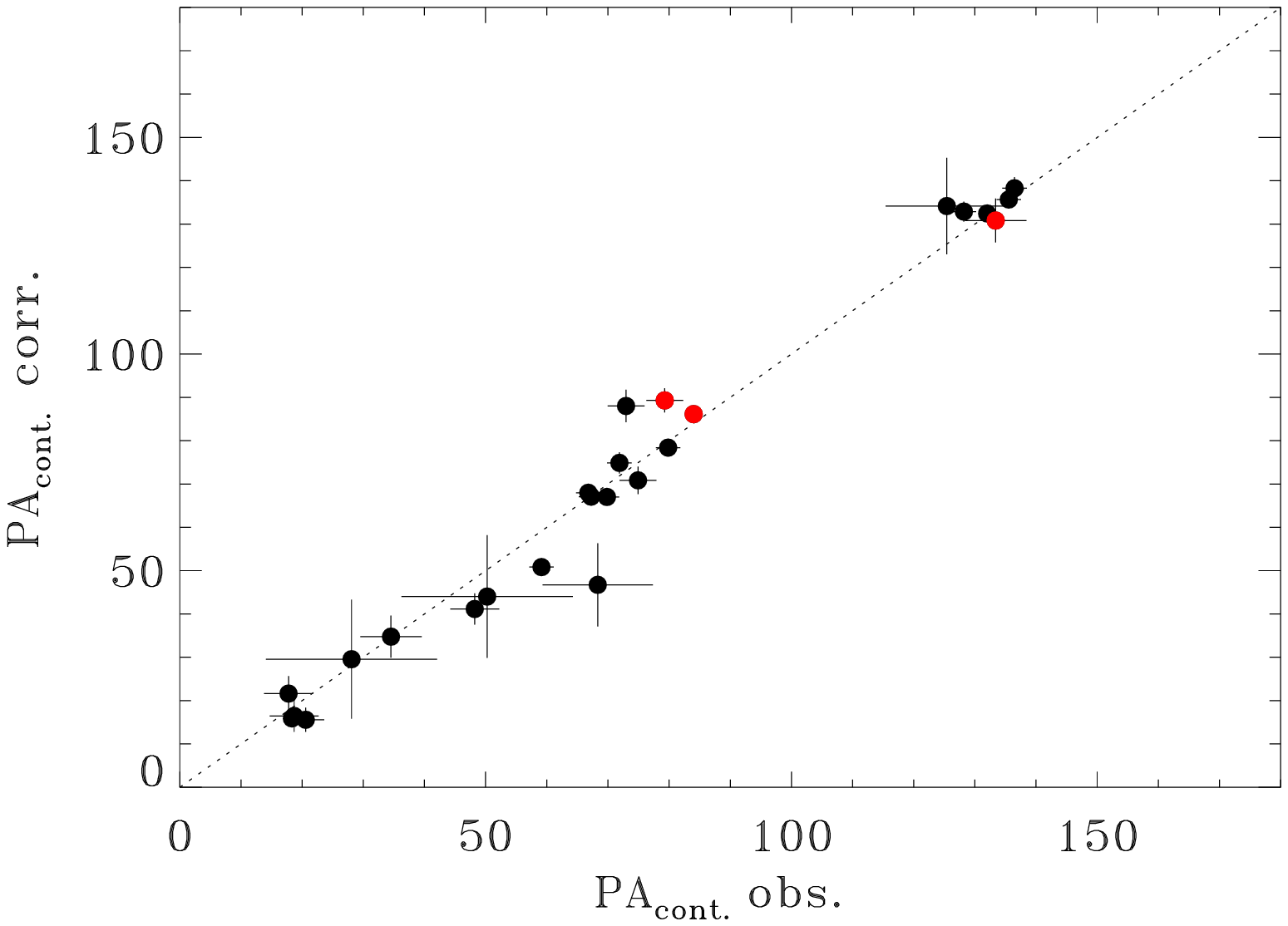,width=1.0\linewidth}
  \caption{Top: comparison between the observed continuum polarization
    and the expected contribution from the ISP. Comparison of the
    observed continuum polarization (middle) and PA (bottom) with the
    values after correction for the ISP contribution. We mark in red
    the three sources (namely J154019, J155444, and J214054) where the
    observed and corrected percentage of polarization differ by more
    than 2$\sigma$. For these objects we applied the ISP correction.}
\label{ispfig}
\end{figure}

In Fig. \ref{ispfig} (top panel) we compare the observed continuum
polarization with the estimated value of the ISP. The other two panels
show instead the effects of the subtraction of the ISP on $P_{\rm
  cont.}$ and PA$_{\rm cont}$. Overall, these measurements are
consistent within the errors. The only exceptions are J154019,
J155444, and J214054 where the observed and corrected percentage of
polarization differ by more than 2$\sigma$. For these objects we
applied the ISP correction.

We report in the Table below the observed and corrected value for these sources.

\begin{table}
\caption{ISP corrections}
\begin{tabular}{l | c r c r r r | r r r r r}
\hline
Name    & $P$ cont. (\%)   &PA cont.    & $P$ corr. (\%)   &PA corr.    \\
\hline                                                                 
J154019 &   1.13$\pm$ 0.11 &  79$\pm$ 3 &   2.03$\pm$ 0.26 &  89$\pm$ 4 \\
J155444 &   1.50$\pm$ 0.06 &  84$\pm$ 1 &   1.23$\pm$ 0.09 &  86$\pm$ 2 \\
J214054 &   0.40$\pm$ 0.07 & 133$\pm$ 5 &   0.67$\pm$ 0.12 & 131$\pm$ 5 \\
\hline                                                                          
\end{tabular}
\label{tabisp}

\medskip
Column description: 1) source name, 2) and 3) observed percentage of
polarization and polarization position angle of the continuum at 6563
\AA, 4) and 5) percentage of
polarization and polarization position angle after ISP subtraction.
\end{table}

\newpage
\clearpage
\section{Supplementary material}

\begin{figure*}
\includegraphics[width=0.49\textwidth]{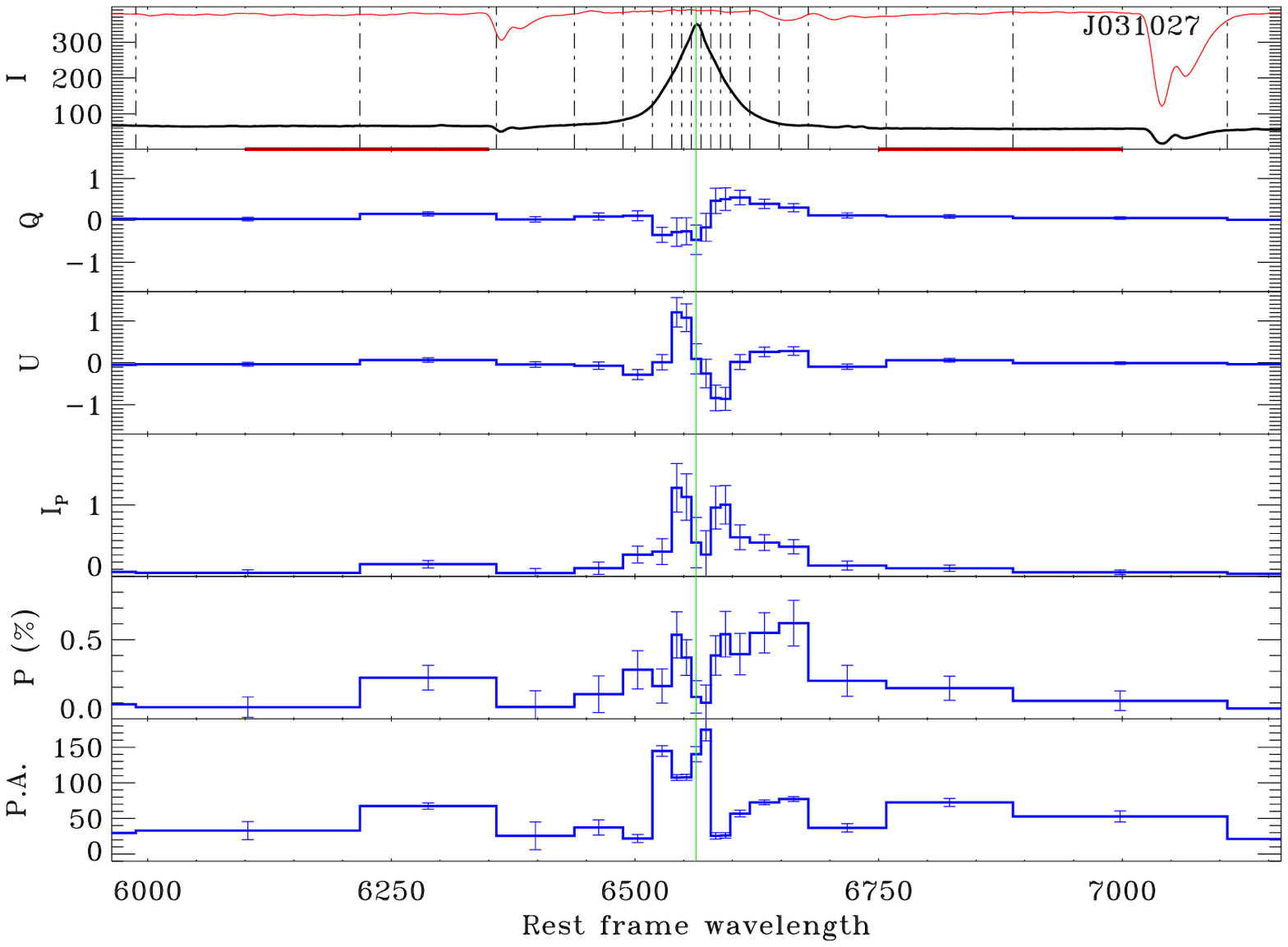}
\includegraphics[width=0.49\textwidth]{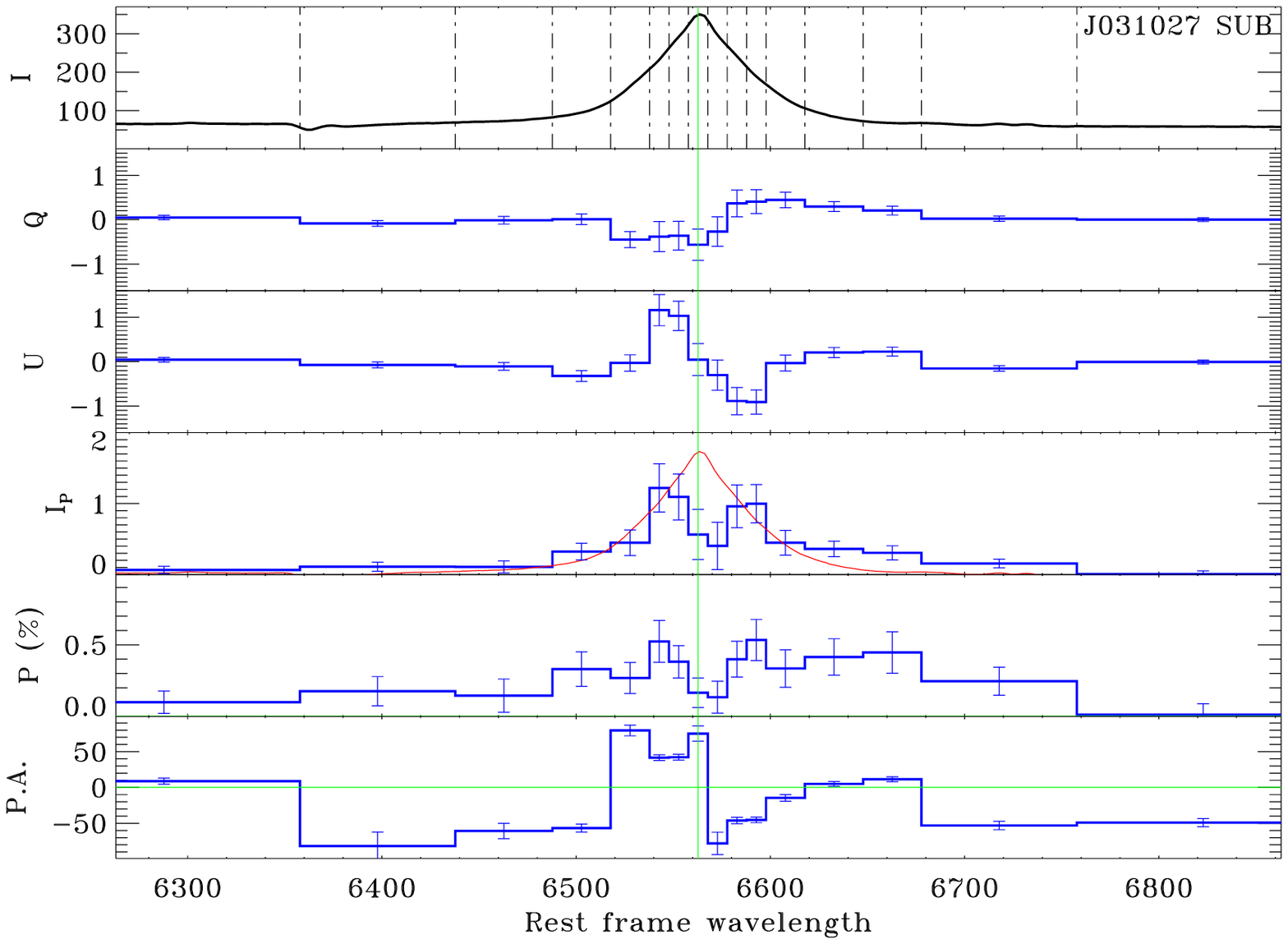}
\bigskip
\includegraphics[width=0.49\textwidth]{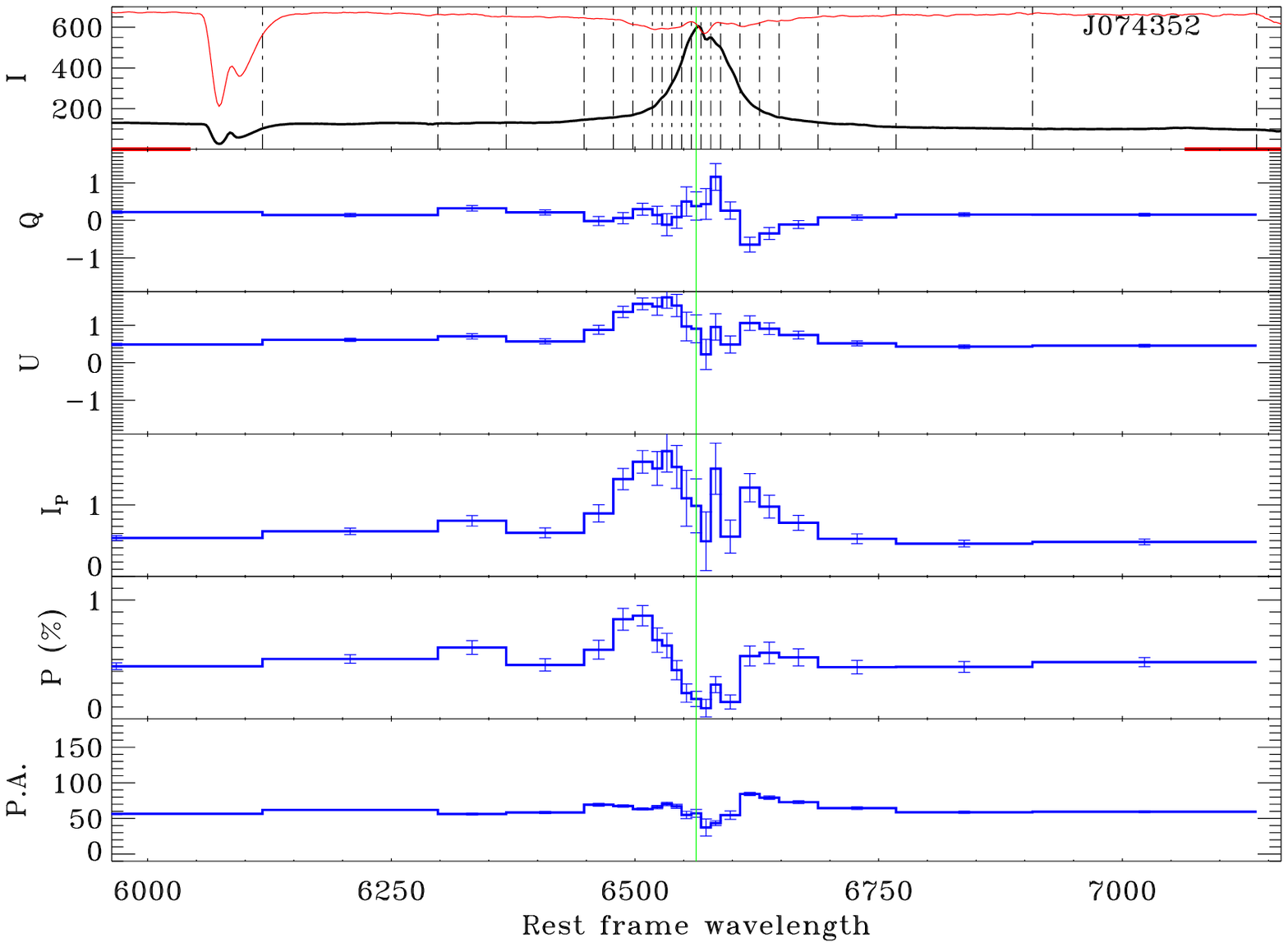}
\includegraphics[width=0.49\textwidth]{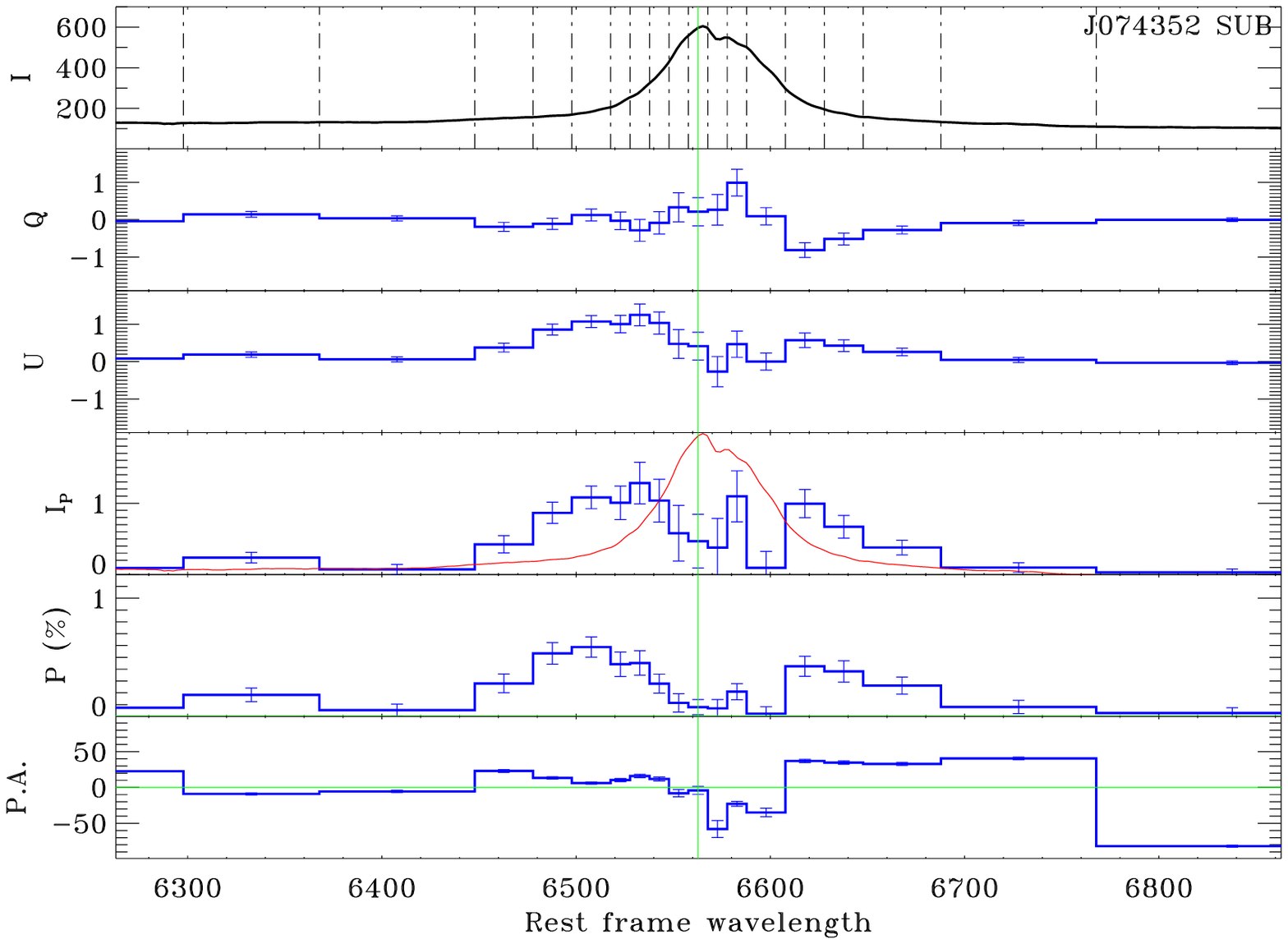}
\small{Left: top three panels: I, Q, and U Stokes parameters shown
  over a spectral region of 1200 \AA. I is shown at full spectral
  resolution, the vertical dashed lines mark the boundaries of the
  regions used for the rebinning. Wavelengths are rest frame in \AA,
  fluxes are in arbitrary units. The green vertical line marks the
  location of the \Ha\ line. The red curve in the top panel represents
  the ratio between the SDSS and the corrected VLT spectra in which
  can be located the telluric absorption bands. The two horizontal red
  lines locate the regions used to estimate the continuum
  polarization. Bottom three panels: polarized flux $I_P$ , percentage
  of polarization $P$ , and polarization position angle P.A.. Right:
  same as the left panels, but after subtraction of the continuum
  polarization and with a reduced spectral region (600 \AA). On the
  polarized flux we overplotted, in red, the total intensity
  spectrum. In the $P$ panel, the horizontal line shows the continuum
  $P$ at the BLR center. Angles are measured with respect to the
  continuum $P.A.$.}
\label{spectra}
\end{figure*}

\begin{figure*}
\includegraphics[width=0.49\textwidth]{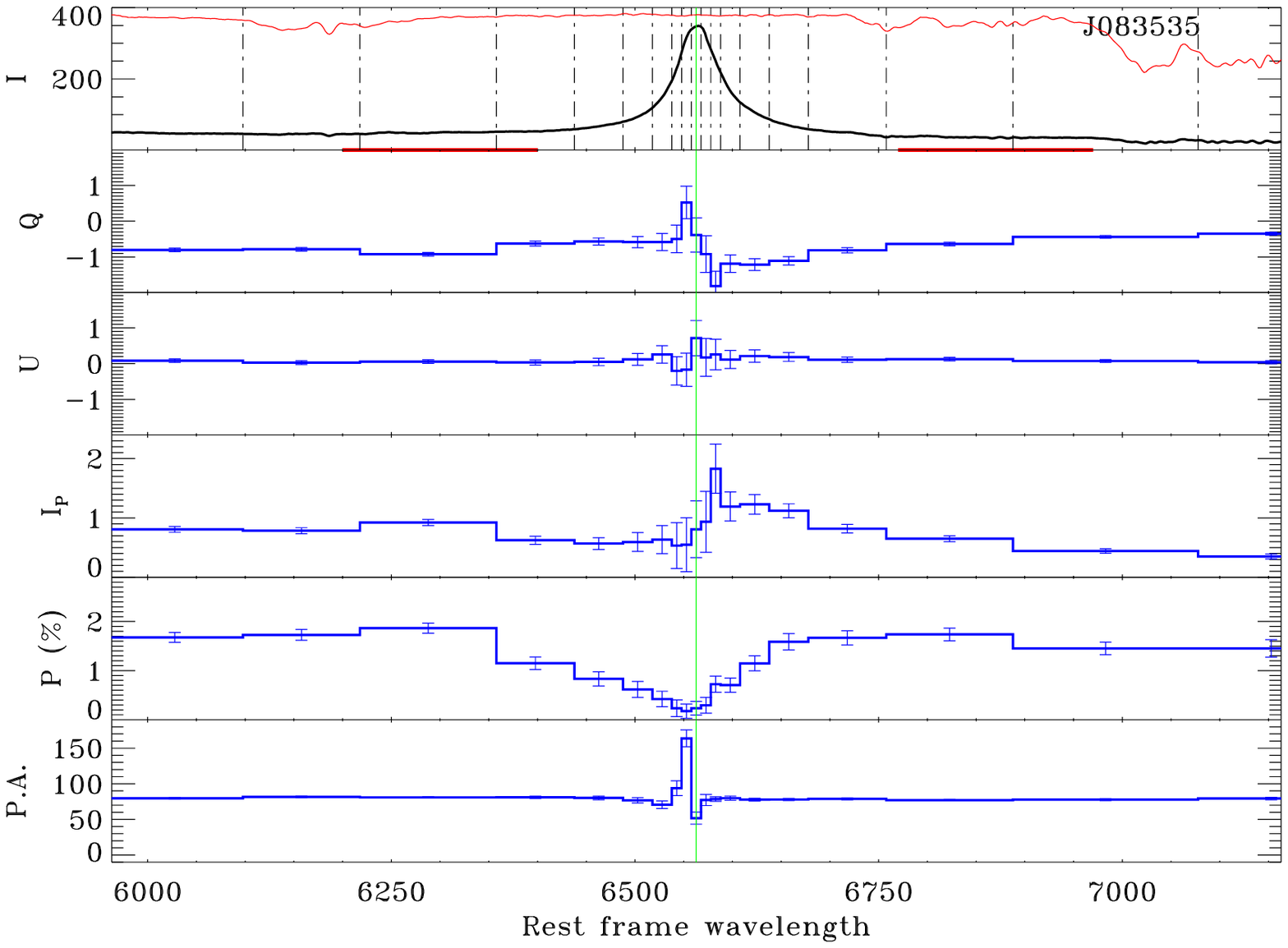}
\includegraphics[width=0.49\textwidth]{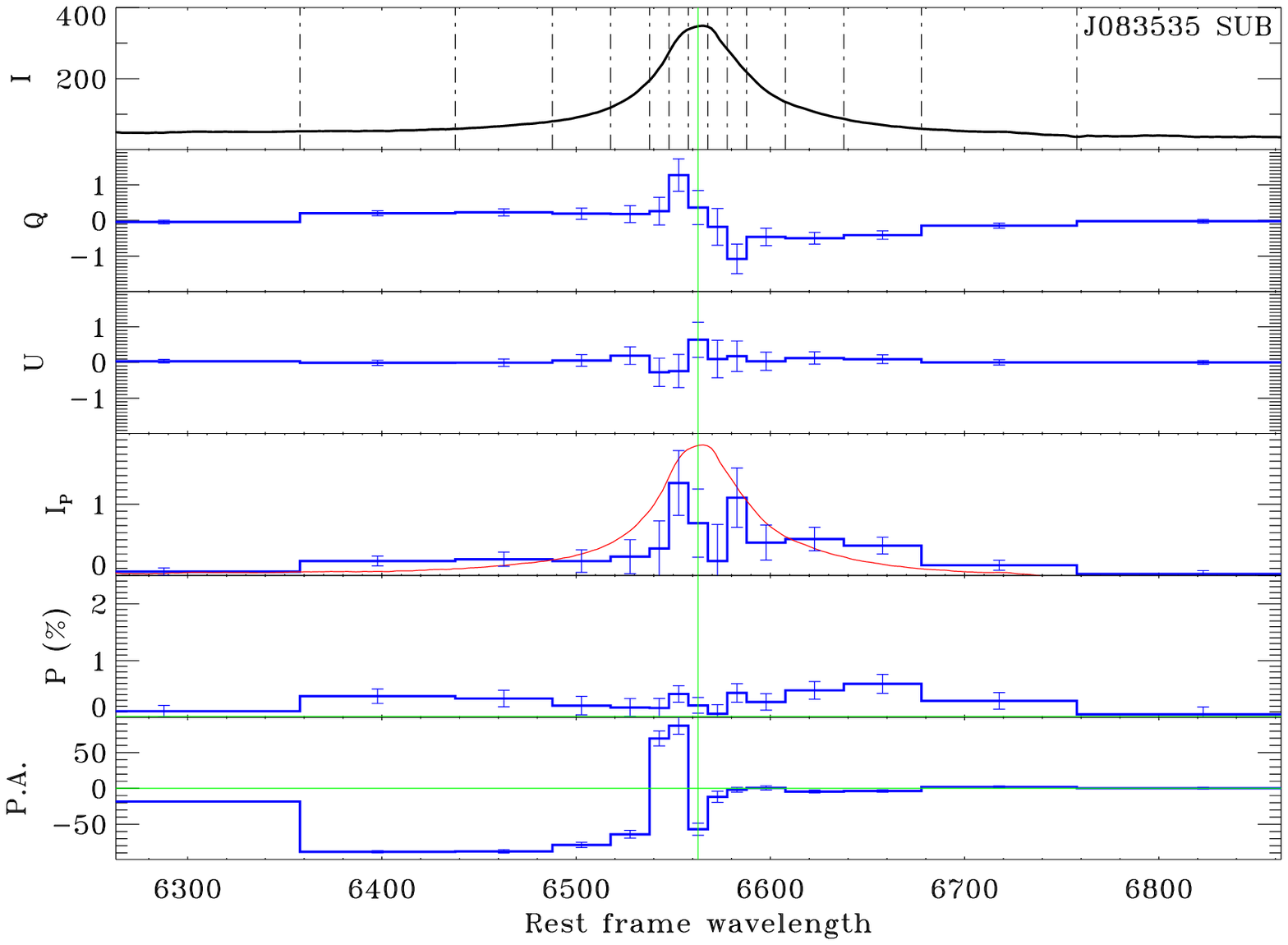}
\bigskip
\includegraphics[width=0.49\textwidth]{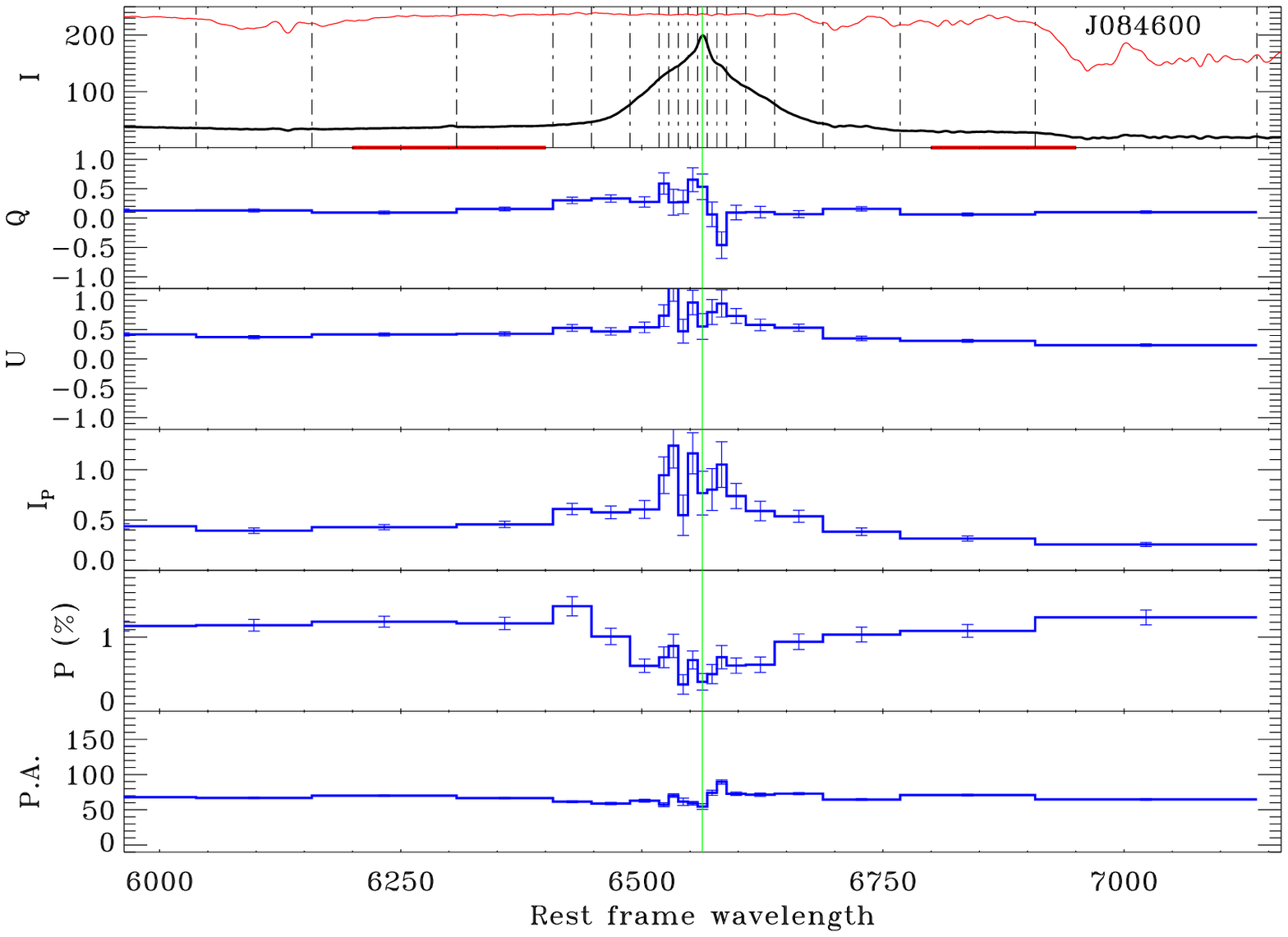}
\includegraphics[width=0.49\textwidth]{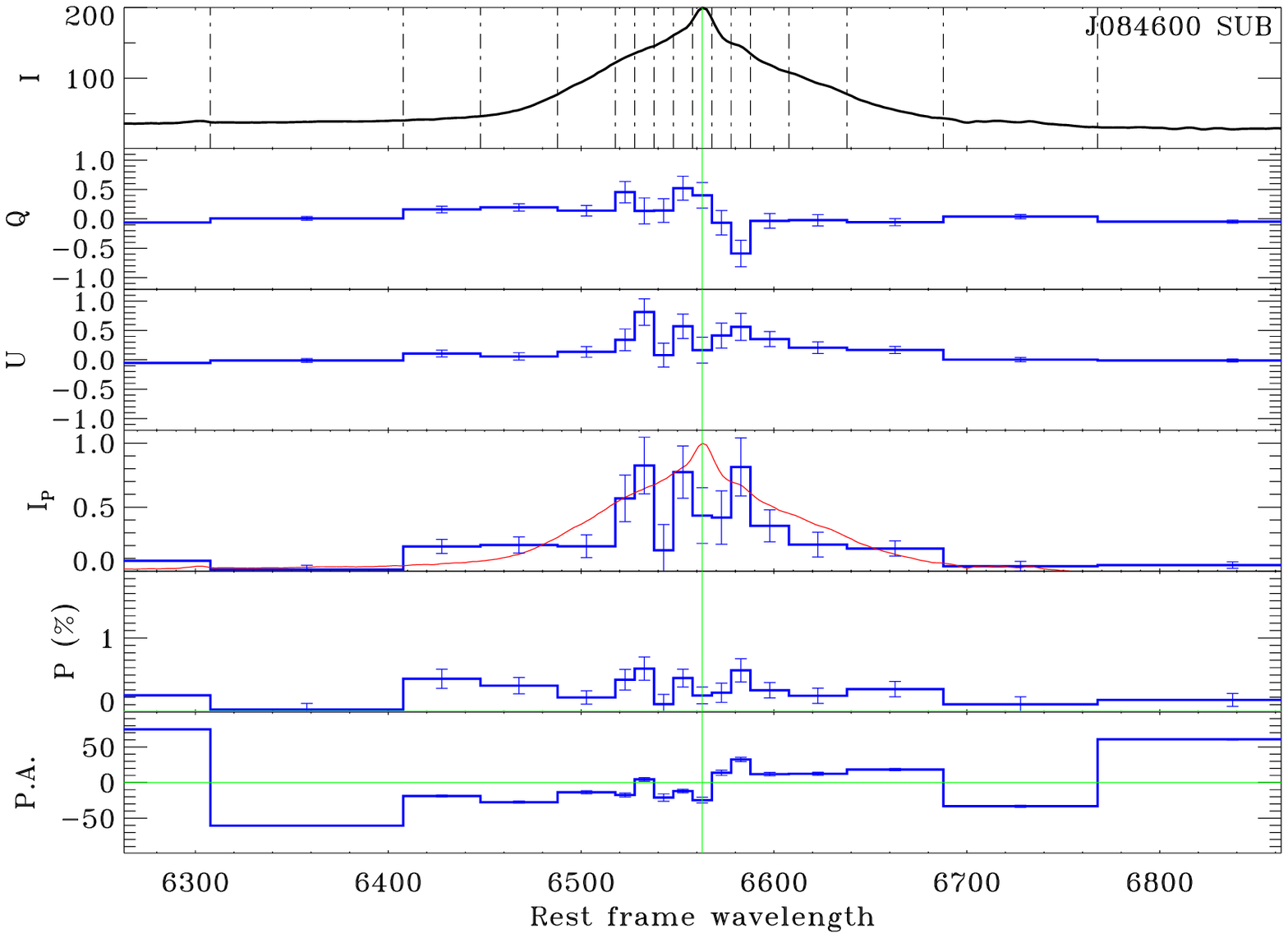}
\bigskip
\includegraphics[width=0.49\textwidth]{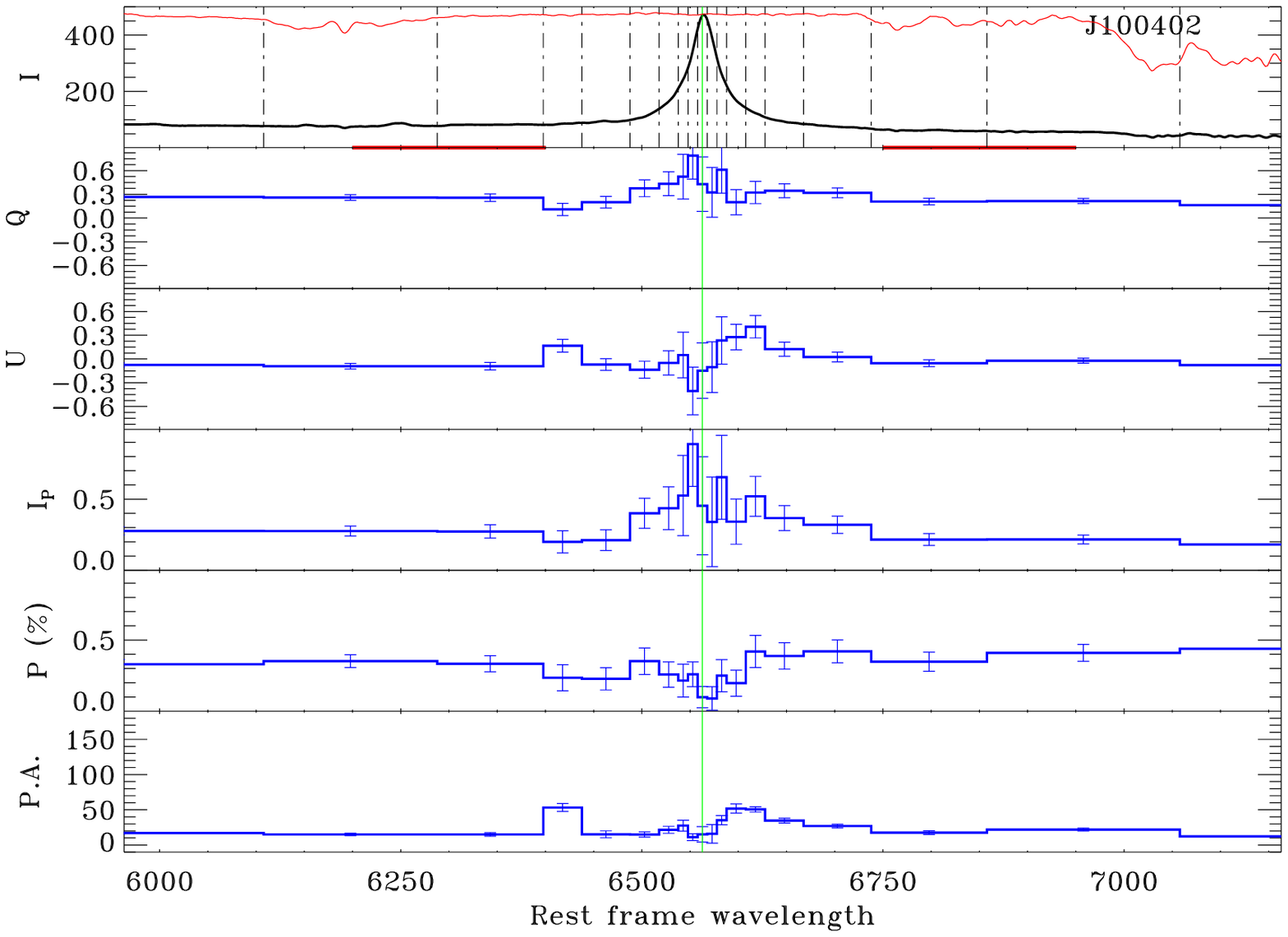}
\includegraphics[width=0.49\textwidth]{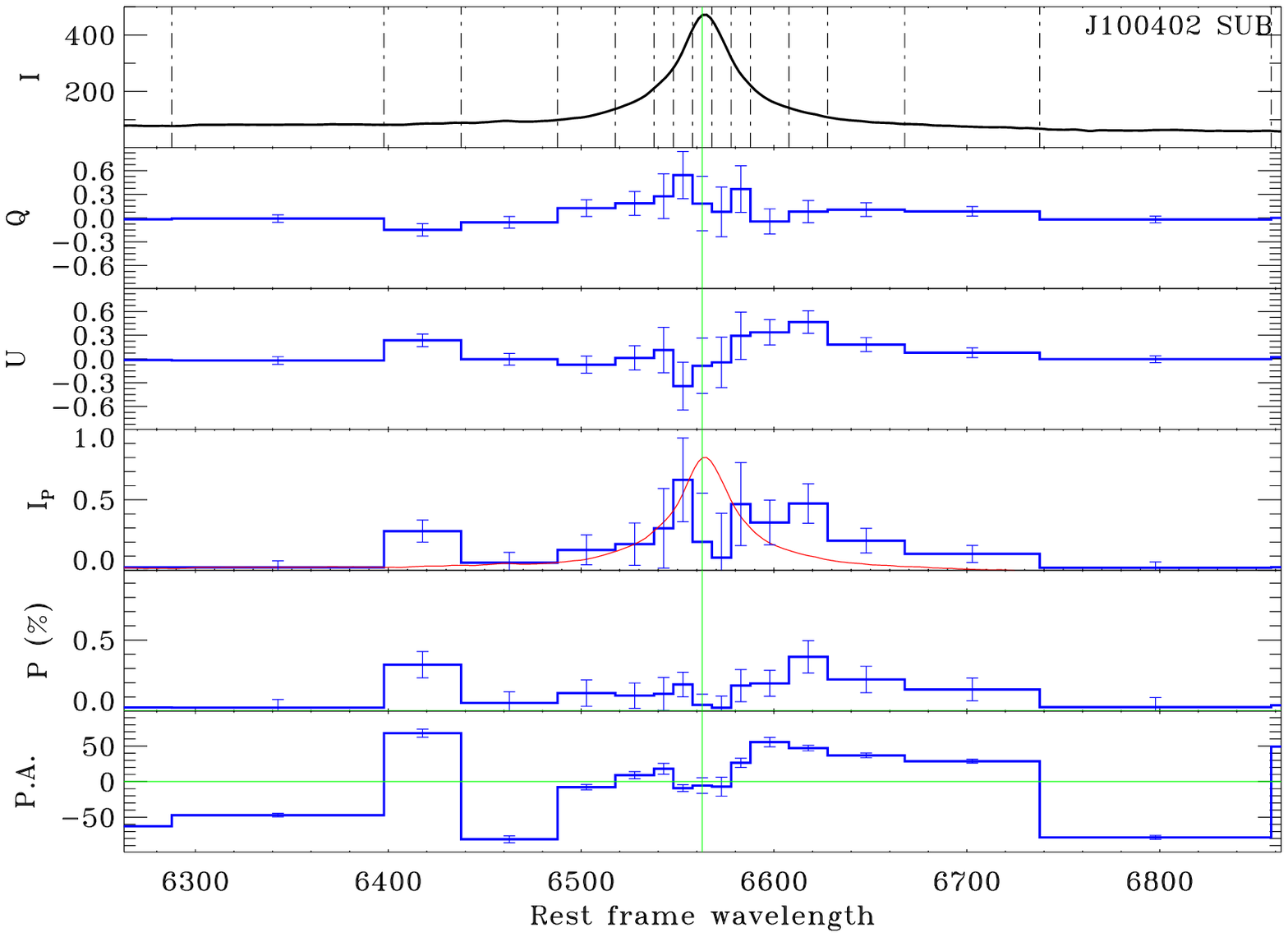}
\end{figure*}

\begin{figure*}
\includegraphics[width=0.49\textwidth]{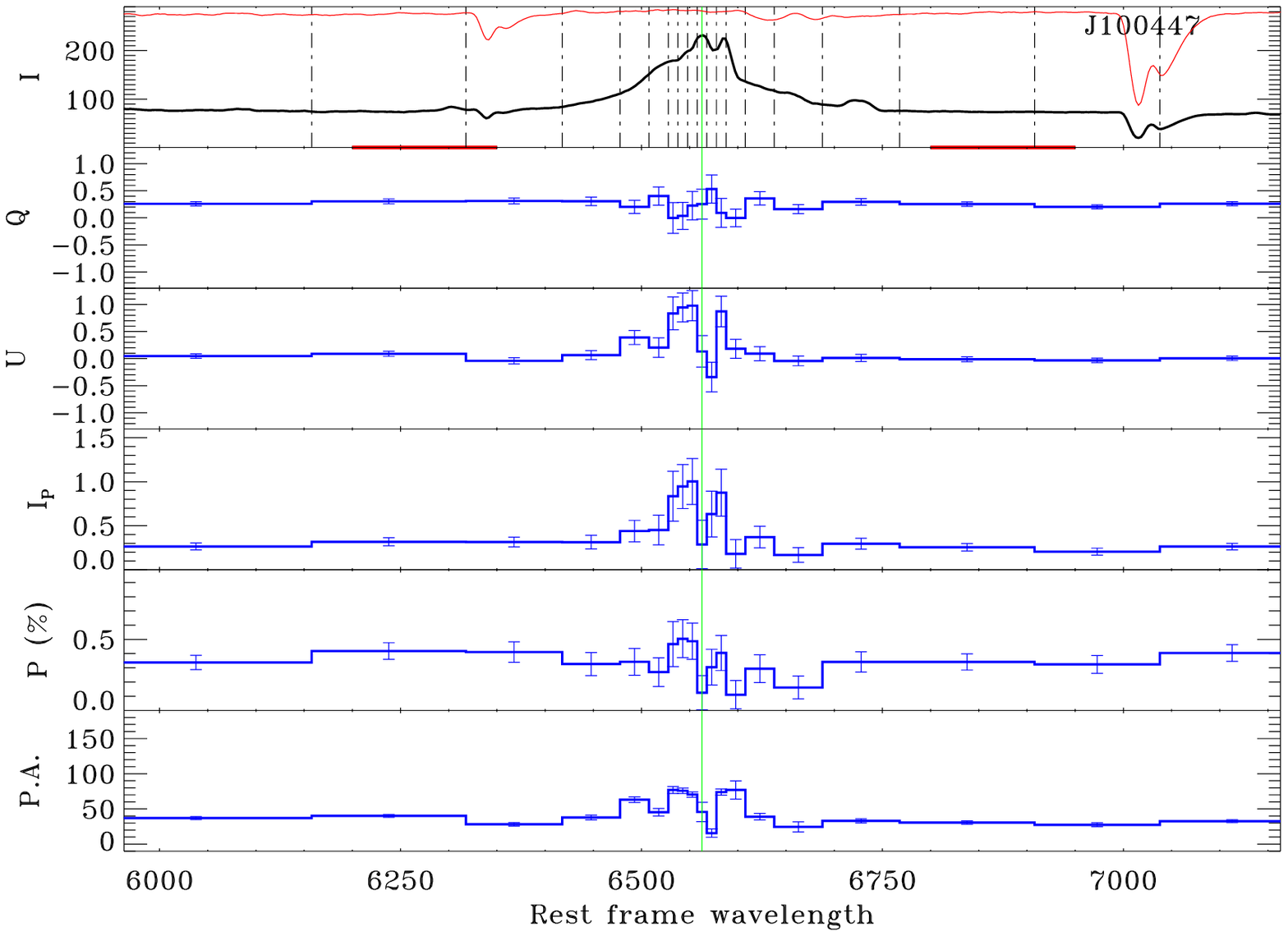}
\includegraphics[width=0.49\textwidth]{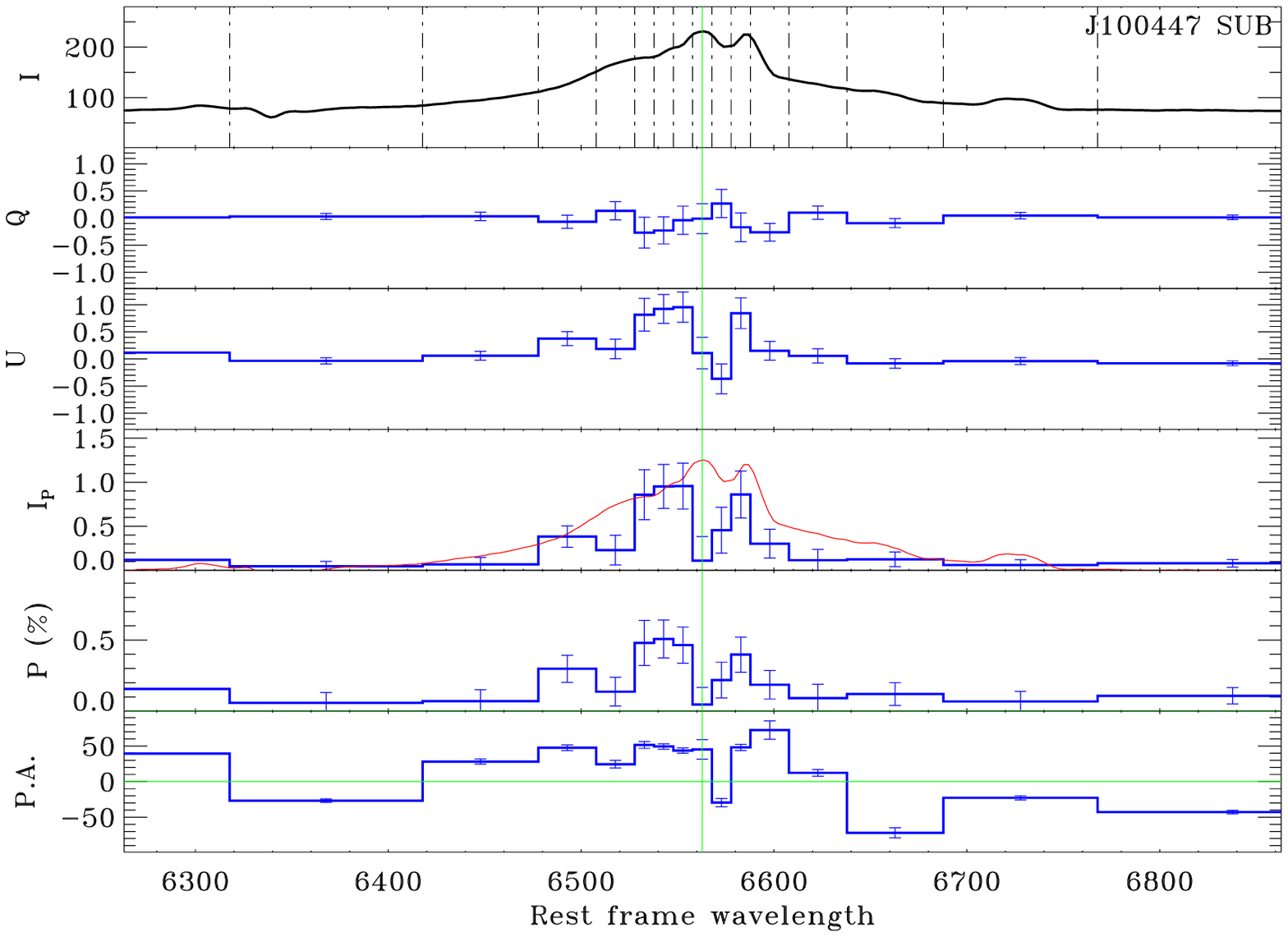}
\bigskip
\includegraphics[width=0.49\textwidth]{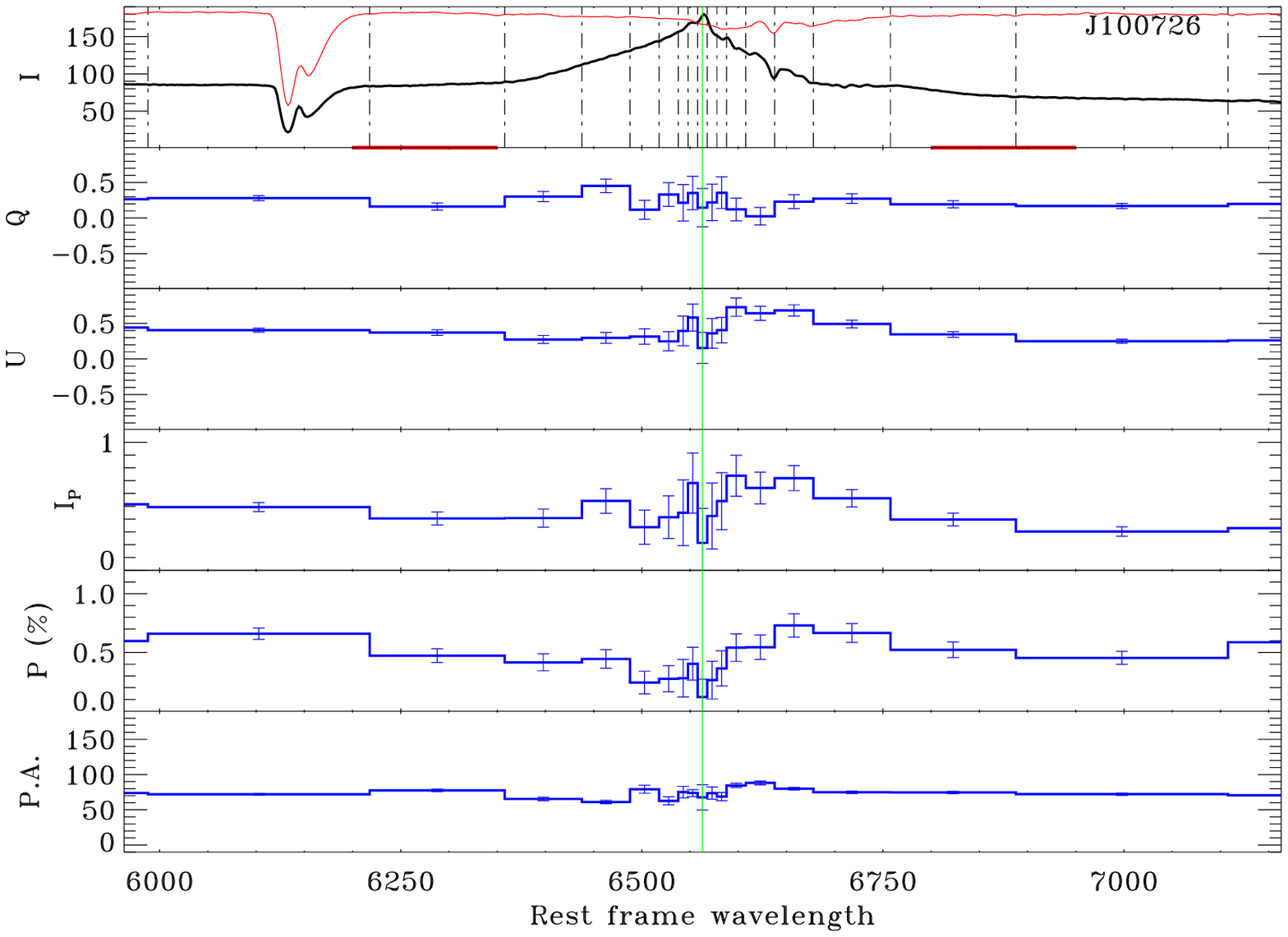}
\includegraphics[width=0.49\textwidth]{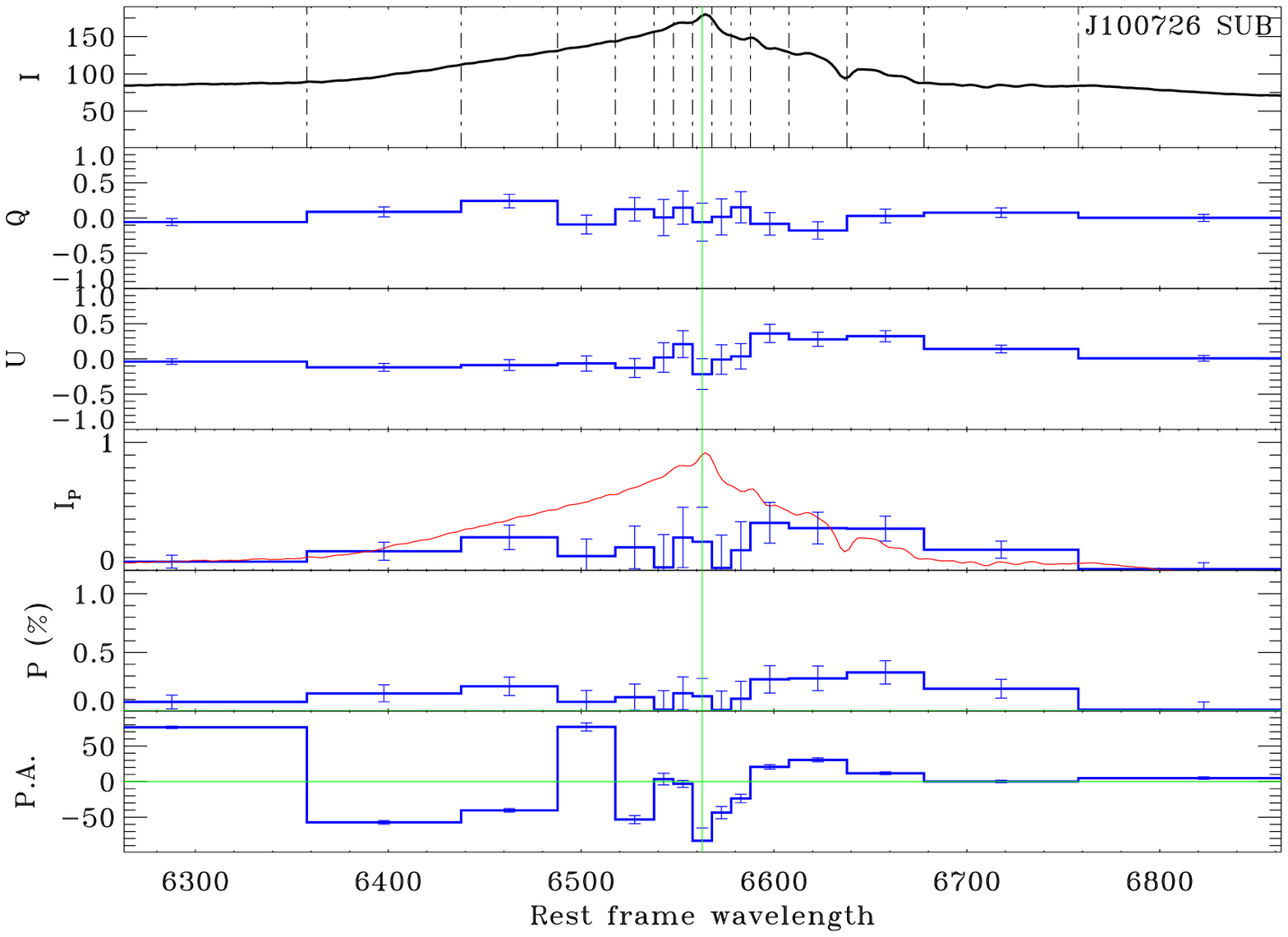}
\bigskip
\includegraphics[width=0.49\textwidth]{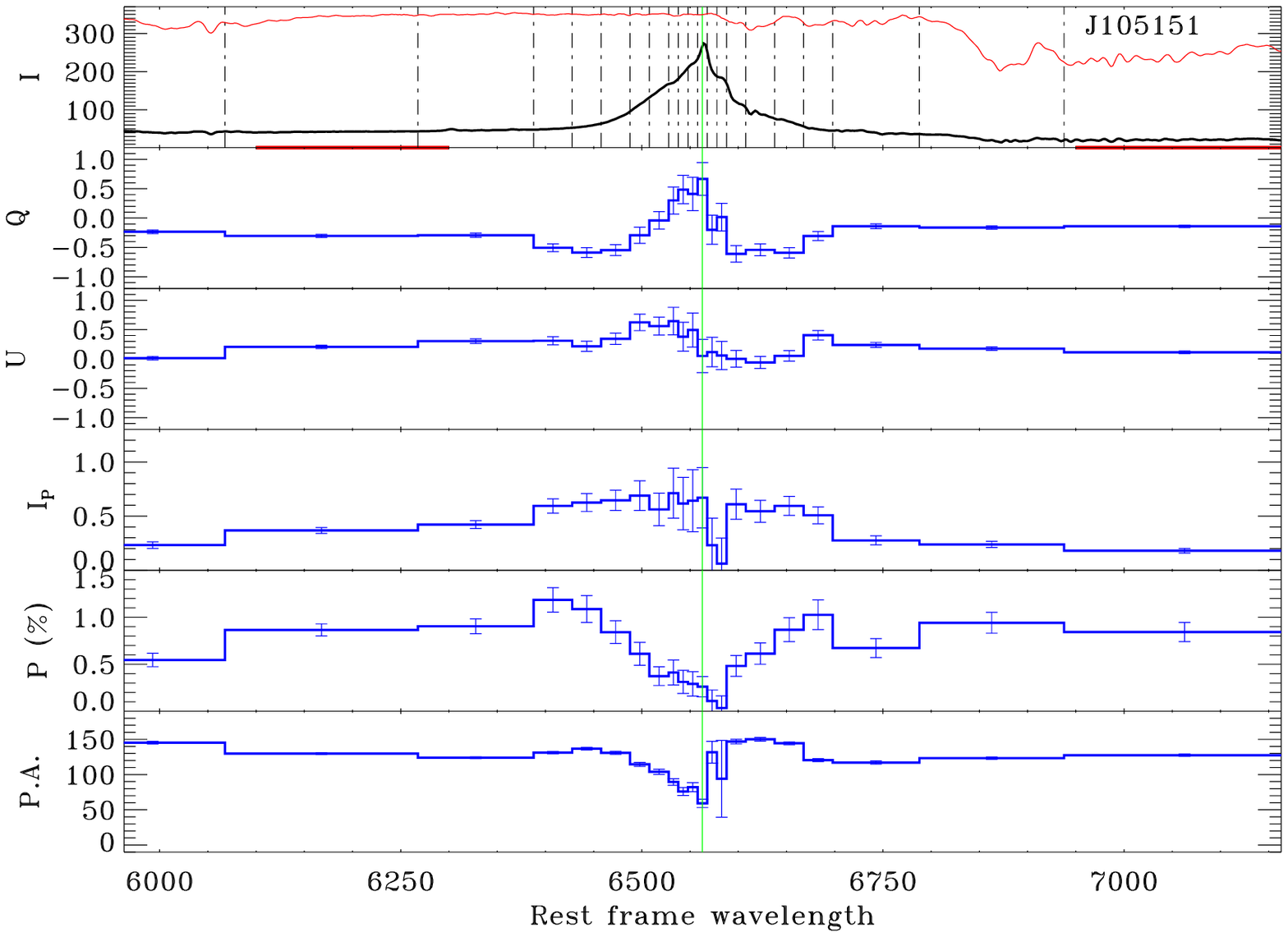}
\includegraphics[width=0.49\textwidth]{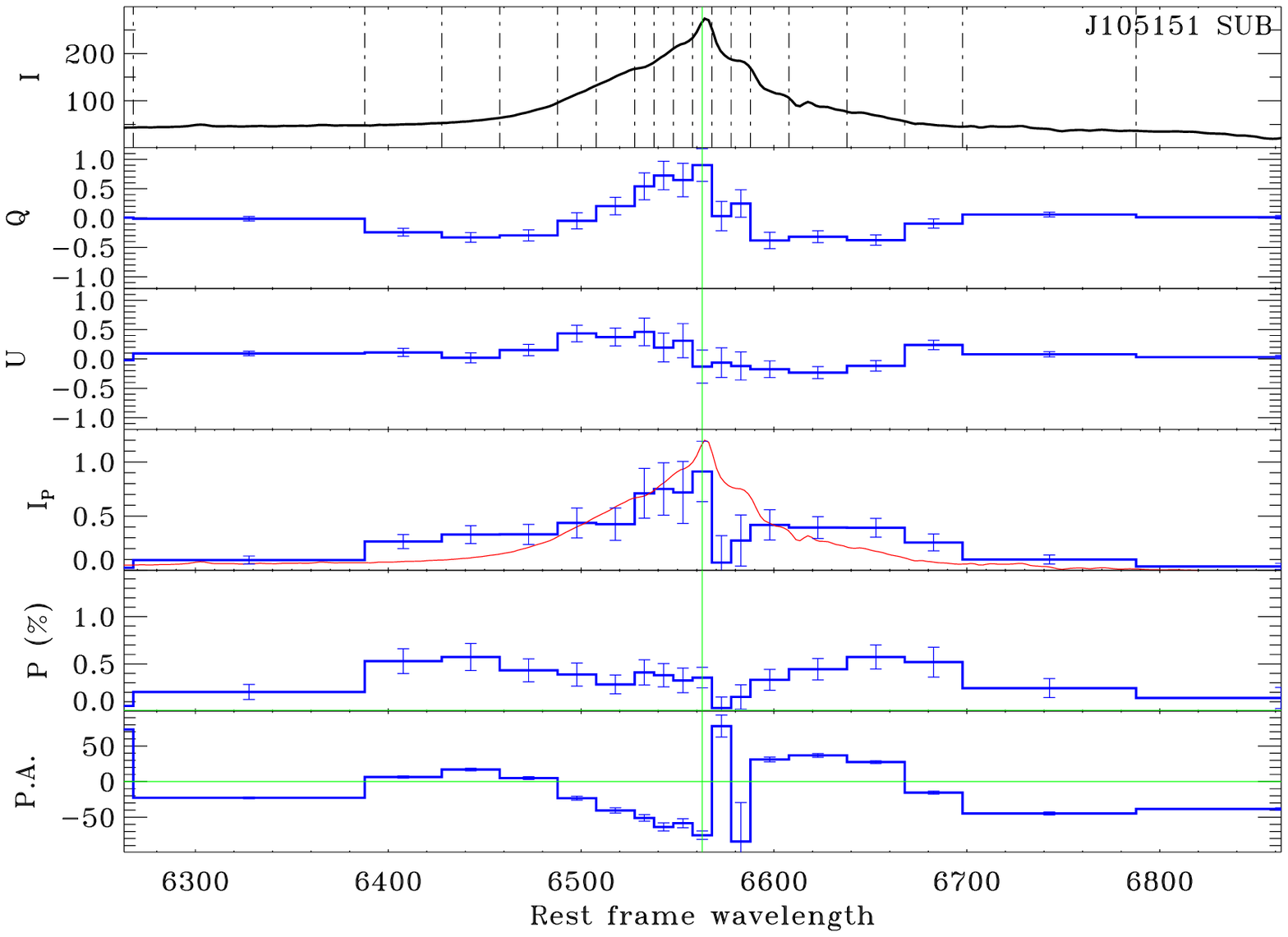}
\end{figure*}

\begin{figure*}
\includegraphics[width=0.49\textwidth]{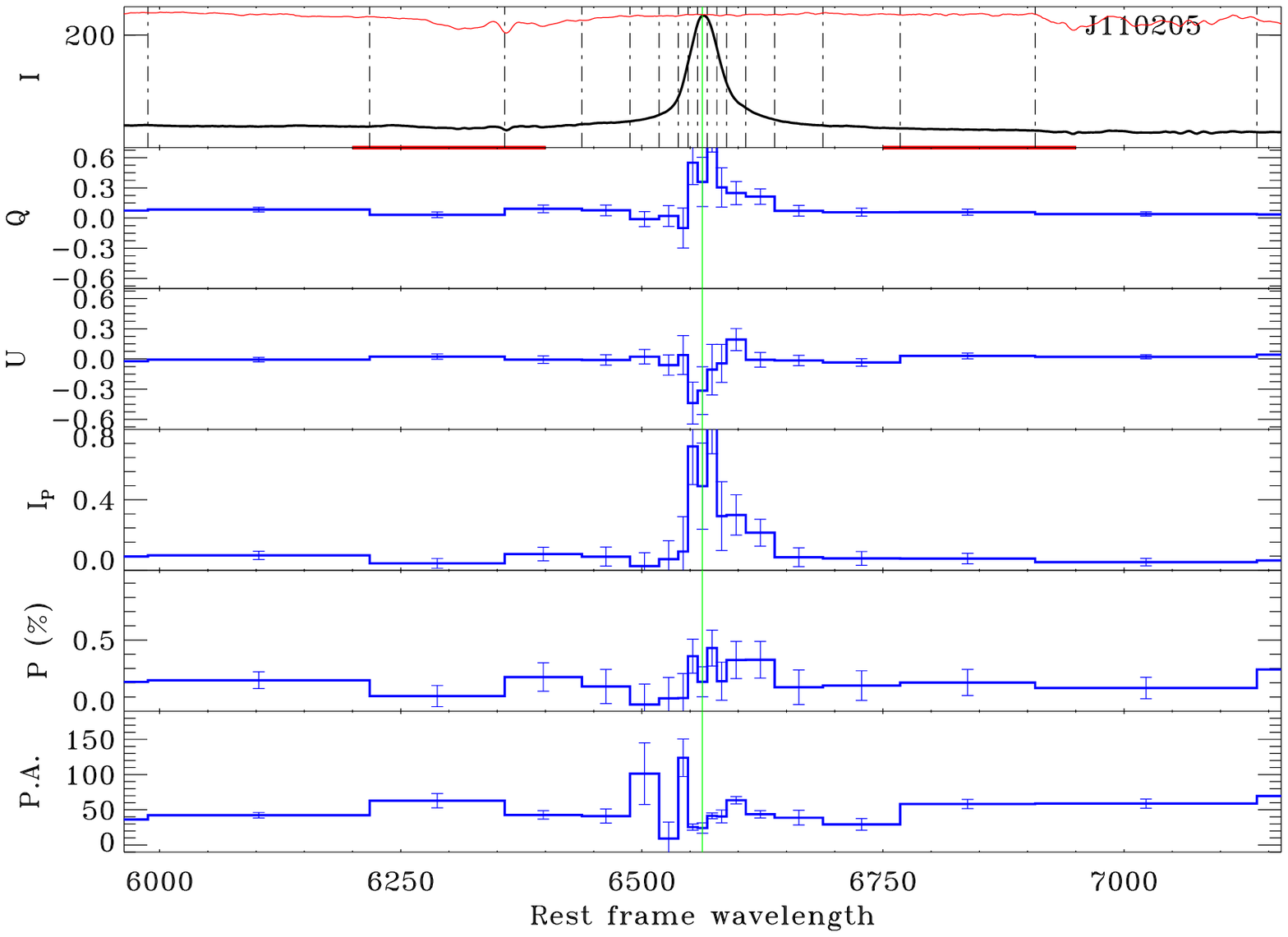}
\includegraphics[width=0.49\textwidth]{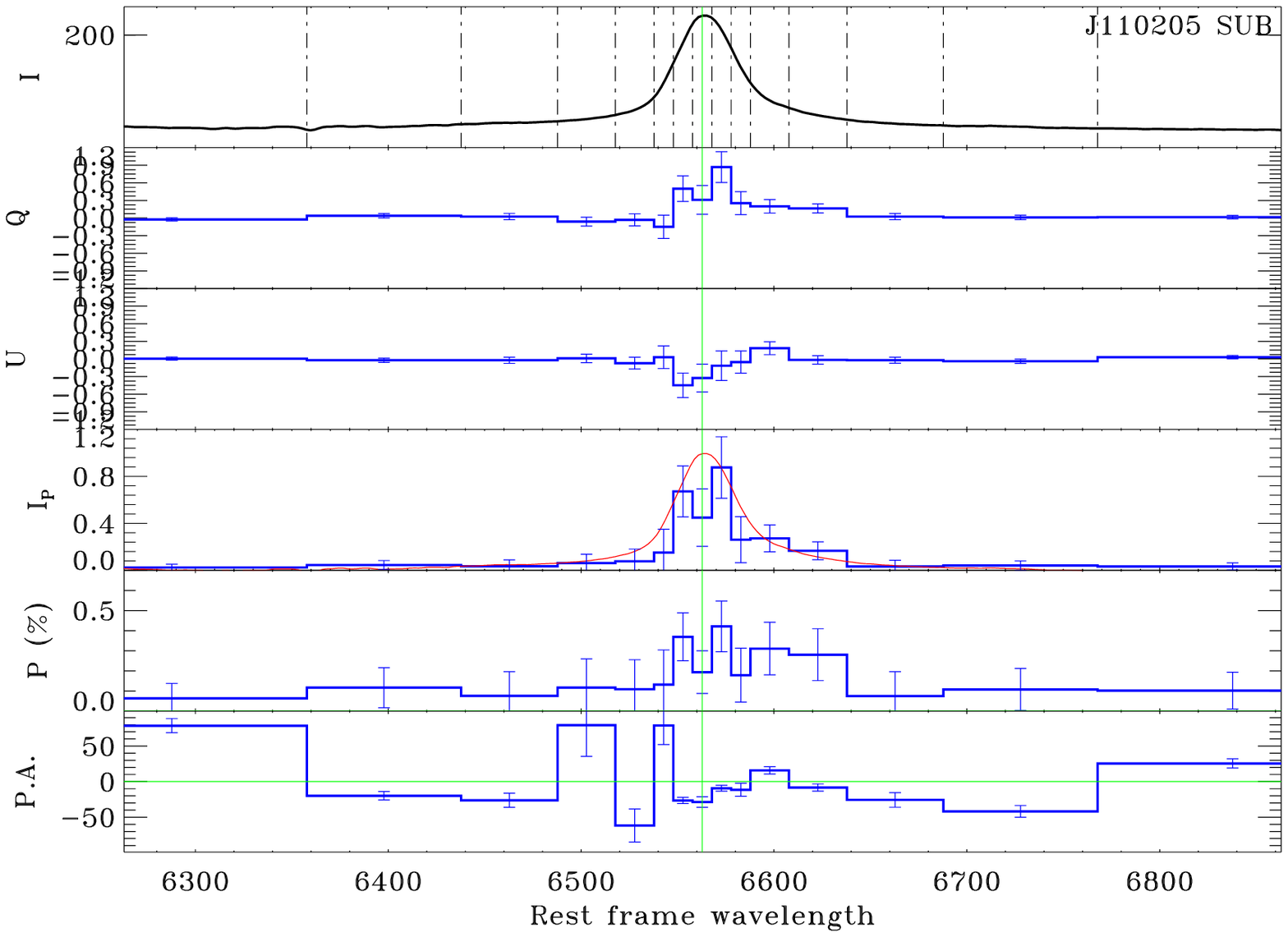}
\bigskip
\includegraphics[width=0.49\textwidth]{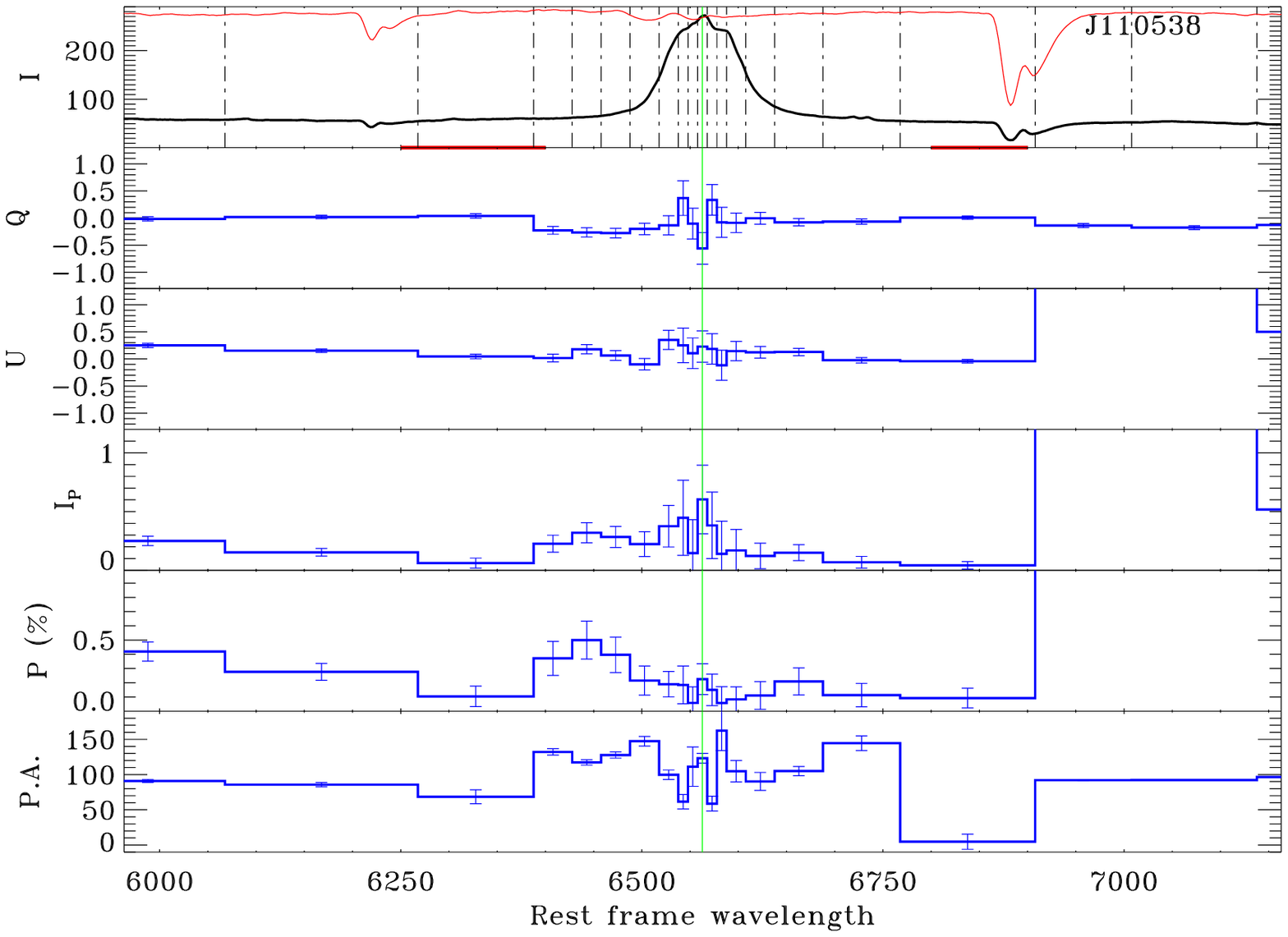}
\includegraphics[width=0.49\textwidth]{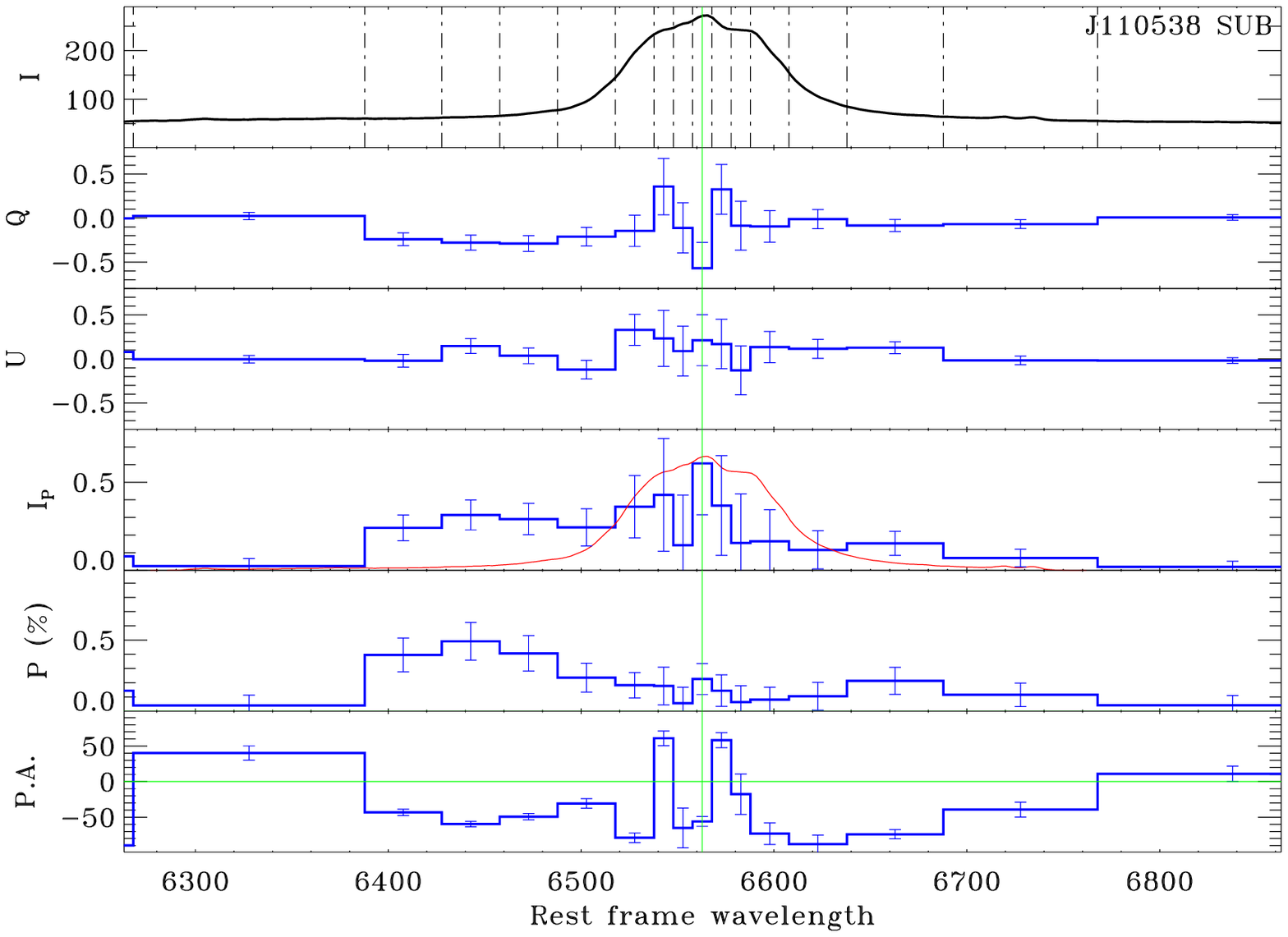}
\bigskip
\includegraphics[width=0.49\textwidth]{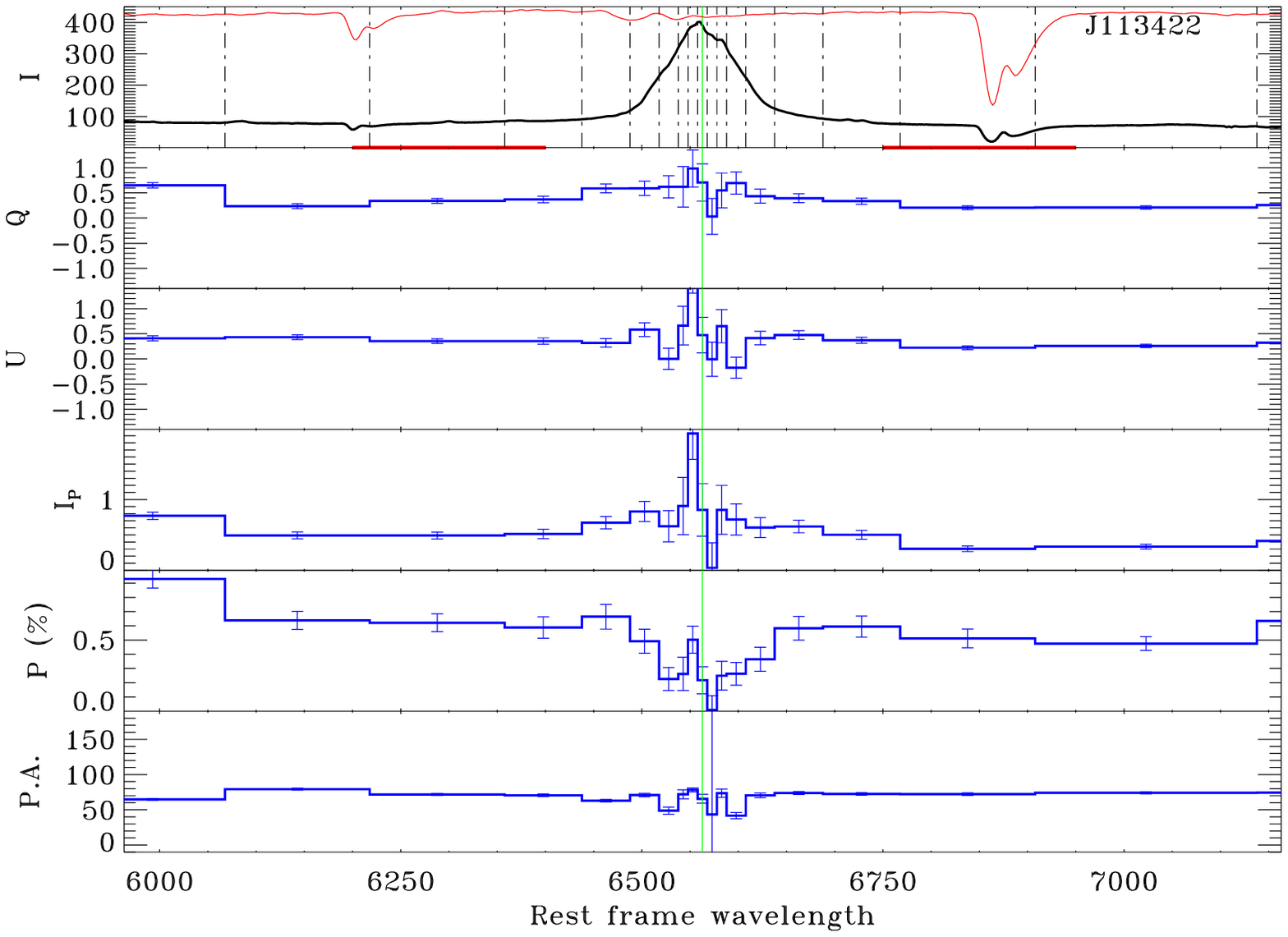}
\includegraphics[width=0.49\textwidth]{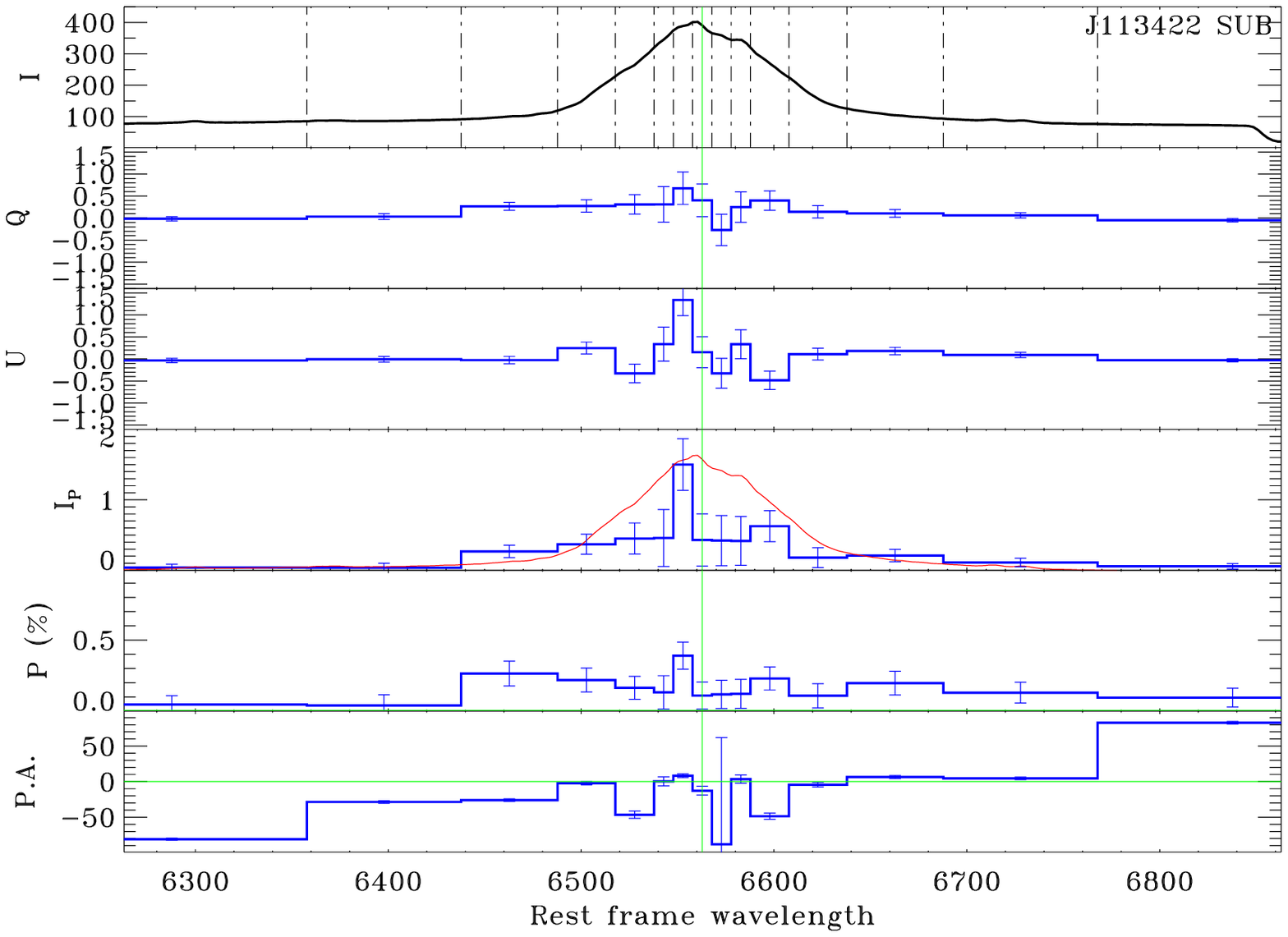}
\end{figure*}

\begin{figure*}
\includegraphics[width=0.49\textwidth]{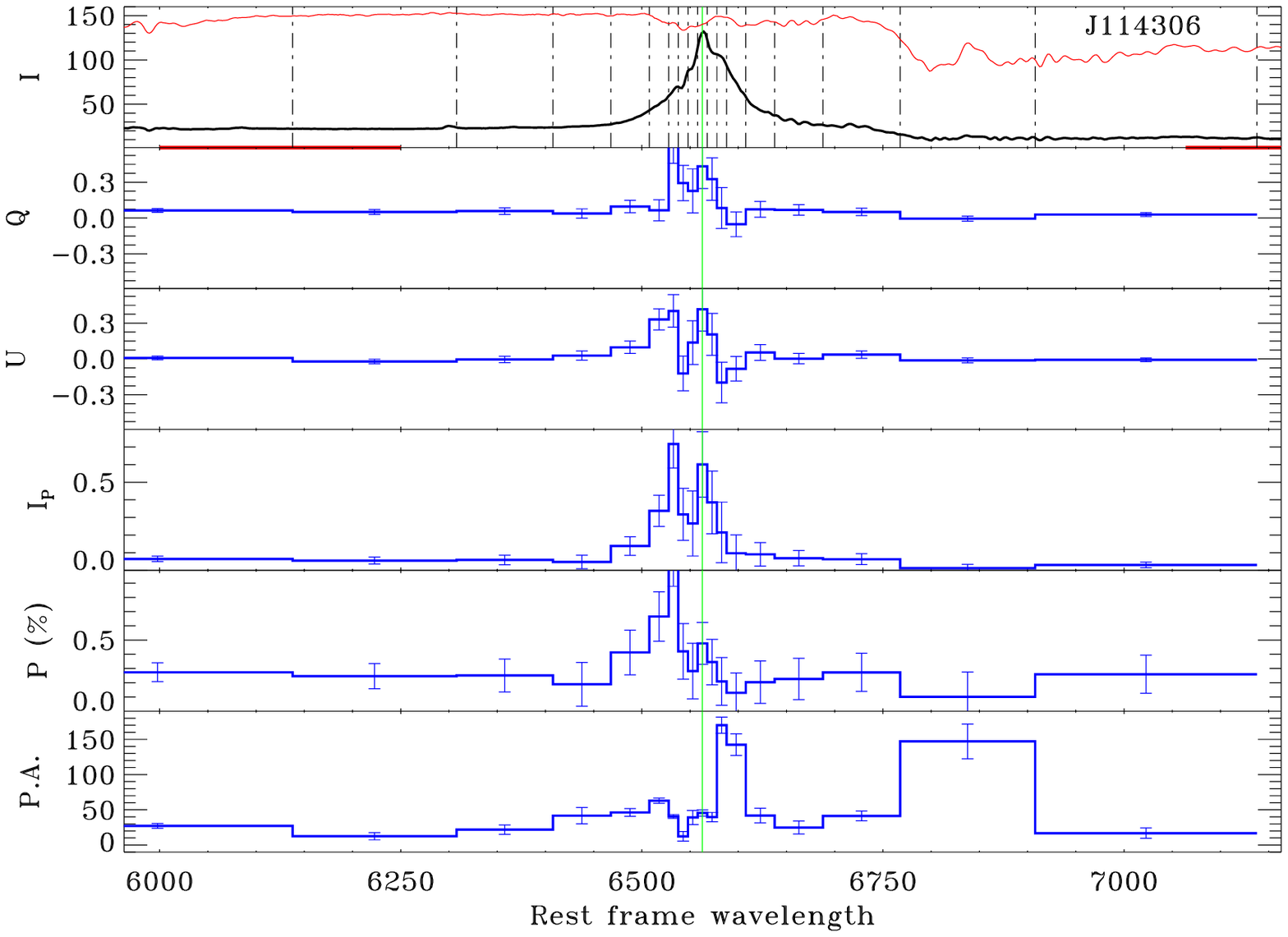}
\includegraphics[width=0.49\textwidth]{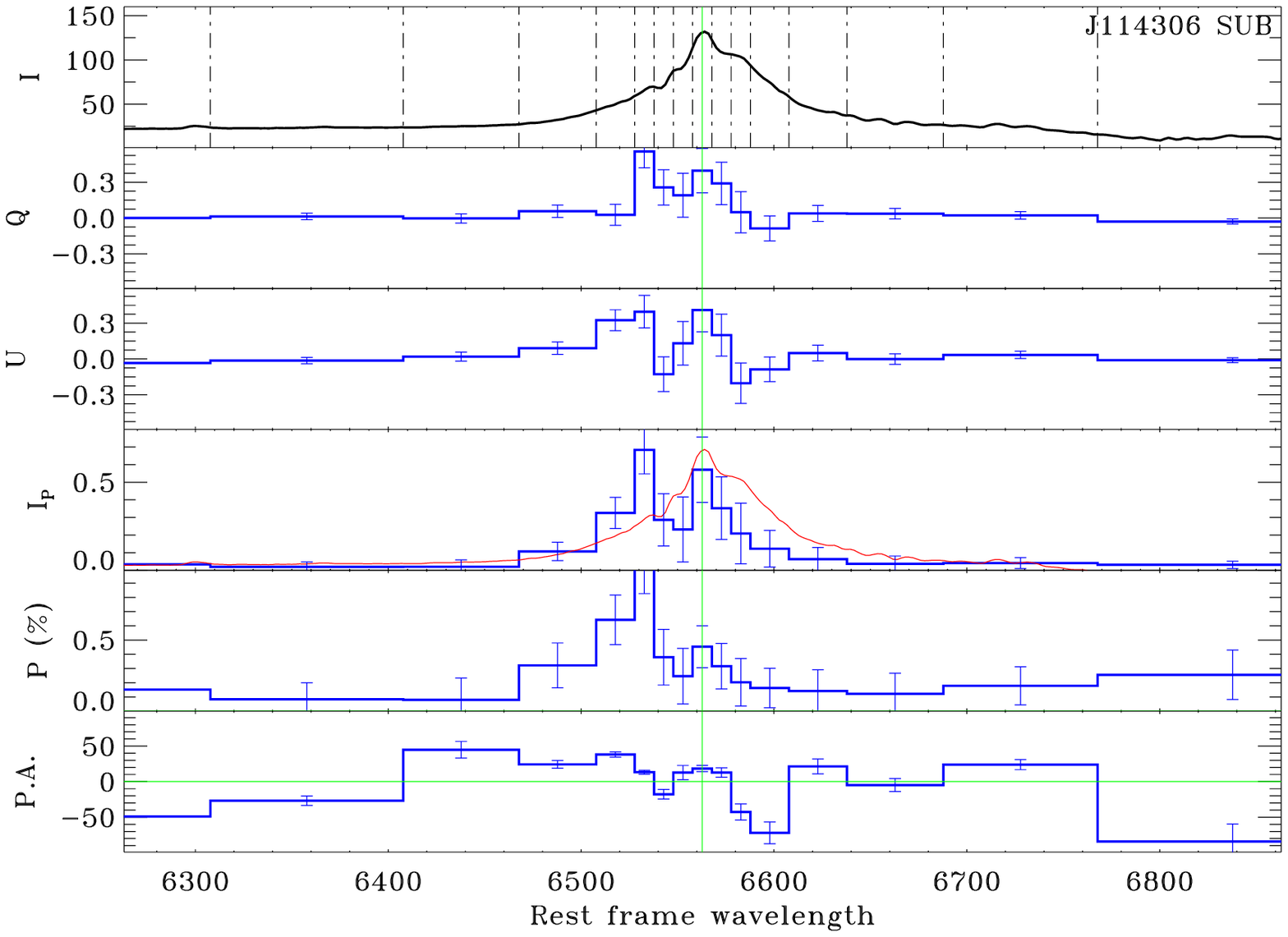}
\bigskip
\includegraphics[width=0.49\textwidth]{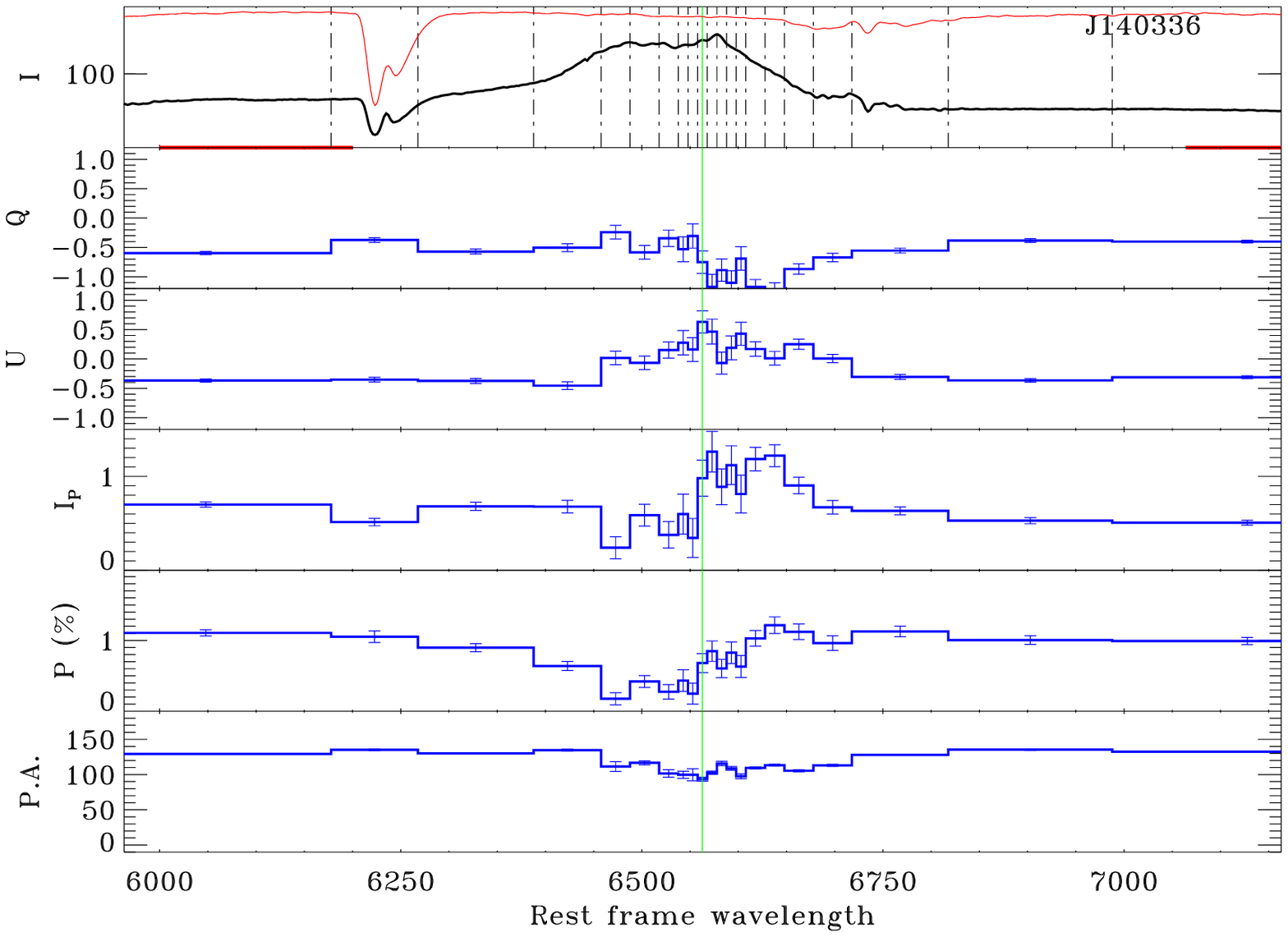}
\includegraphics[width=0.49\textwidth]{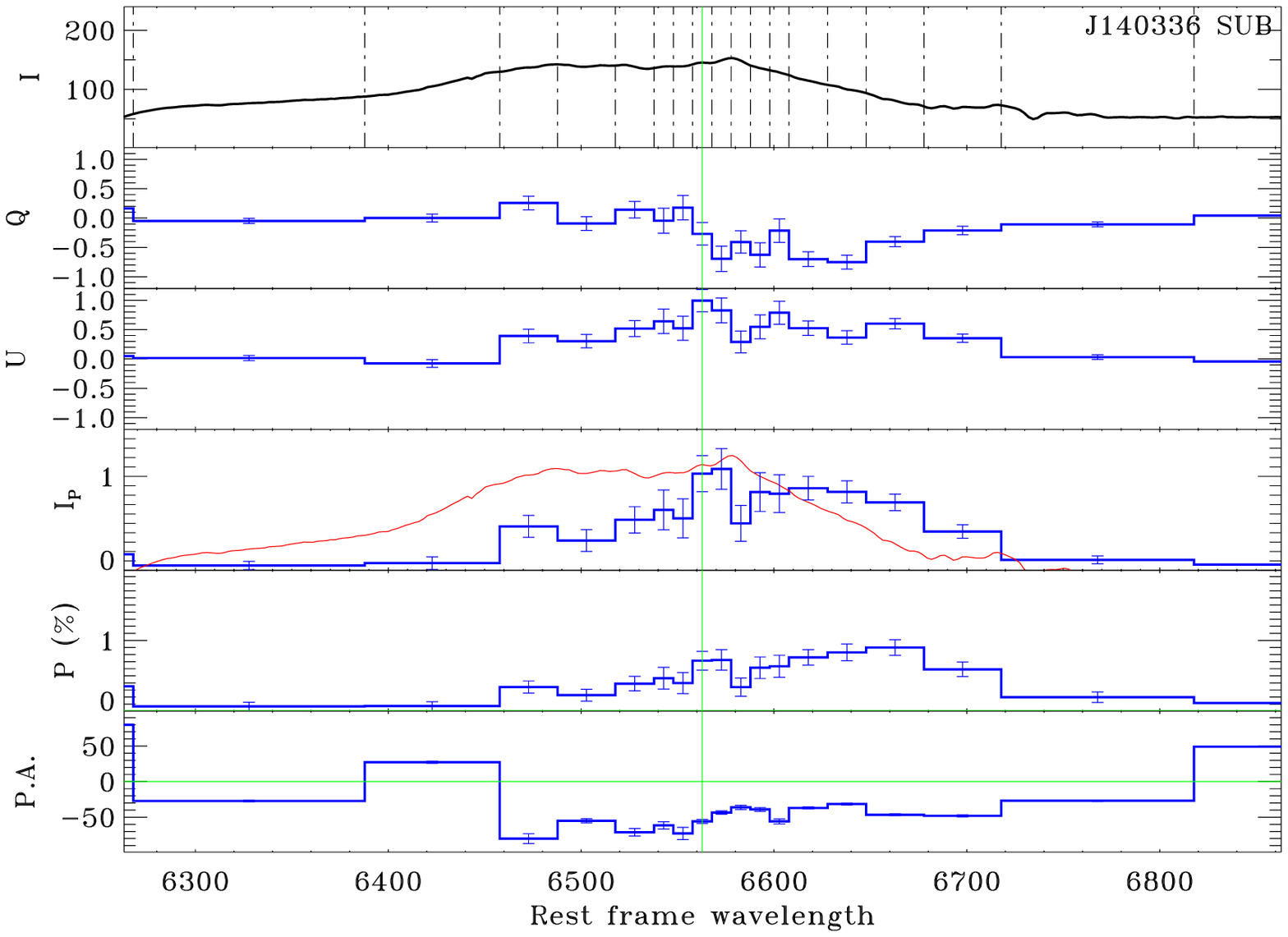}
\bigskip
\includegraphics[width=0.49\textwidth]{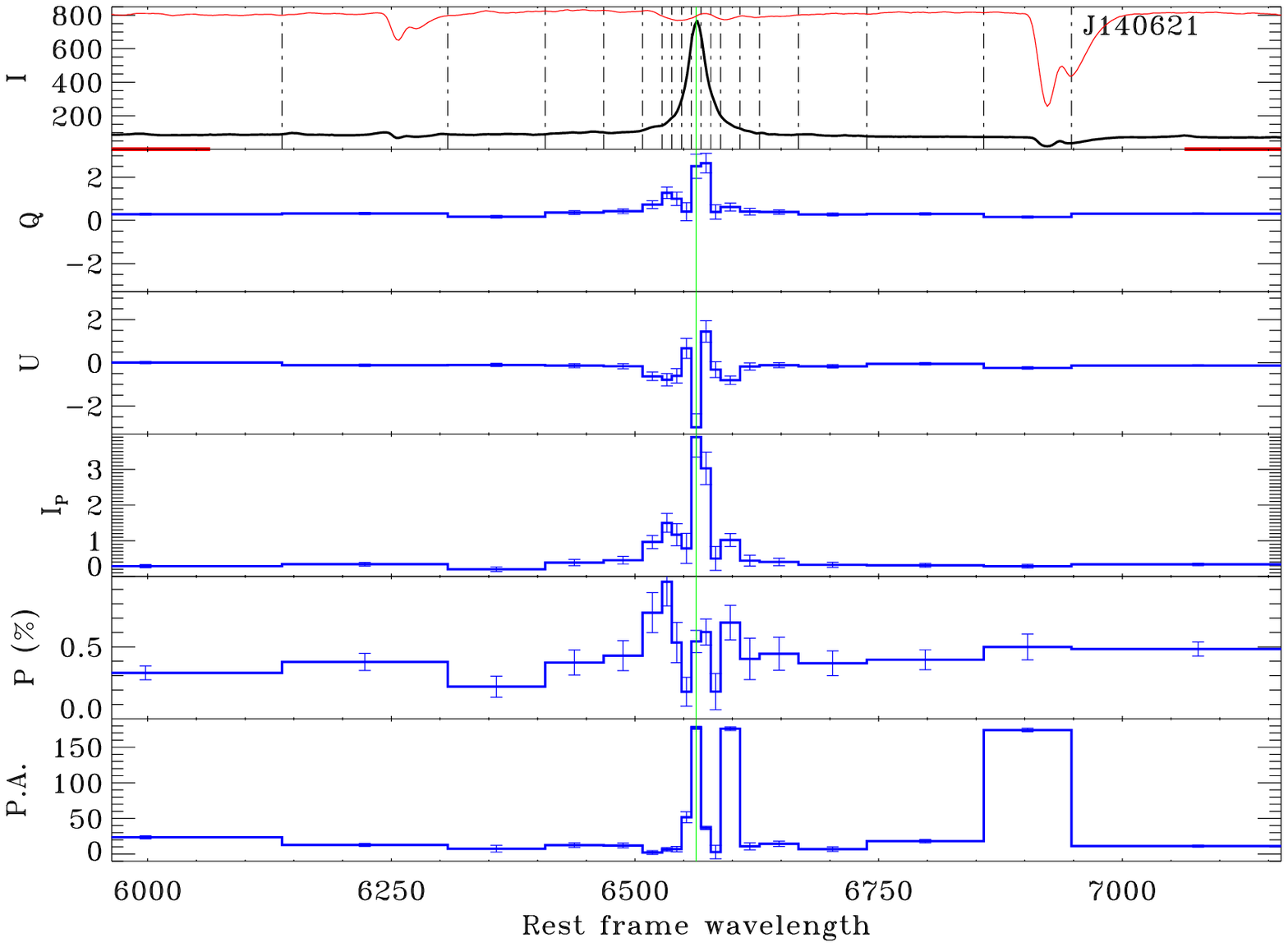}
\includegraphics[width=0.49\textwidth]{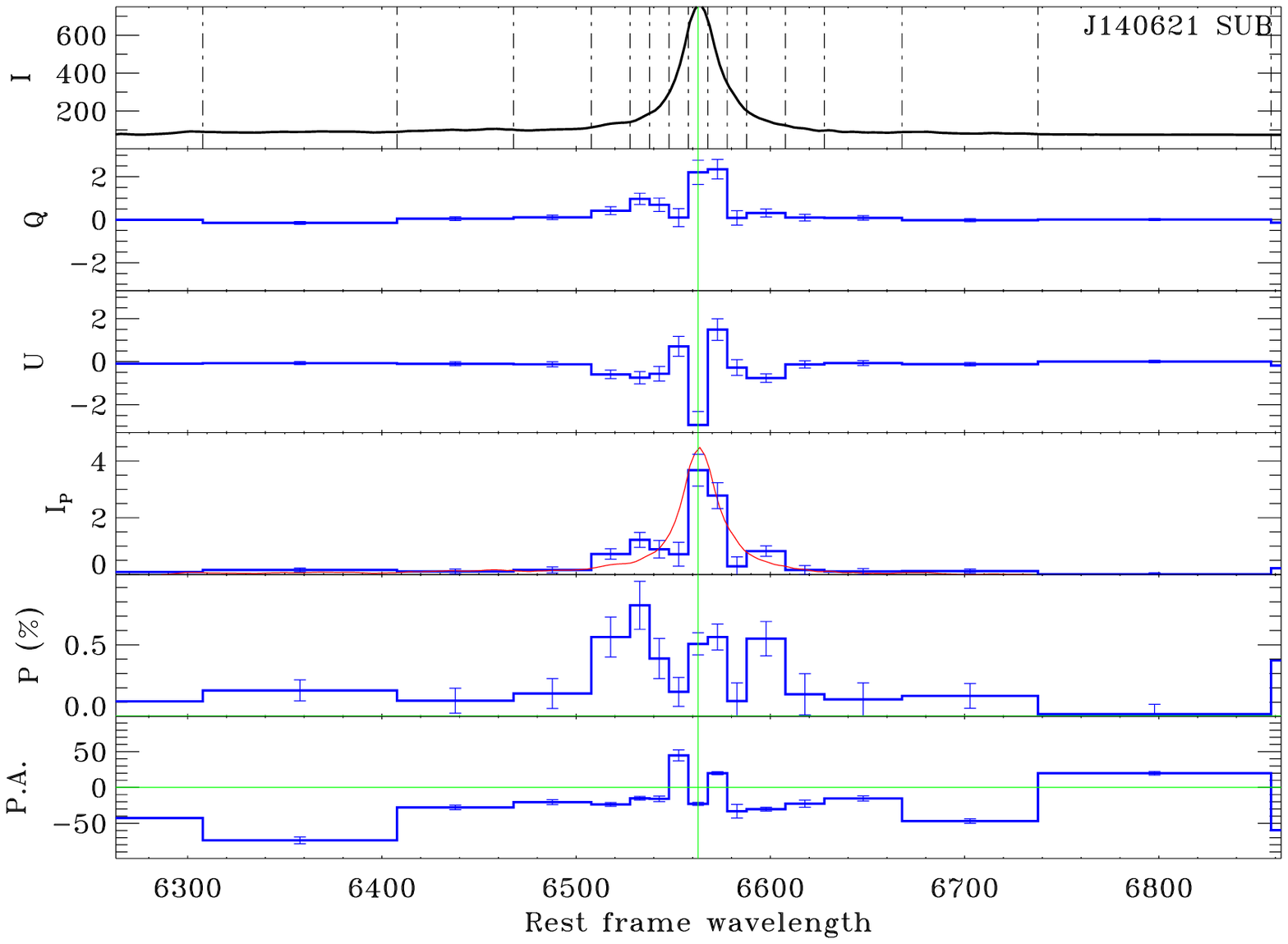}
\end{figure*}

\begin{figure*}
\includegraphics[width=0.49\textwidth]{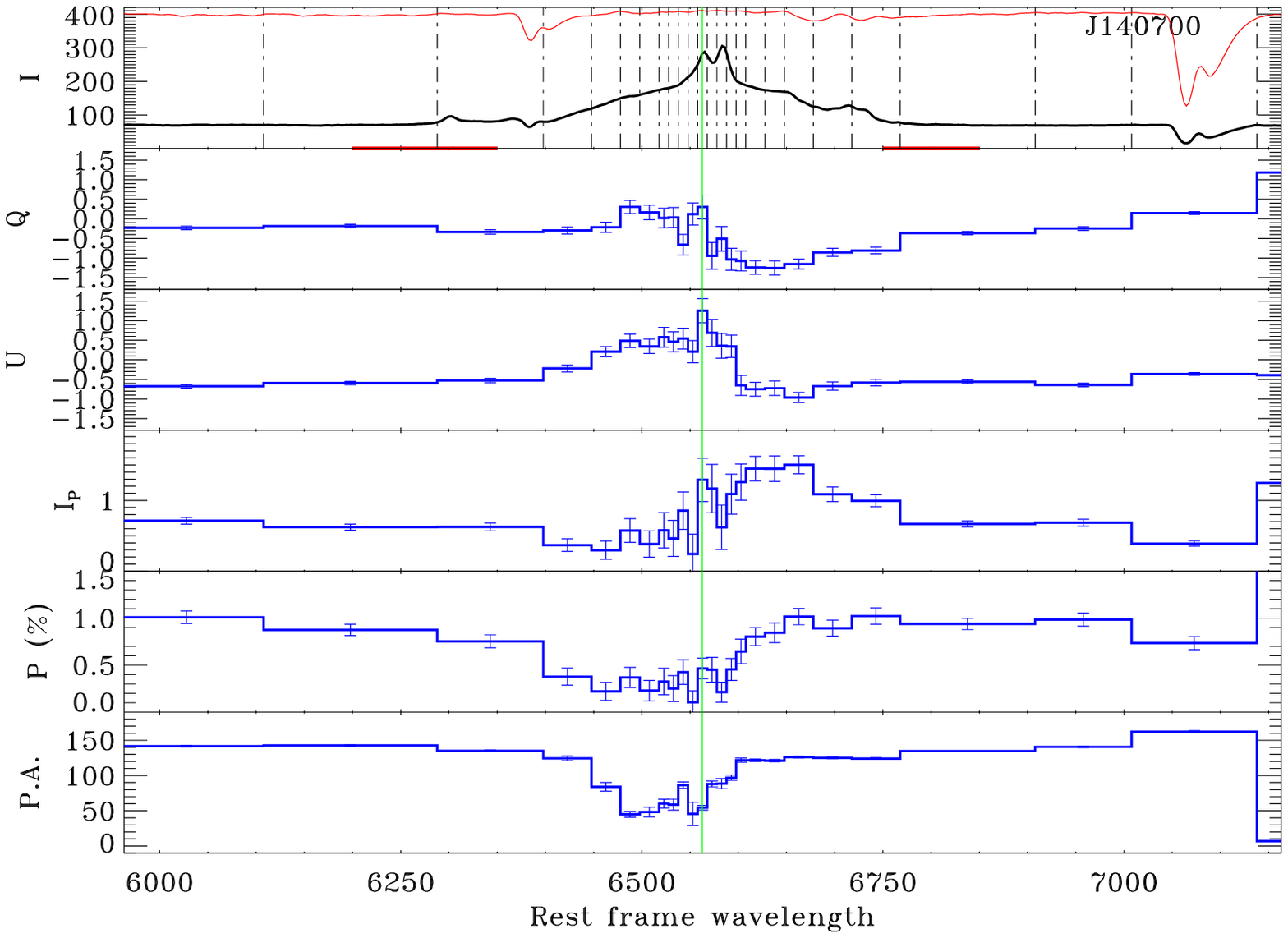}
\includegraphics[width=0.49\textwidth]{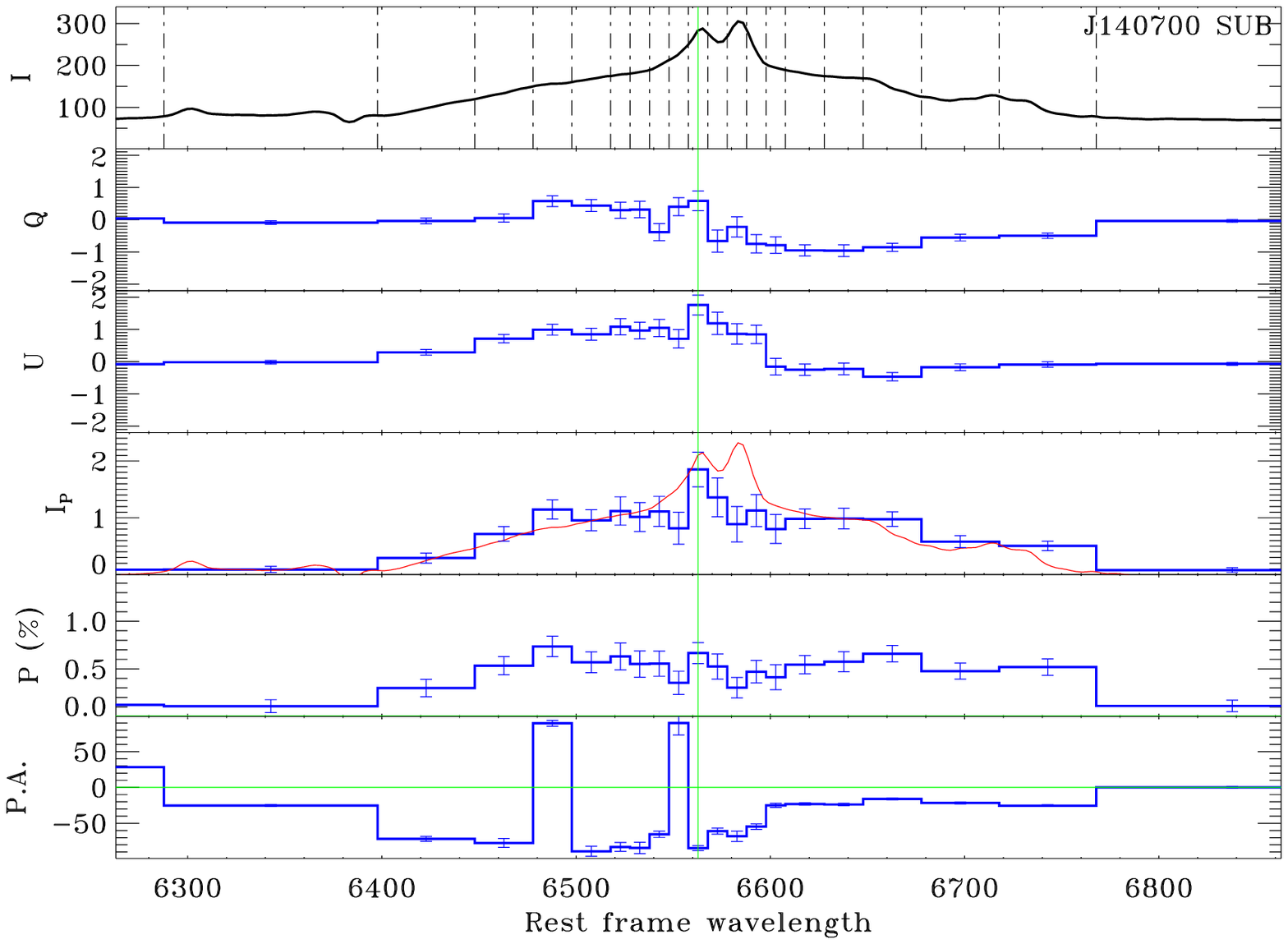}
\bigskip
\includegraphics[width=0.49\textwidth]{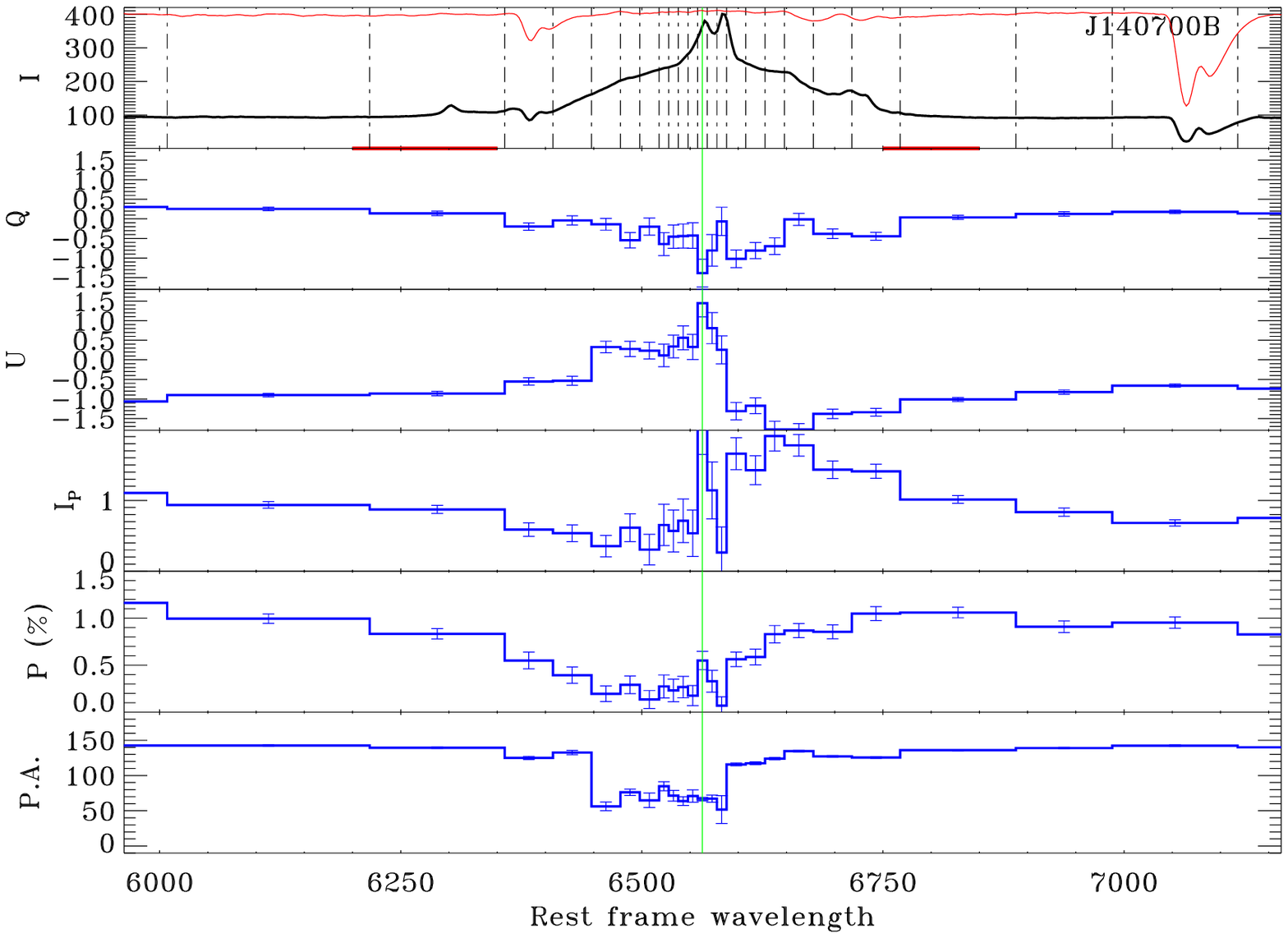}
\includegraphics[width=0.49\textwidth]{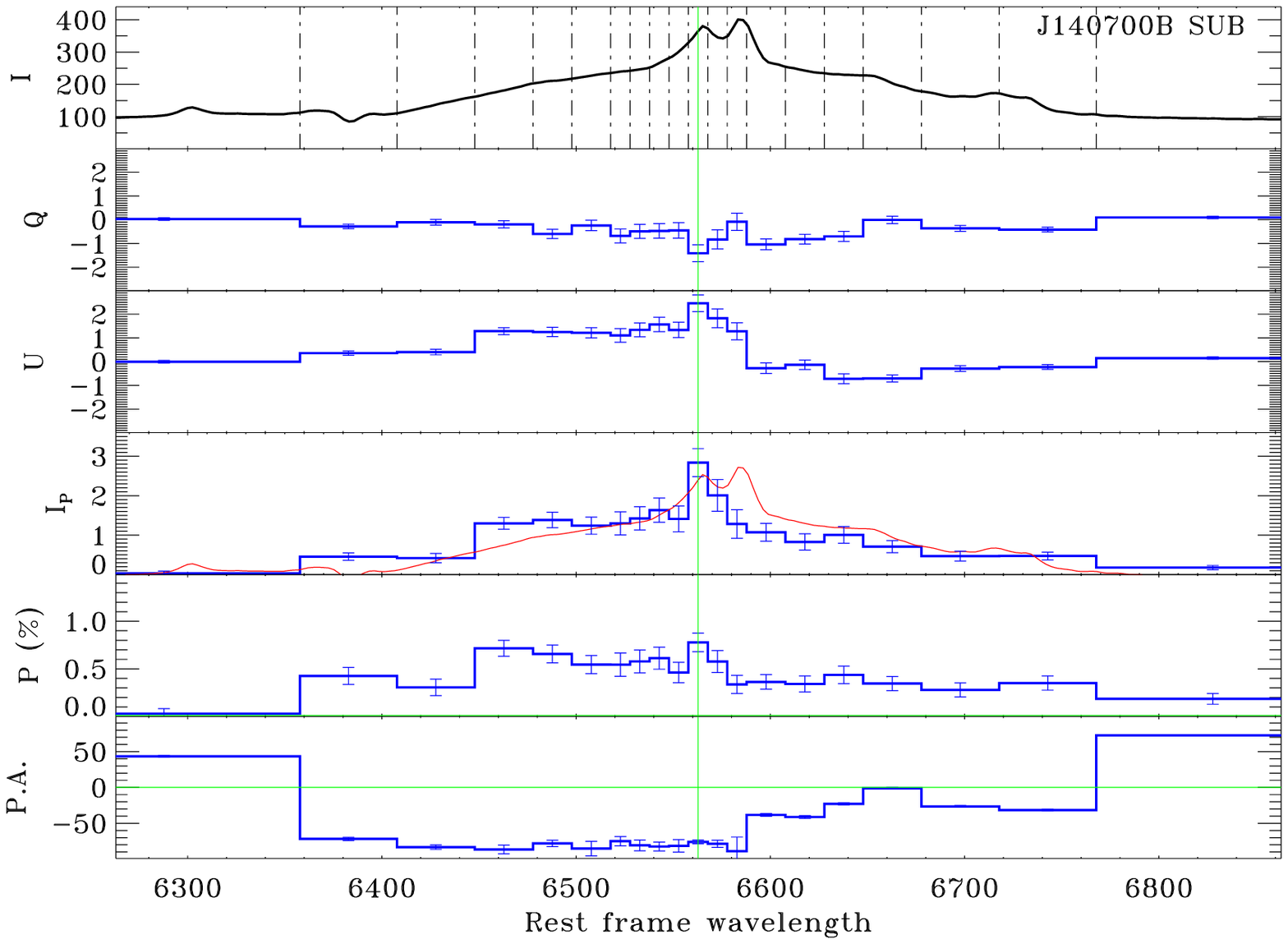}
\bigskip
\includegraphics[width=0.49\textwidth]{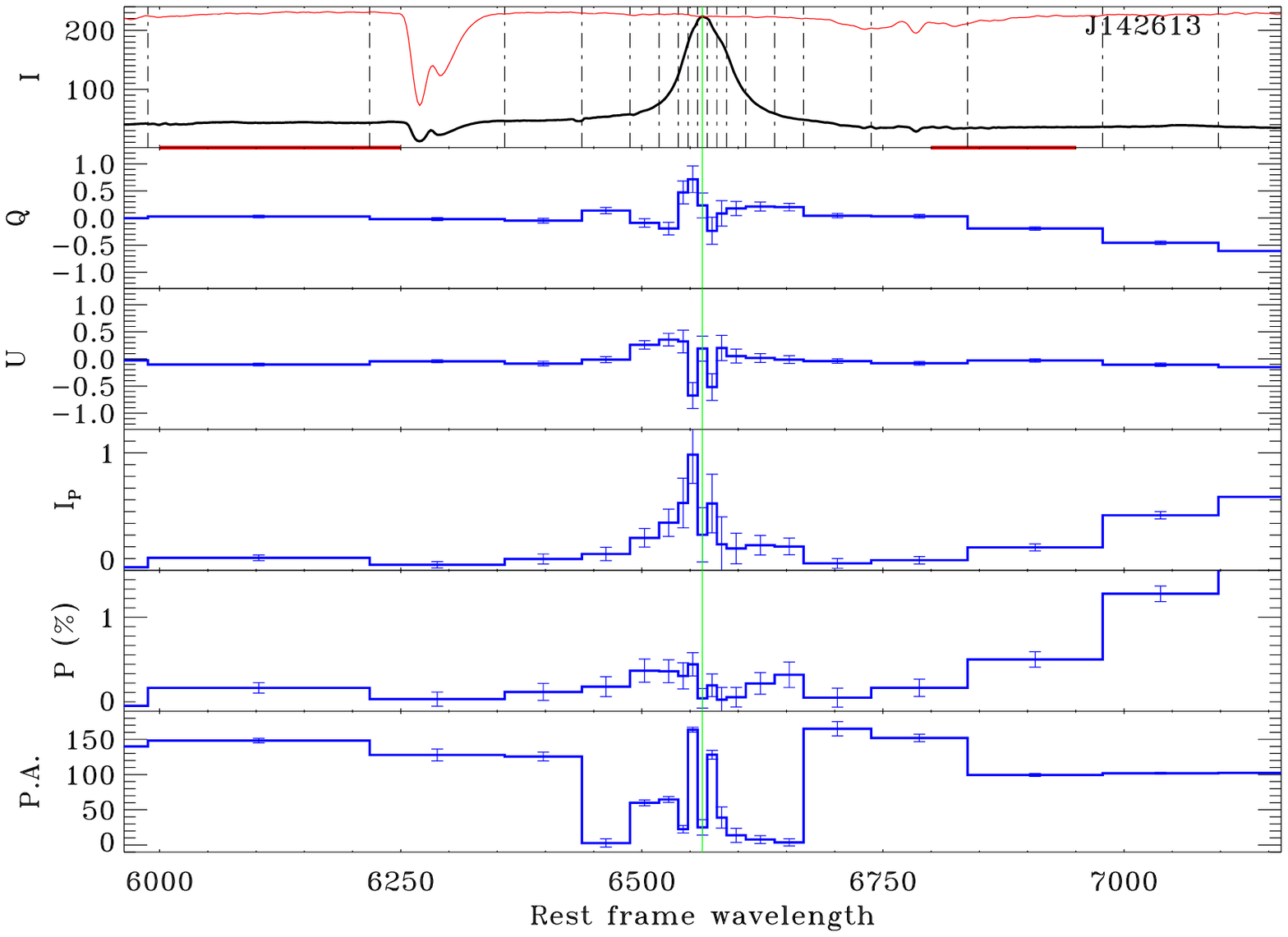}
\includegraphics[width=0.49\textwidth]{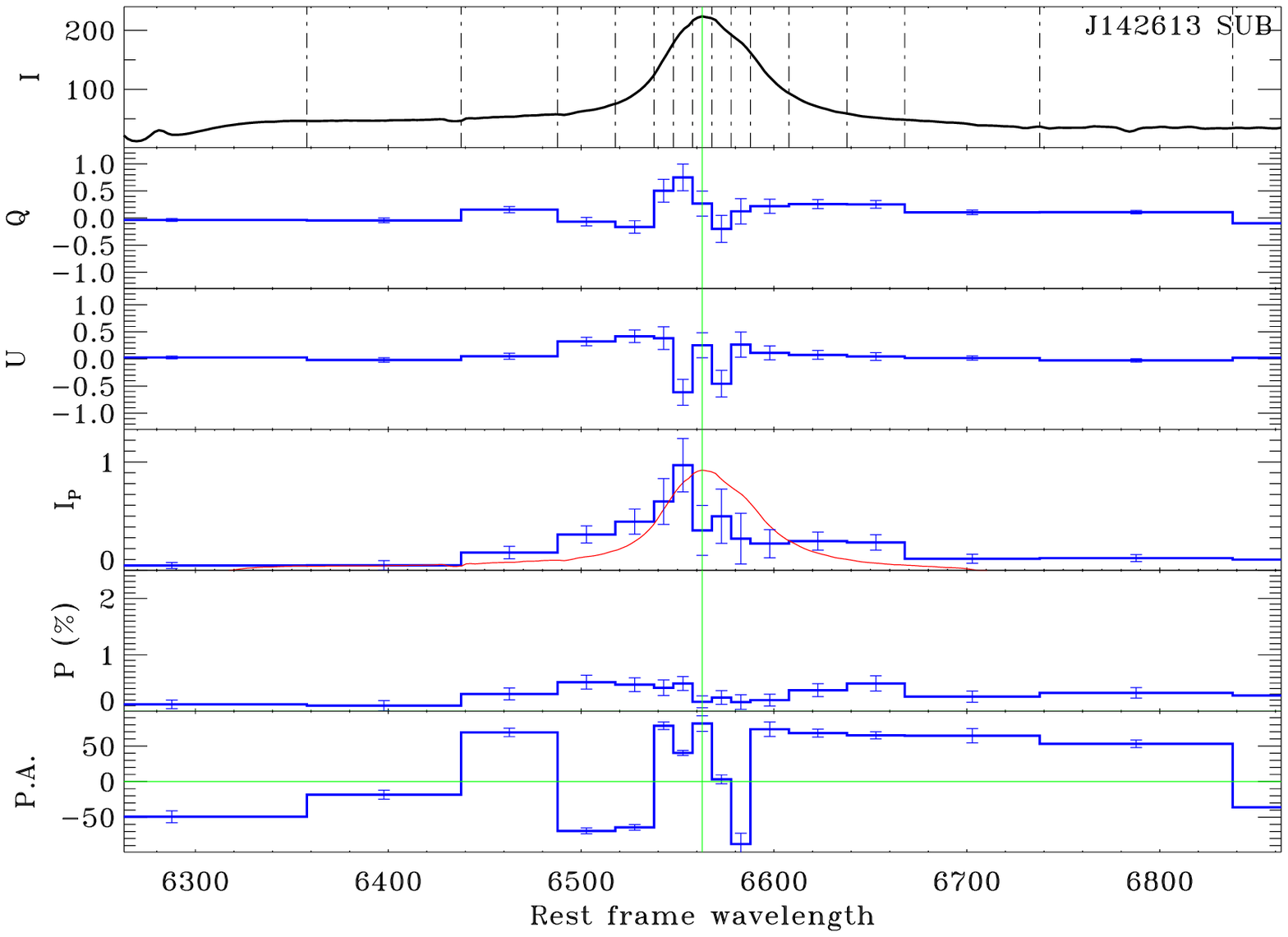}
\end{figure*}

\begin{figure*}
\includegraphics[width=0.49\textwidth]{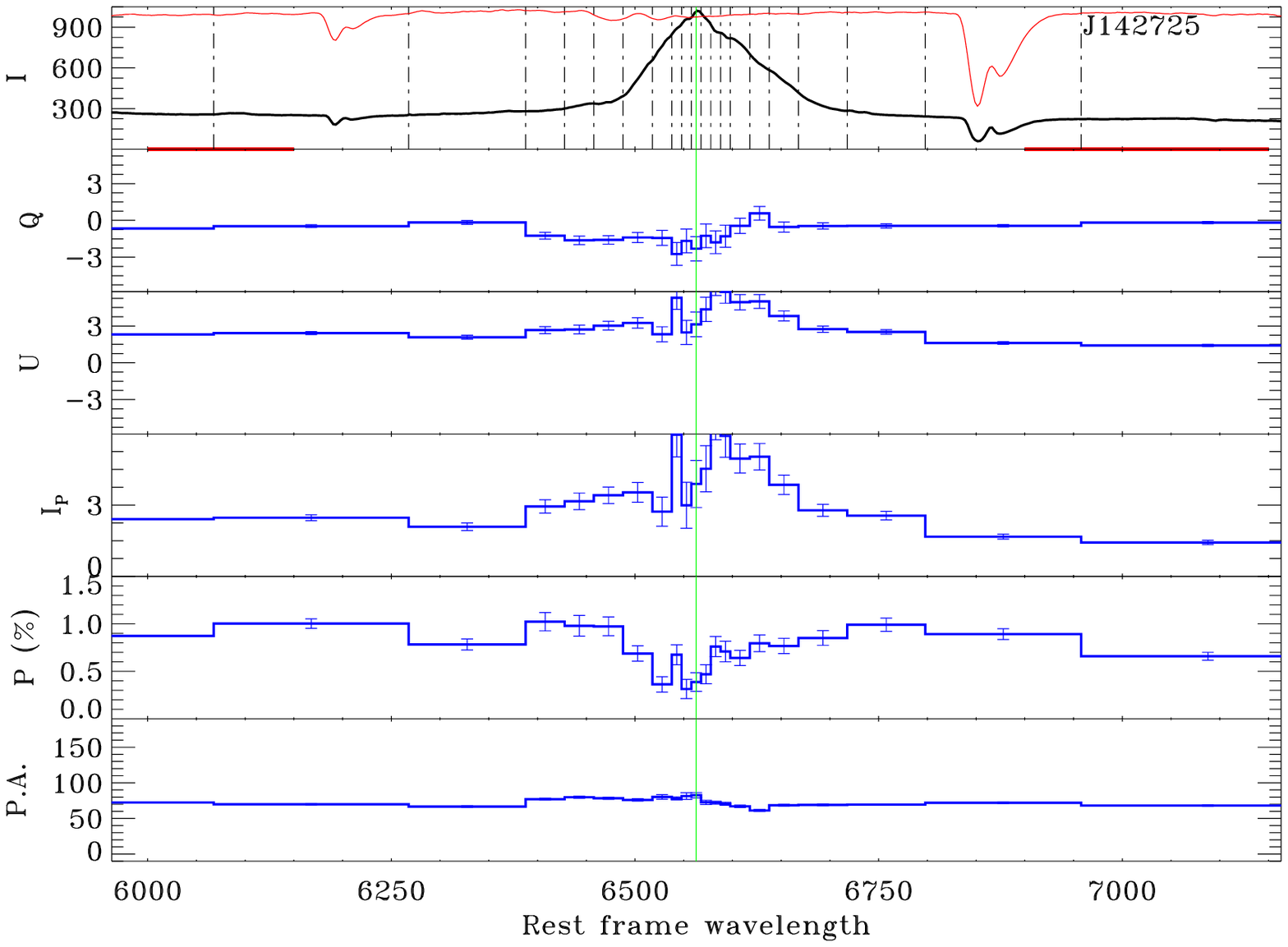}
\includegraphics[width=0.49\textwidth]{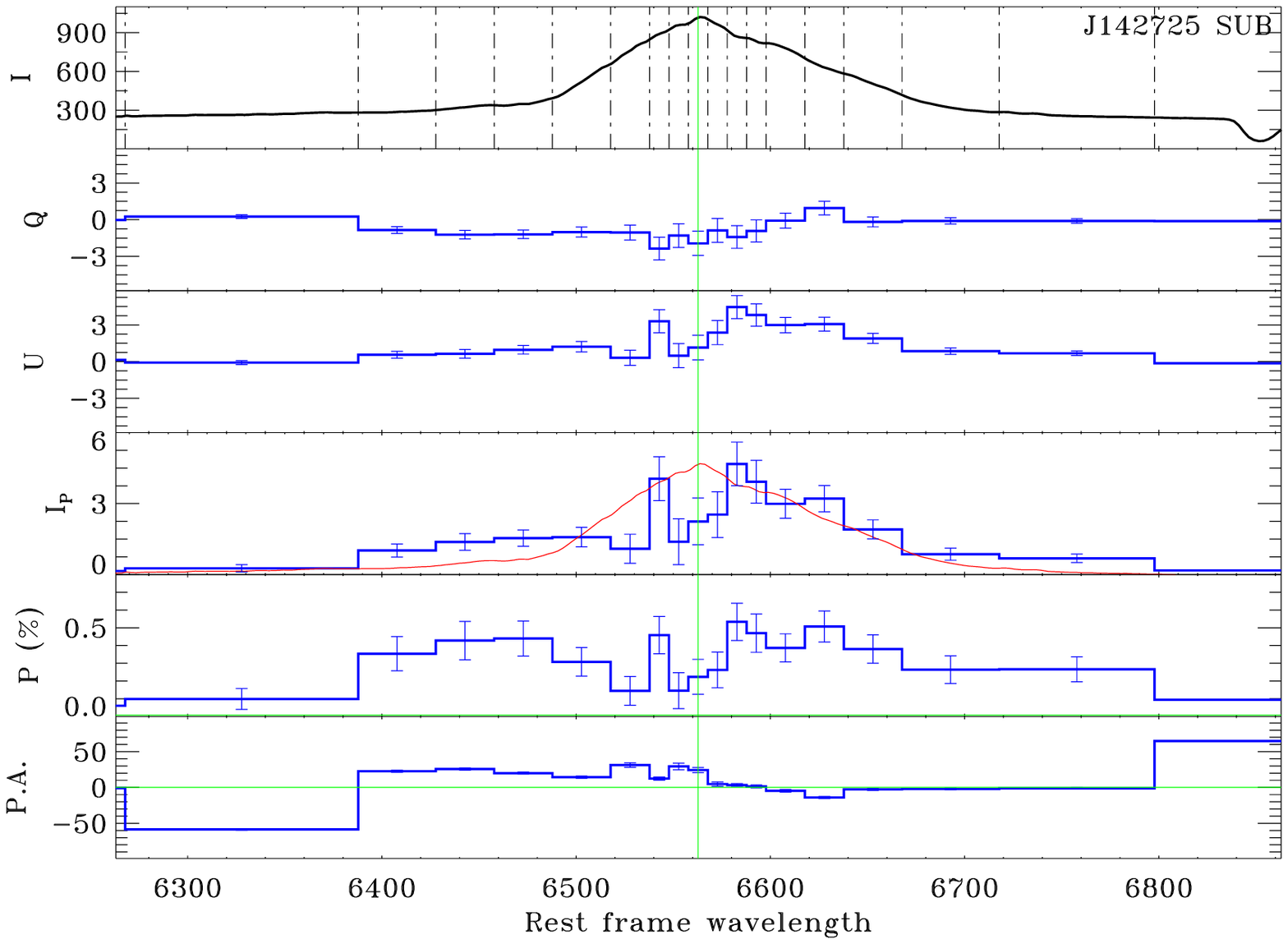}
\bigskip
\includegraphics[width=0.49\textwidth]{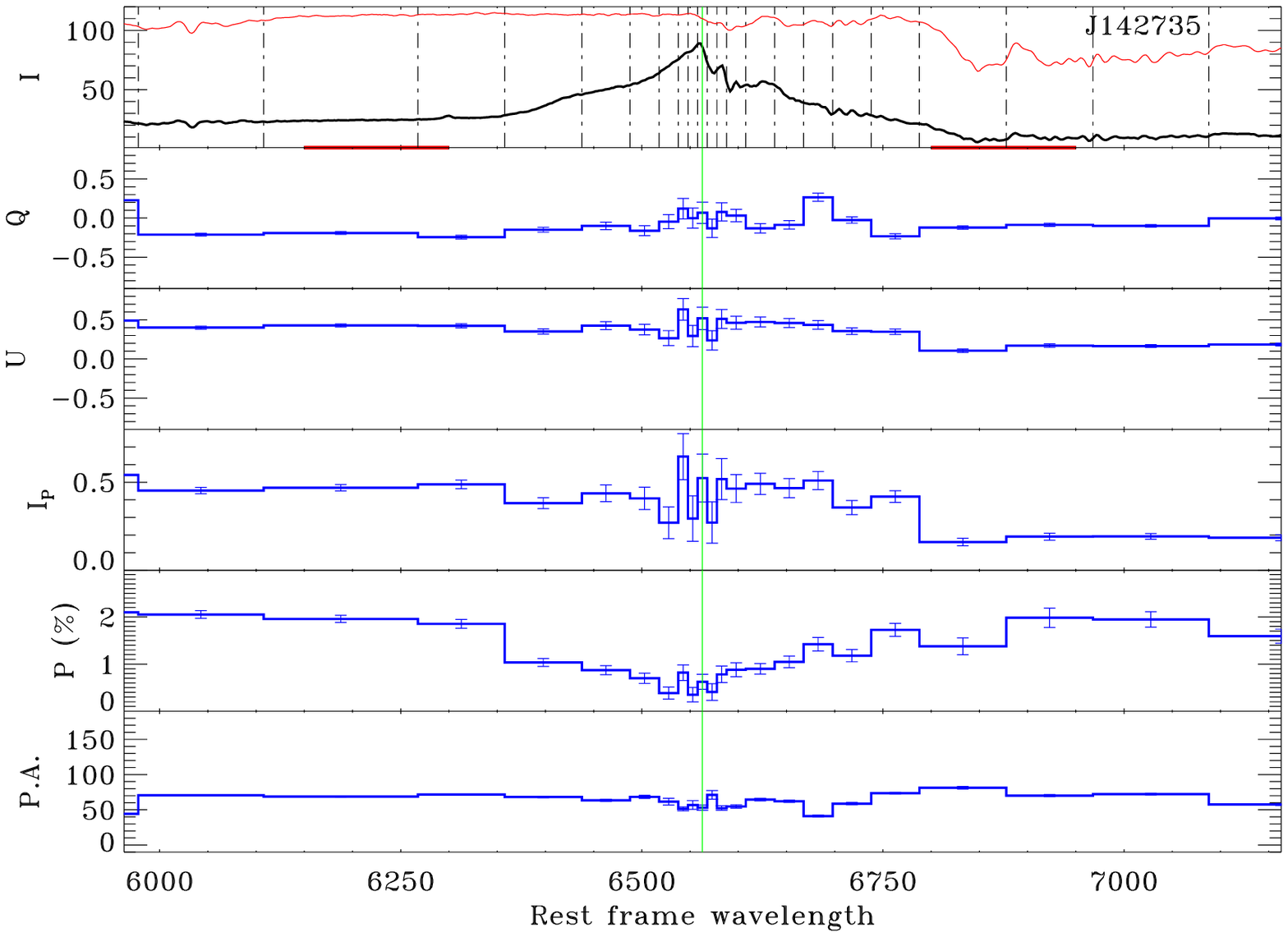}
\includegraphics[width=0.49\textwidth]{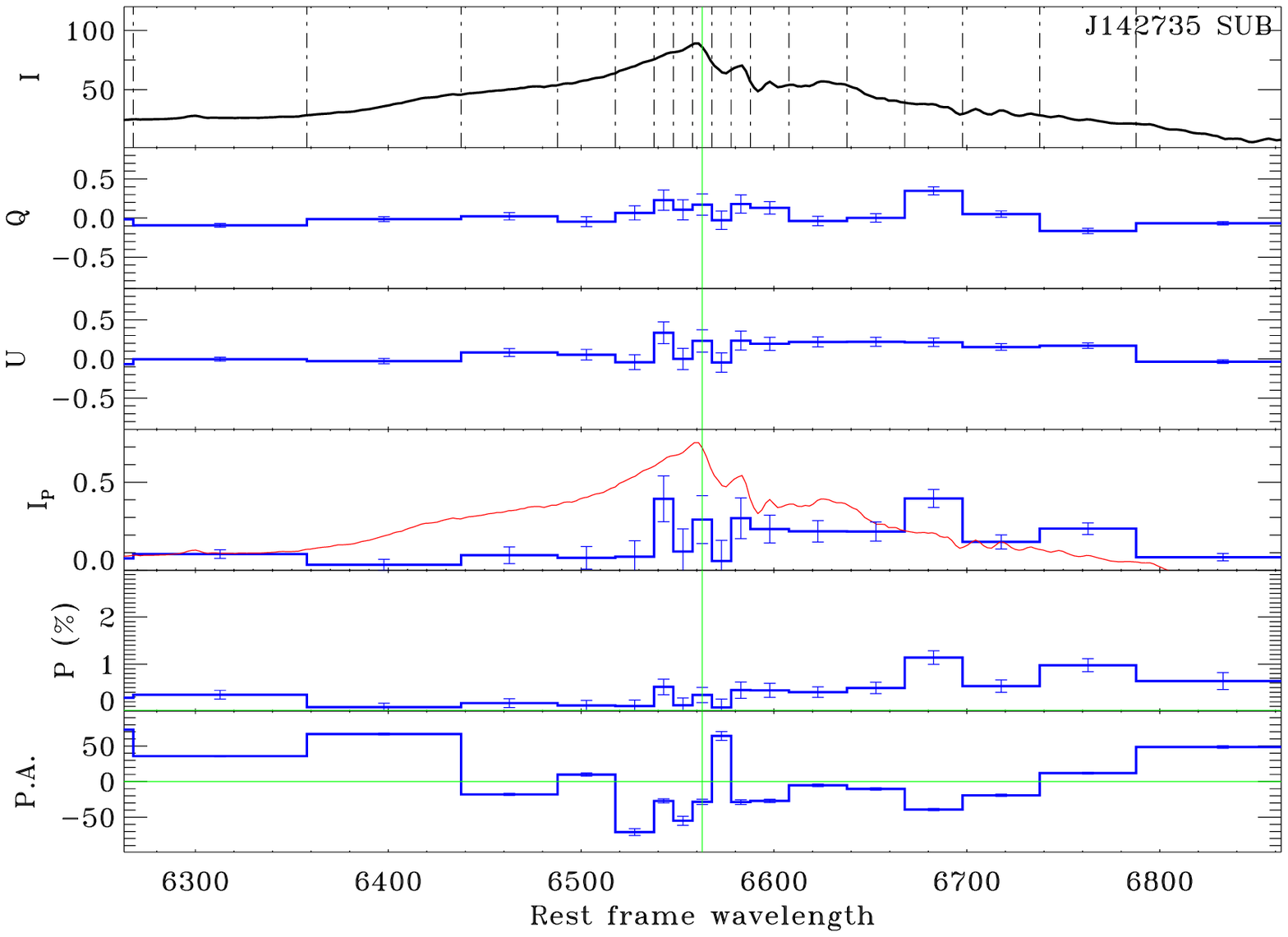}
\bigskip
\includegraphics[width=0.49\textwidth]{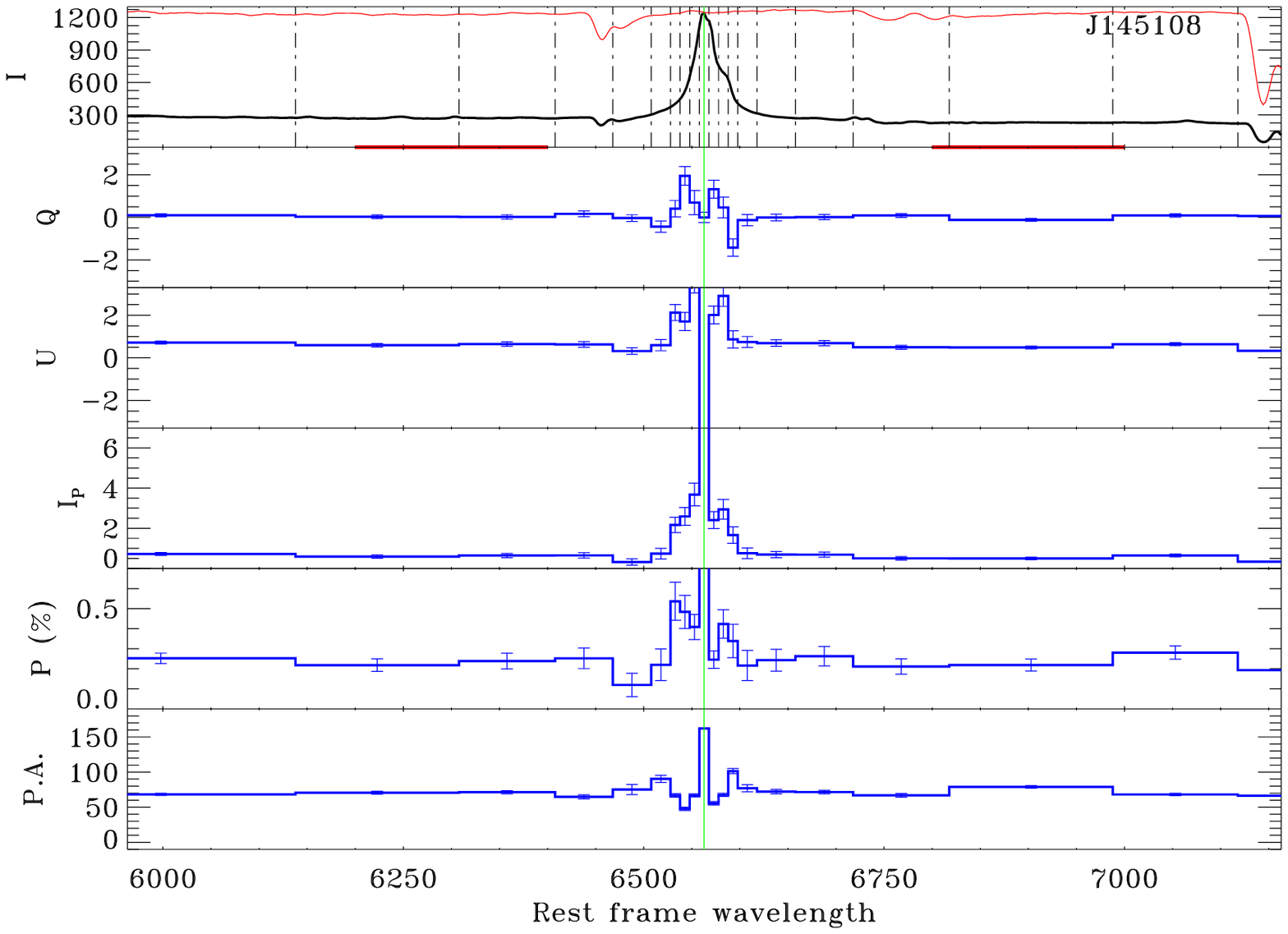}
\includegraphics[width=0.49\textwidth]{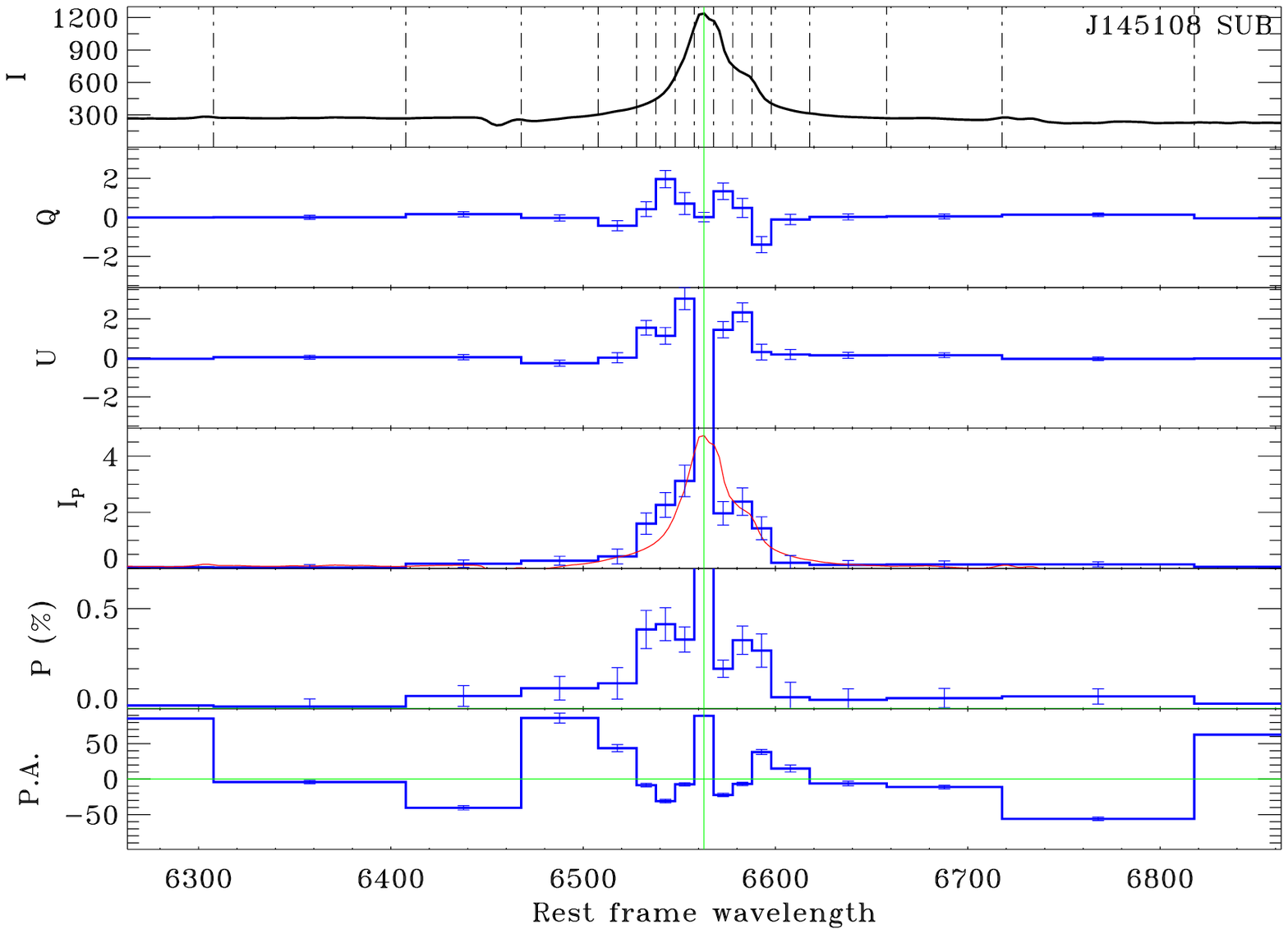}
\end{figure*}

\begin{figure*}
\includegraphics[width=0.49\textwidth]{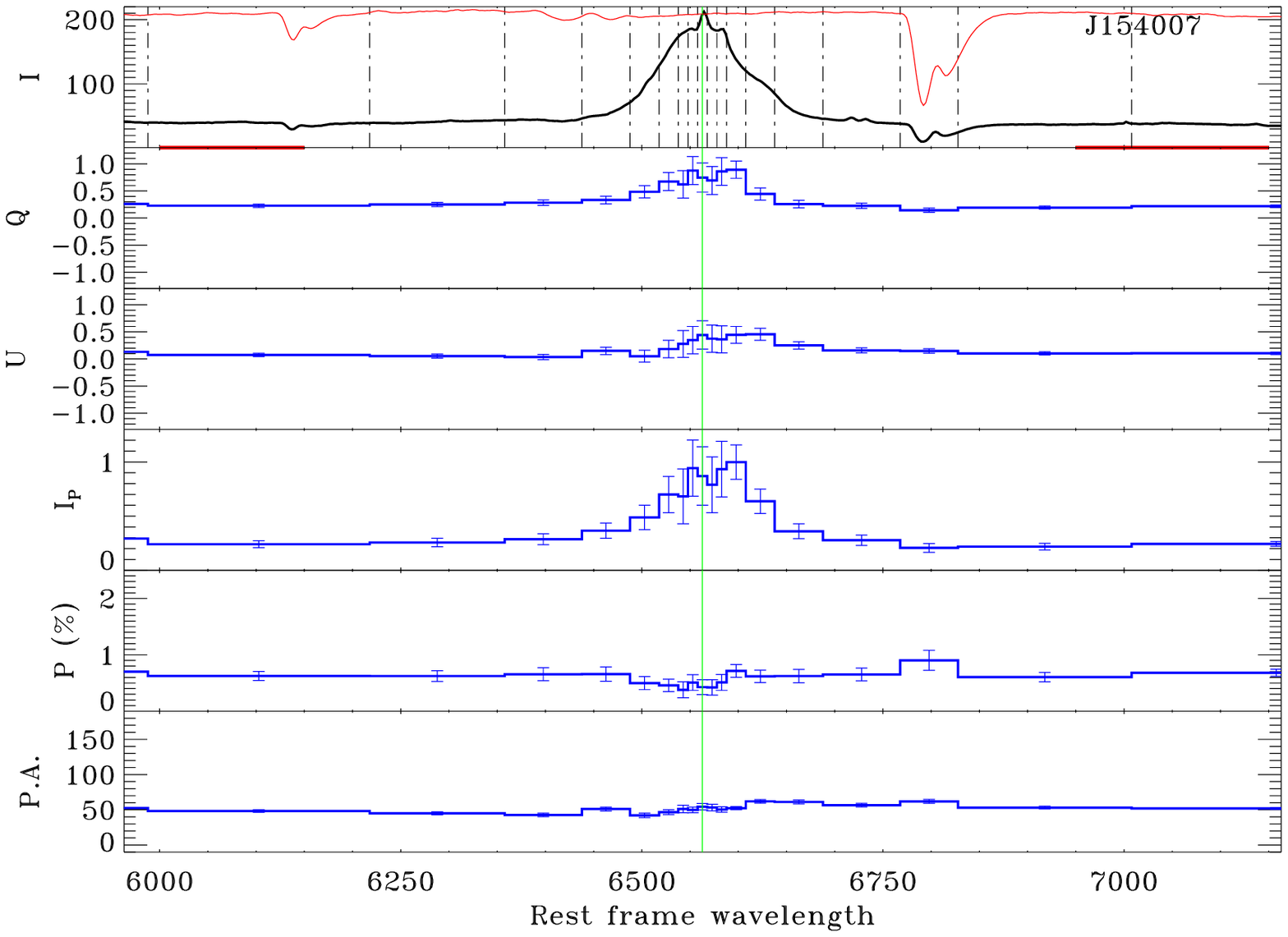}
\includegraphics[width=0.49\textwidth]{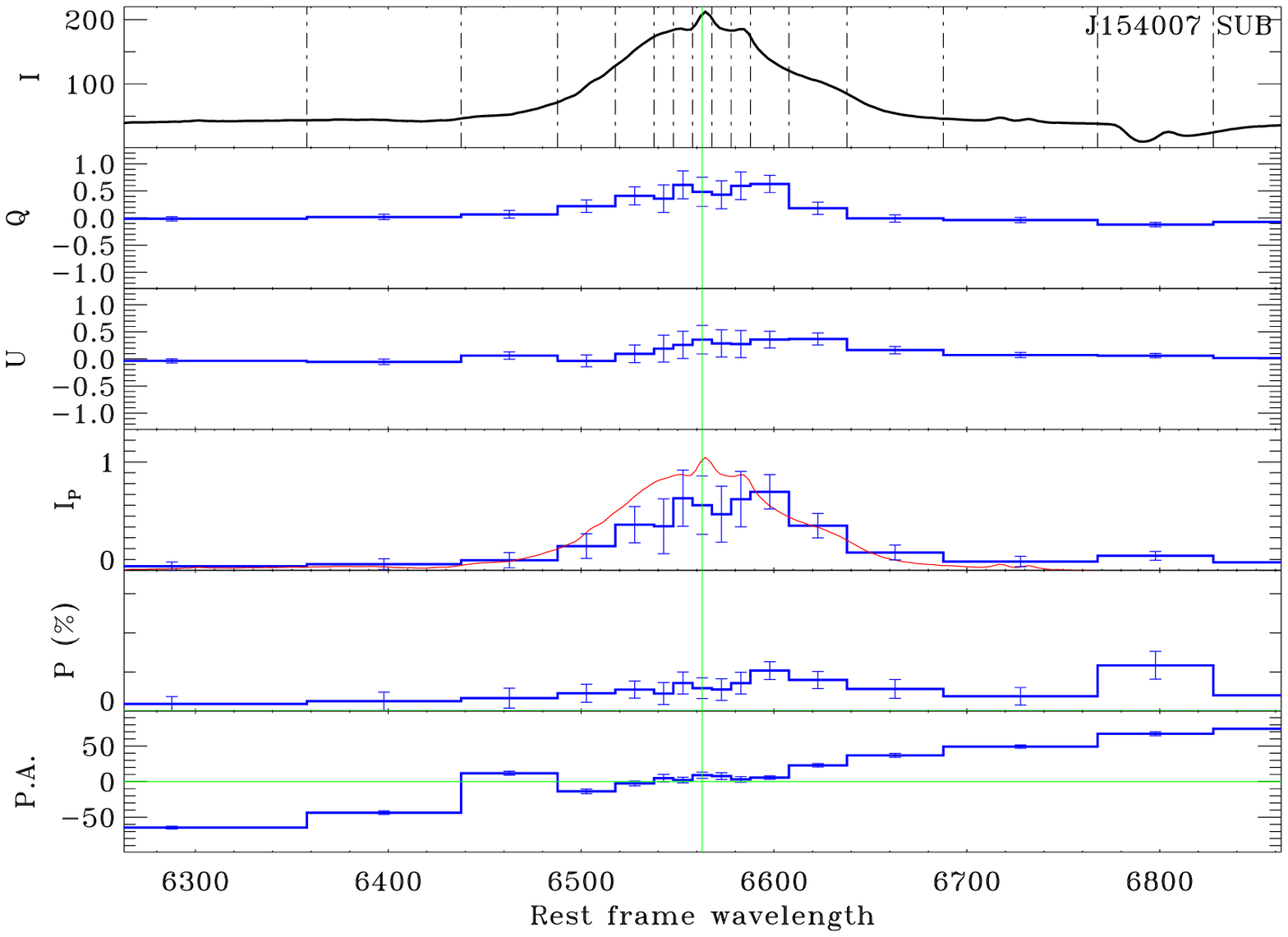}
\bigskip
\includegraphics[width=0.49\textwidth]{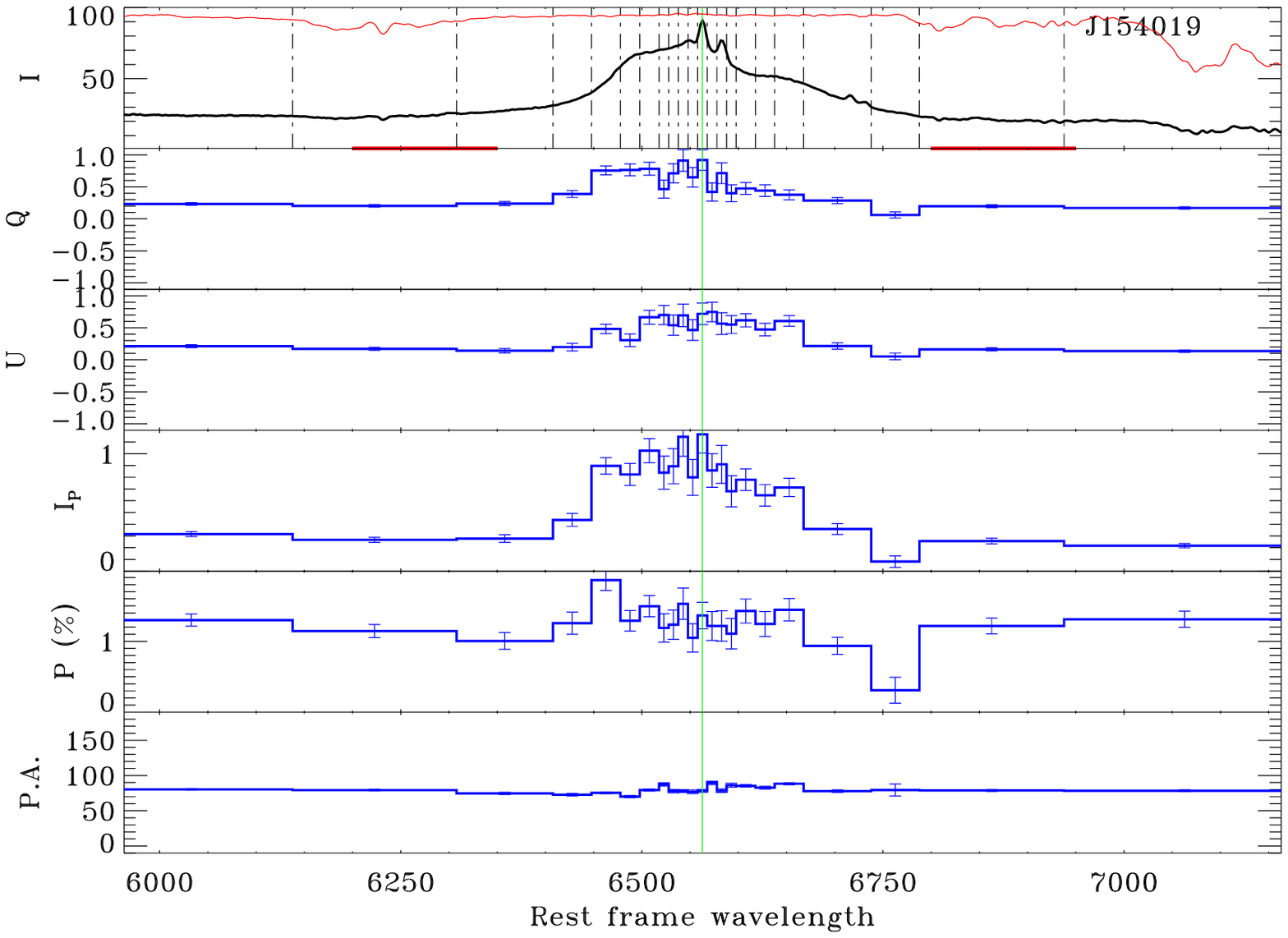}
\includegraphics[width=0.49\textwidth]{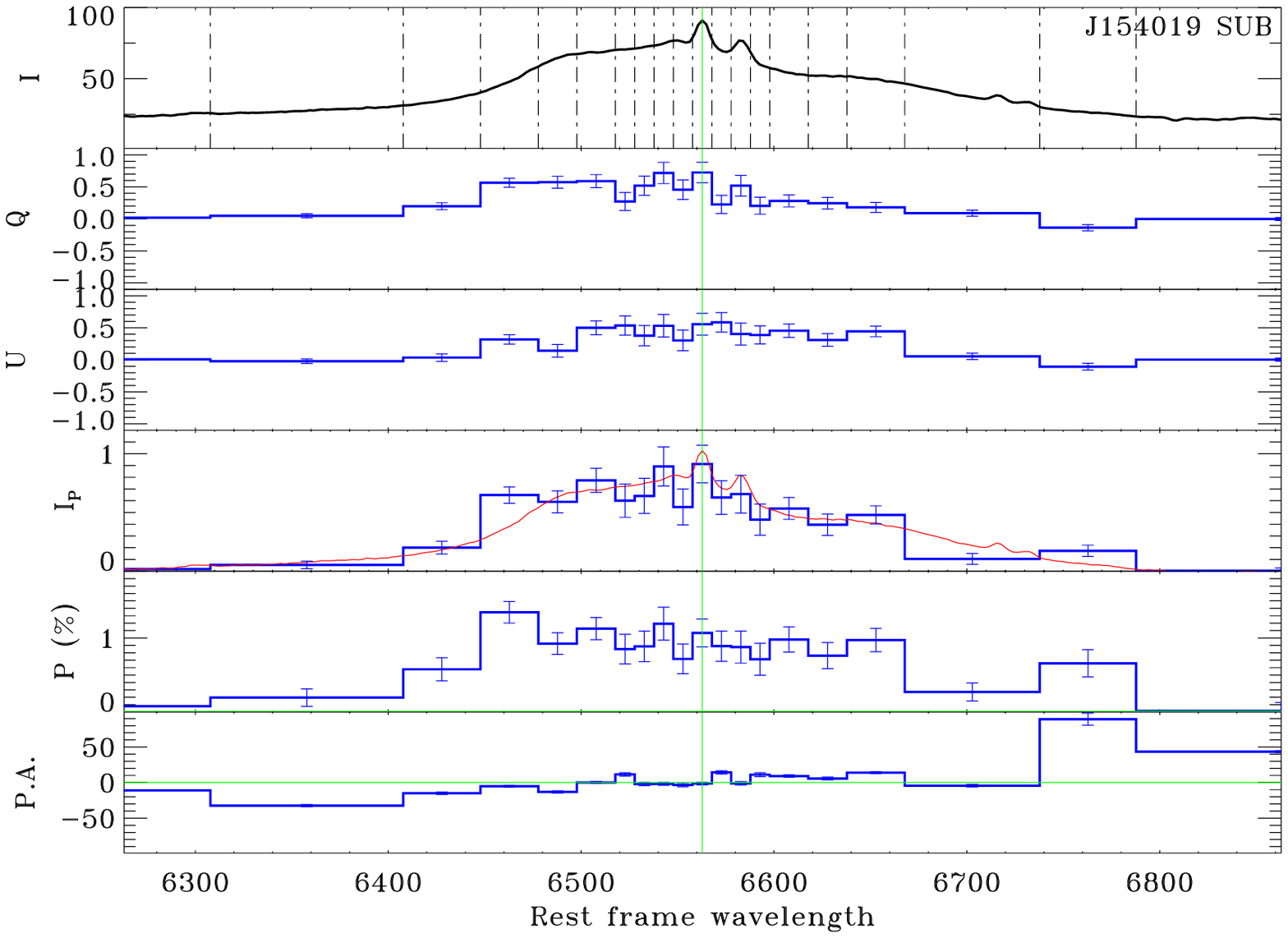}
\bigskip
\includegraphics[width=0.49\textwidth]{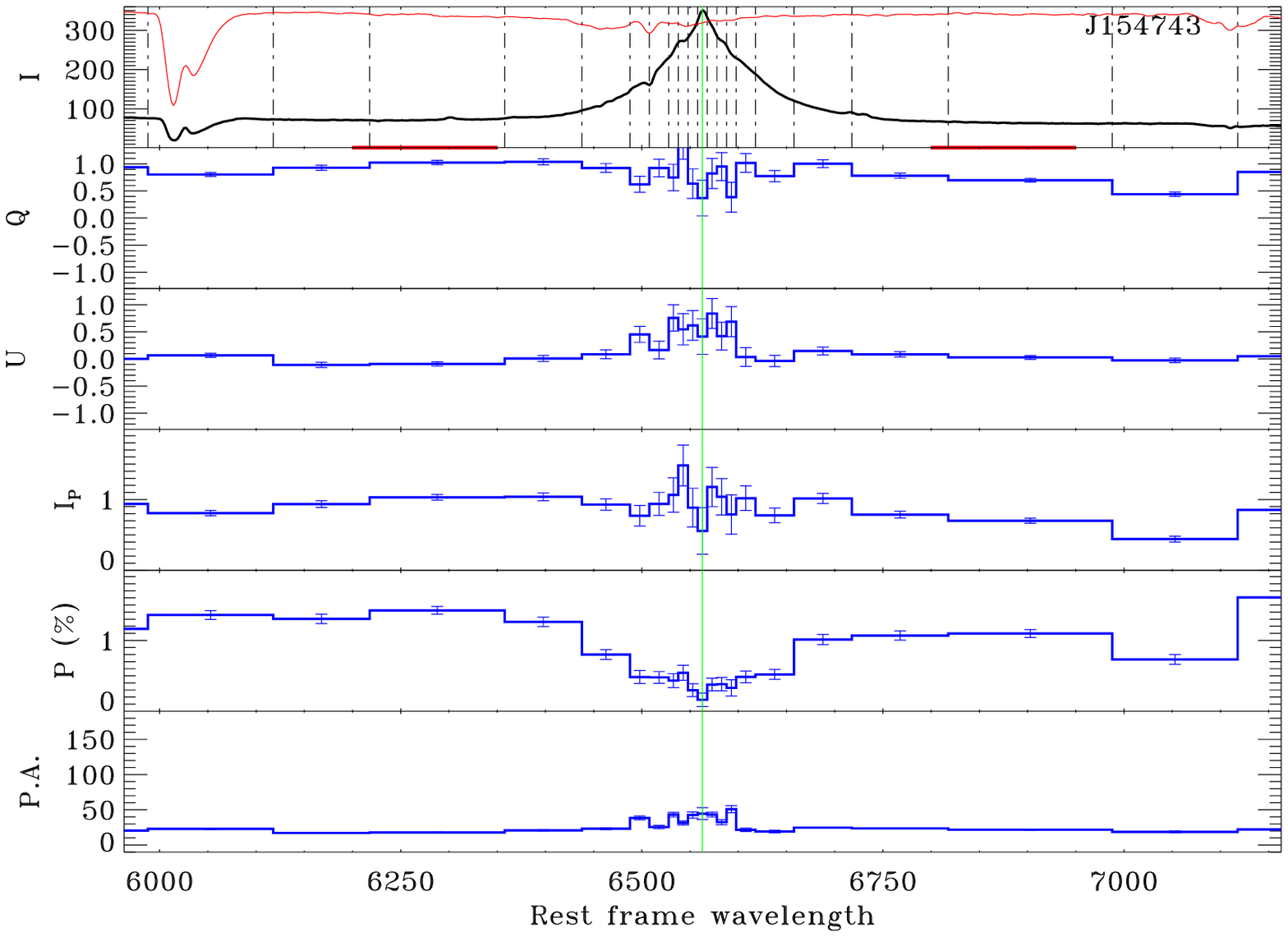}
\includegraphics[width=0.49\textwidth]{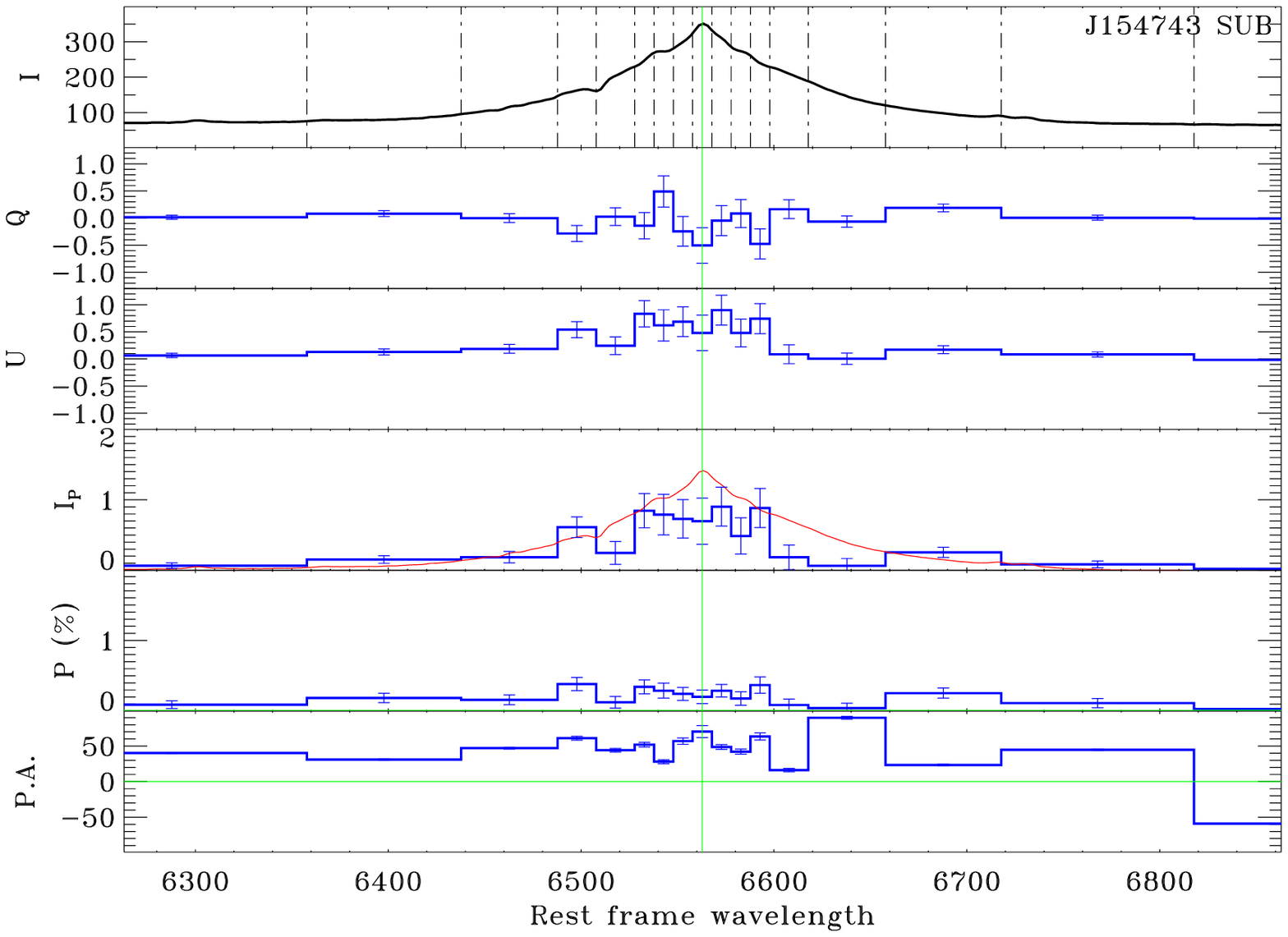}
\end{figure*}

\begin{figure*}
\includegraphics[width=0.49\textwidth]{J155444.ps}
\includegraphics[width=0.49\textwidth]{J155444cont.ps}
\bigskip
\includegraphics[width=0.49\textwidth]{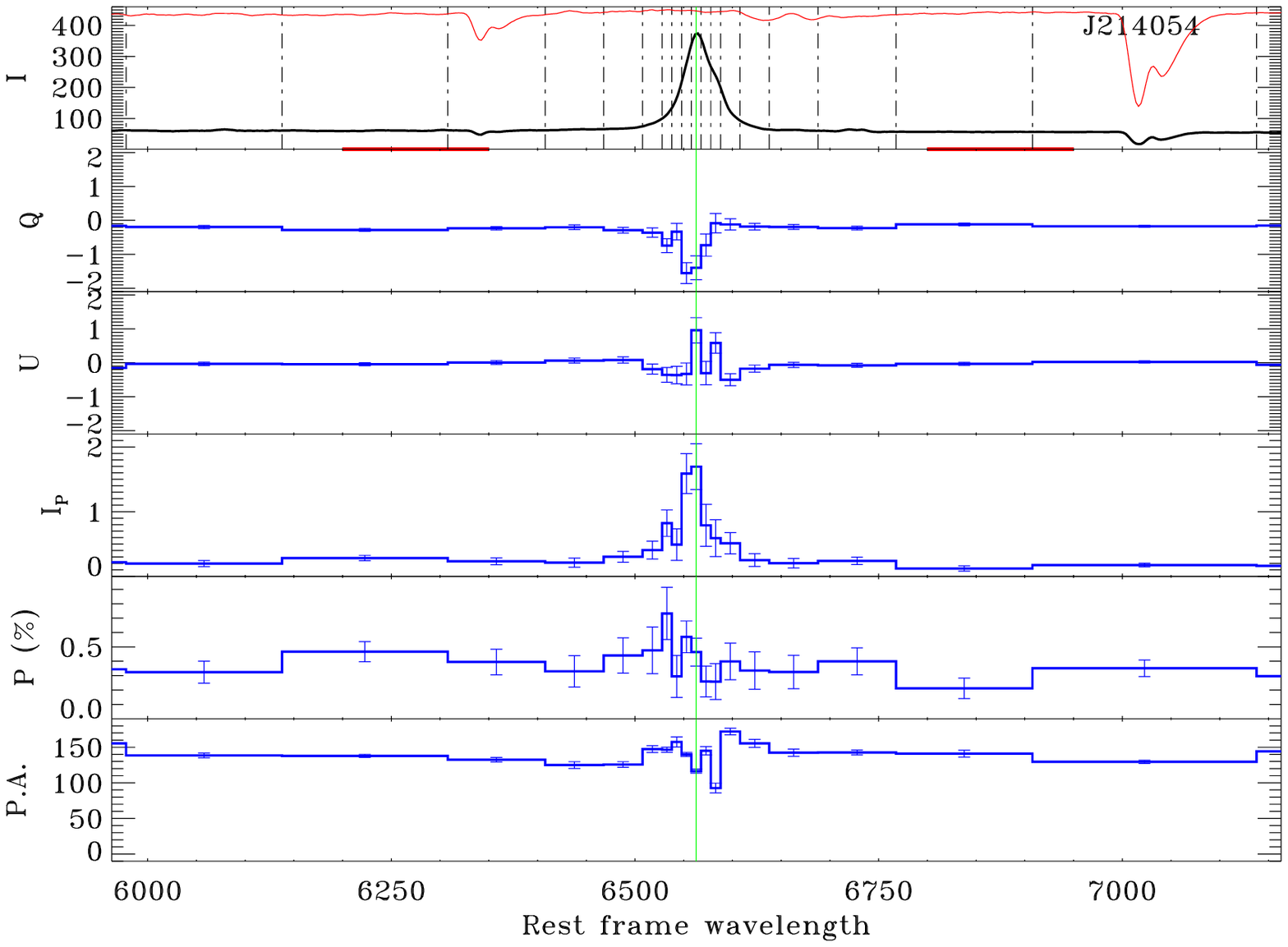}
\includegraphics[width=0.49\textwidth]{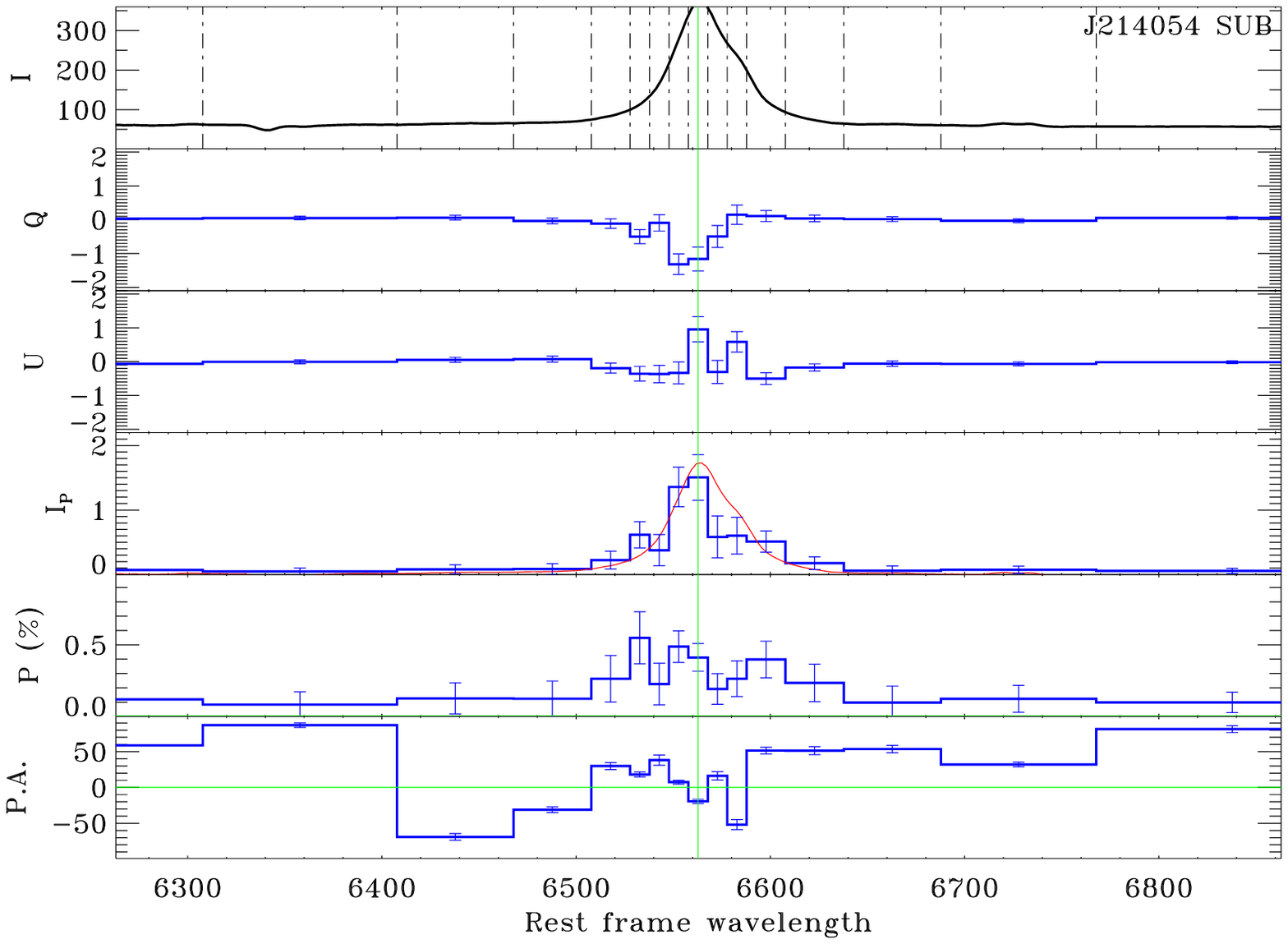}
\bigskip
\includegraphics[width=0.49\textwidth]{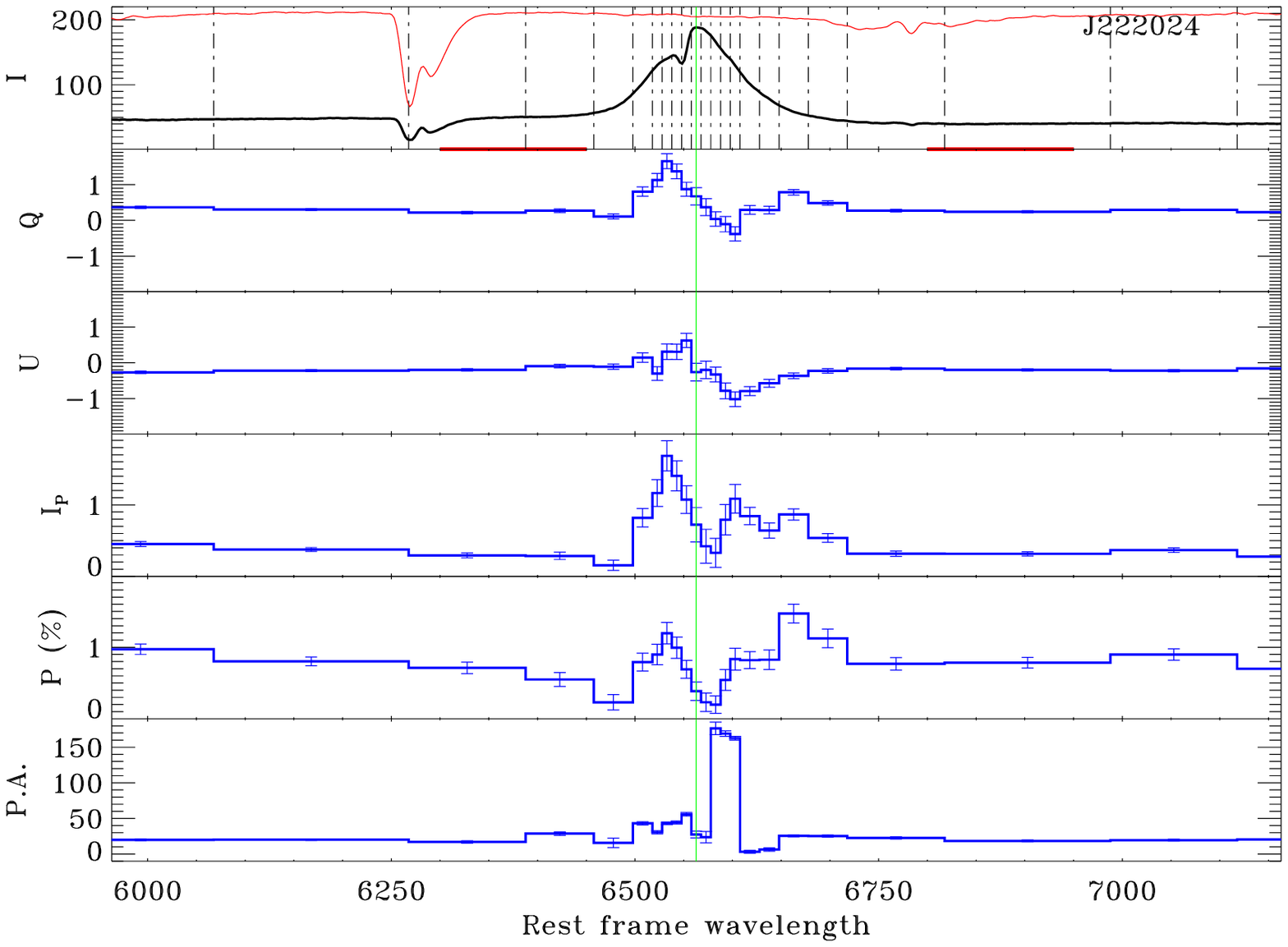}
\includegraphics[width=0.49\textwidth]{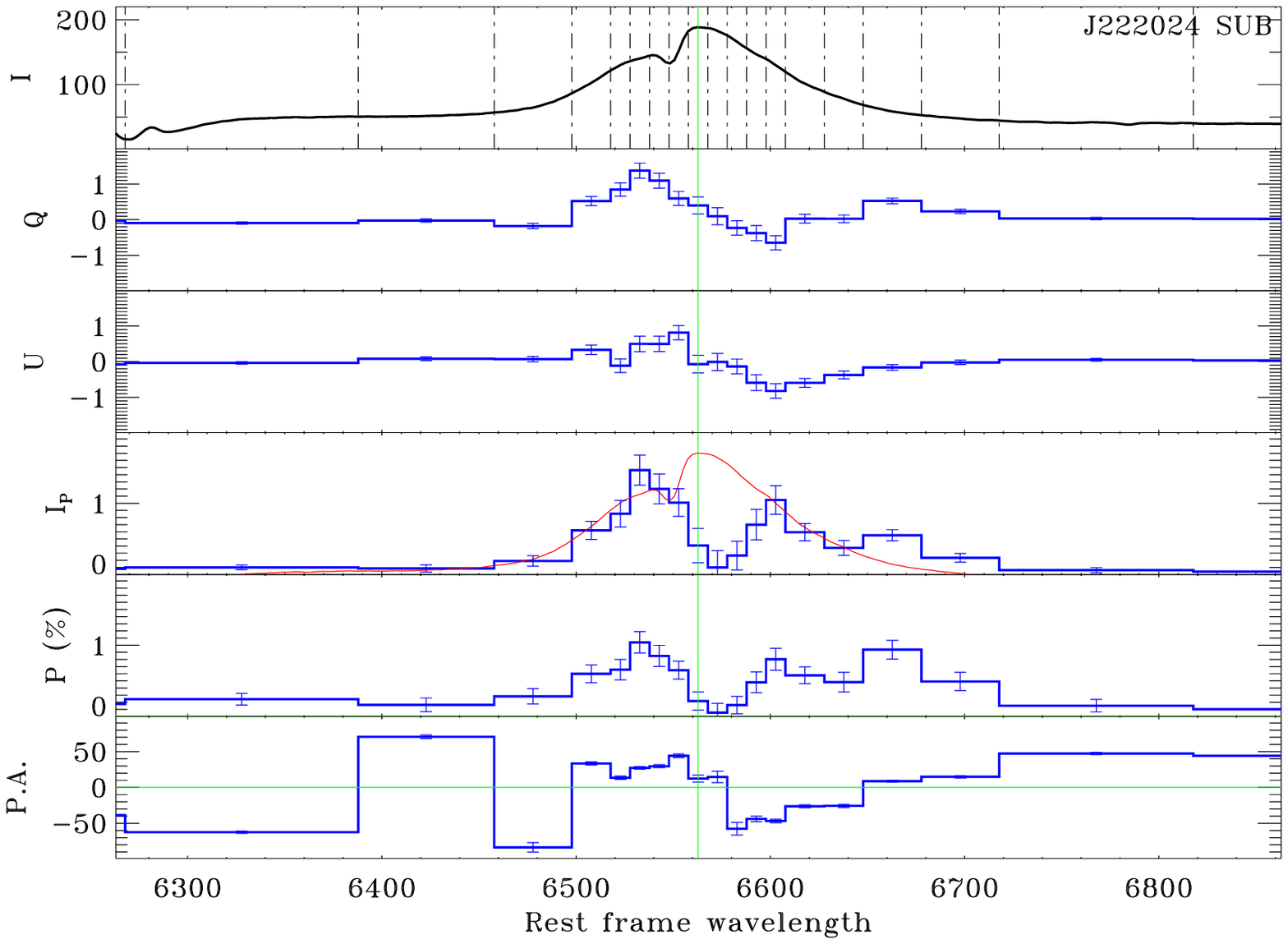}
\end{figure*}

\end{document}